\documentclass{jfm}

\usepackage{siunitx}[=v2] 
\usepackage{upgreek} 
\usepackage{mathtools} 
\usepackage{bigints} 
\usepackage{textcomp, gensymb} 

\usepackage{cleveref}

\DeclareSIUnit{\wtpercent}{wt.\%}
\DeclareSIUnit{\volpercent}{vol.\%}
\DeclareSIUnit\atm{atm}

\usepackage{subcaption}

\usepackage{xcolor}
\usepackage[normalem]{ulem}

\begin{document}

\newtheorem{lemma}{Lemma}
\newtheorem{corollary}{Corollary}

\shorttitle{Dynamics of interacting evaporating droplets on compliant substrates} 
\shortauthor{A. Malachtari and G. Karapetsas} 

\title{Dynamics of the interaction of a pair of thin evaporating droplets on compliant substrates}

\author
 {
  A. Malachtari
  \and
  G. Karapetsas
  \corresp{\email{gkarapetsas@auth.gr}}
  }

\affiliation
{
Department of Chemical Engineering, Aristotle University of Thessaloniki, Thessaloniki 54124, Greece
}

\maketitle

\begin{abstract}
The dynamics of the interaction of a system of two thin volatile liquid droplets resting on a soft viscoelastic solid substrate are investigated theoretically. The developed model fully considers the effect of evaporative cooling and the generated Marangoni stresses due to the induced thermal gradients, while also accounting for the effect of the gas phase composition and the diffusion of vapour in the atmosphere of the droplets. Using the framework of lubrication theory, we derive evolution equations for both the droplet profile and the displacement of the elastic solid, which are solved in combination with Laplace's equation for the vapour concentration in the gas phase.  A disjoining-pressure/precursor-film approach is used to describe contact-line motion. The evolution equations are solved numerically, using the finite-element method, and we present a thorough parametric analysis to investigate the physical properties and mechanisms that affect the dynamics of droplet interactions. The results show that the droplets interact through both the soft substrate and the gas phase. In the absence of thermocapillary phenomena, the combined effect of non-uniform evaporation due to the increased vapour concentration between the two droplets and elastocapillary phenonema determines whether the drop-drop interaction is attractive or repulsive. The Marangoni stresses suppress droplet attraction at the early stages of the drying process and lead to longer droplet lifetimes. For substrates with intermediate stiffness, the emergence of spontaneous symmetry breaking at late stages of evaporation is found. The rich dynamics of this complex system is explored by constructing a detailed map of the dynamic regimes.

\end{abstract}


\section{Introduction}
The dynamic wetting behavior of liquid droplets on soft solid substrates has received a great deal of attention lately, due to its relevance in diverse applications \citep{Bico_elastocapillarity_2018} ranging from biology, i.e. the inhibition of the dispersal of cancer cells \citep{douezan_wetting_2012} or the control of medicine dispersal on tissues, to the control of the spreading of the deposited particles over a compliant substrate after the evaporation of ink-jetted microdroplets \citep{park_control_2006} and microfabrication of materials in technology \citep{kong_3d_2014,bonaccurso_fabrication_2005,pericet-camara_arrays_2007}.

The evaporation of droplets on rigid substrates has been widely studied over the years, underlining various aspects of evaporation such as droplet
lifetimes \citep{stauber_et_al_2014,stauber_et_al_2015}, the impact of capillary flow on the coffee stain effect \citep{deegan_capillary_1997}, or the effect of substrate properties \citep{erbil_evaporation_2012}. A key concept in droplet evaporation is the so-called shielding effect, where neighboring droplets interact with each other, since the presence of the vapour from adjacent droplets reduces the evaporation rate and increases the droplet lifetime in comparison to those of a single isolated droplet \citep{fabrikant_membranes_1985, Wray_et_al_2020,wray_contactline_2021,masoud_evaporation_2021}. 
On the contrary, the study of droplet evaporation on compliant solid substrates is insubstantial so far.

The unbalanced vertical forces acting on the contact line of the liquid result in a local deformation of the compliant substrate affecting droplet shape and, ultimately, the dynamics of the flow \citep{Andreotti_wetting_2020}. This unique attribute of the compliant substrates is responsible for the creation of a macroscopic surface protrusion, most widely referred to as a wetting ridge \citep{shanahan_spreading_1988, park_visualization_2014, Chen_et_al_2020}. The structure of the wetting ridge is an immediate outcome of the balance between capillary and elastic forces, while a key role of the solid surface tension has been recently identified \citep{jerison_deformation_2011,style_static_2012,marchand_capillary_2012,style_universal_2013}. Besides slowing down the contact line, the wetting ridge can also constitute the reason for periodic stick–slip behavior, with a periodic depinning of the contact line \citep{kajiya_advancing_2012, lopes_influence_2013,yu_experimental_2013,kajiya_liquid_2014,
karpitschka_liquid_2016,van_gorcum_dynamic_2018,
mokbel_stick-slip_2022}. Starting from a rigid substrate, when we increase the substrate softness we observe an initial strong increase in the wetting ridge size while the droplet footprint is kept relatively constant, which is replaced by a strong depression of the substrate under the droplet in much softer substrates \citep{charitatos_soft_2020,henkel_gradient-dynamics_2021}. The arising solid angle is governed by a balance of surface tensions \citep{style_static_2012,marchand_capillary_2012,style_universal_2013}. The possible strain dependence of the solid surface tension (i.e., the Shuttleworth effect) \citep{andreotti_soft_2016,xu_direct_2017,pandey_singular_2020,Henkel_Shuttleworth_2022} gives rise to another complexity.

Depending on a combination of capillarity and bulk elasticity, adjacent droplets on solid substrates can interact, leading either to droplet attraction and even coalescence on thick substrates, or to droplet repulsion on thinner substrates \citep{hernandez-sanchez_symmetric_2012, karpitschka_liquid_2016,leong_droplet_2020}. 
\cite{karpitschka_liquid_2016} described the droplet interaction when deposited on soft solids as the "inverse Cheerios effect" with direct reference to the liquid-on-solid analog of the so-called "Cheerios effect" \citep{vella_cheerios_2005}; the latter refers to solid particles attraction when floating on liquids, mediated by surface tension forces, and it has been named after the sticking of breakfast cereals either to the walls of a bowl or to each other. However, there are substantial differences between the "Cheerios effect" and the "inverted Cheerios effect" regarding the driving force and the mechanism which mediates the interaction. The shape of the liquid interface in the "Cheerios effect" is specified by the balance between surface tension and gravity, while the interaction is driven by a change in gravitational potential energy. On the contrary, in the "inverse Cheerios effect" there is no gravity involved and the shape of the solid interface is specified by elastocapillarity \citep{karpitschka_droplets_2015}.

Despite the innumerable studies concerning droplet coalescence \citep{eggers_coalescence_1999,aarts_hydrodynamics_2005}, not many of them consider the role that a compliant substrate might play in the process. More recently, \cite{henkel_gradient-dynamics_2021} investigated the two basic coarsening modes of two droplets on soft substrates without evaporation; the volume or mass transfer mode (also referred to as drop collapse mode, diffusion-controlled ripening or Ostwald ripening) and the translation mode (also referred to as coalescence, collision or migration mode). On the one hand, on the mass transfer mode, material is transferred from the small drop to the larger one until the smaller drop has completely vanished, while the centers of mass of each droplet remain constant. This mass transfer might occur either through the vapour phase in case of volatile droplets, or through an adsorption layer in case of non-evaporating partially wetting fluids. On the other hand, on the translation mode, there is droplet migration towards each other, until their contact lines touch, leading to coalescence \citep{henkel_gradient-dynamics_2021}.   

When compared to rigid substrates, early experimental studies \citep{lopes_evaporation_2012,pu_water_2012, lopes_influence_2013,yu_experimental_2013,chuang_evaporation_2014,gerber_wetting_2019} highlighted the faster evaporation of droplets observed on softer substrates, due to the longer pinning of the contact line throughout evaporation, caused by the formation of the wetting ridge. In addition, some of these experimental studies \citep{lopes_influence_2013,yu_experimental_2013} revealed that while on the initial stages of evaporation, the droplet appears to remain pinned, while the contact angle is decreased. After the droplet depins, the opposite behavior is observed, i.e. the contact angle remains constant and the contact radius decreases. Then, at the late stages of droplet evaporation, the contact angle slightly increases, before ultimately decrease until the droplet completely evaporates.

As it has been established in studies for rigid substrates, the dynamics of the contact line plays a crucial role on droplet evaporation \citep{lopes_evaporation_2012,pu_water_2012, lopes_influence_2013,yu_experimental_2013,chuang_evaporation_2014,gerber_wetting_2019} and in order to model the contact line motion an approach, followed by several researchers, has been to introduce the effect of disjoining pressure, by assuming the presence of an adsorbed precursor film ahead of the droplet, which is stabilized by the action of intermolecular interactions between the two interfaces. The presence of the precursor film is evident in experimental studies with microscopic techniques \citep{kavehpour_microscopic_2003,xu_molecular_2004,hoang_dynamics_2011} and constitutes the reason for the levelled transition from the contact line to the flat gas-solid interface, circumventing the singularity arising from the shear stress \citep{wang_dynamics_2021}, due to the contradiction between the non-zero displacement on the contact line and the no-slip boundary condition on the liquid-solid interface on the same point. This approach has been widely implemented for the modelling of not only perfectly wetting fluids \citep{bonn_wetting_2009}, but also partially wetting fluids \citep{schwartz_hysteretic_1998,schwartz_simulation_1998,gomba_regimes_2010}. A similar approach has been also used by \cite{charitatos_droplet_2021} to examine droplet evaporation
on partially wetted soft viscoelastic substrates.

To account for the droplet evaporation, two approximations have been mostly used \citep{wilson_evaporation_2023}. In the so-called one-sided model, the attention is solely drawn to the liquid phase, since vapour viscosity, density and thermal conductivity are considered negligible. In this approach, evaporation is limited by the rate on which the molecules of the solvent are headed from the liquid towards the gas phase \citep{burelbach_nonlinear_1988}. On that principle, the work of \cite{moosman_evaporating_1980}, \cite{ajaev_steady_2001} and later \cite{ajaev_spreading_2005} considered an adsorbed thin film ahead of the evaporating liquid, with non-zero film thickness, which is in thermodynamic equilibrium with both the solid and the gas phases. The work of Ajaev has been the grounding principle for many researchers studying qualitatively droplet spreading and evaporation of more complex systems, such as droplets with nanoparticles \citep{matar_dynamic_2007}, the deposition of particles while in the presence of surfactants \citep{karapetsas_evaporation_2016}, as well as the evaporation of droplets which consist of ethanol-water or other binary mixtures \citep{williams_spreading_2021}, or the vapour adsorption of hygroscopic aqueous solution droplets \citep{wang_dynamics_2021}. 
Concerning droplet evaporation on compliant substrates, the only theoretical work so far is the work of \cite{charitatos_droplet_2021}, where a one-sided model is developed to study the evaporation of a single droplet. 

However, when the evaporation is diffusion-limited and, hence, the vapour phase can not be considered irrelevant, quantitative results can only be achieved by employing a two-sided approximation \citep{sultan_evaporation_2005, Schofield_2020, Wray_et_al_2020}. Typically, one-sided models consider that the evaporation flux is only a function of pressure and temperature differences in the liquid-gas interface and the transport equations in the liquid phase are the only prerequisite for the system modelling. On the contrary, the diffusion-limited model includes the simultaneous solution of a diffusion equation concerning the vapor gas phase concentration. The second approach entails a higher computational cost but provides a more accurate description of phase change phenomena \citep{deegan_capillary_1997,hu_analysis_2005,
masoud_analytical_2009,cazabat_evaporation_2010,
mikishev_instabilities_2013,larson_transport_2014}. A common assumption to reduce the computational cost, is to consider that droplet retains a spherical-cap shape. This assumption, though, is not always safe, since the droplet shape might be significantly distorted by forces, such as gravitational forces (e.g. evaporation on inclined substrates), the effect of Marangoni or elastic stresses, etc. \cite{Hartmann_transition-limited_2023} recently developed a long-wave model of a sessile shallow droplet of evaporating partially wetting fluid on a rigid substrate, which, similarly to the earlier work of \cite{sultan_evaporation_2005}, captured the transition between the diffusion-limited and the one-sided model.

Several authors have also investigated the effect of Marangoni stresses on droplet spreading and evaporation. Marangoni flows can be induced by thermal gradients, a variation in the concentration of surfactants, or by the presence of binary mixtures \citep{dunn_spreading_2009,williams_spreading_2021,wang_dynamics_2021}. The coffee-ring effect may be suppressed by the action of Marangoni stresses 
\citep{karapetsas_evaporation_2016,seo_coffee_2017}, since they counter the outward liquid flow facilitated by contact-line pinning. Additionally, the coalescence of merging droplets with different surface tensions (such as water and ethanol) has been shown to be strongly delayed \citep{Karpitschka_quantitative_2010,Chen_early_2021}. Concerning evaporating droplets, \cite{Talbot_evaporation_2012}, \cite{Schofield_lifetimes_2018} and \cite{Dunn_conductivity_2009} showed that the thermal effects can significantly extend droplet lifetime.

There have been different approaches in the literature to describe soft substrate wetting, ranging from the development of long-wave models \citep{Kumar_dewetting_2004,matar_nonlinear_2005, Gielok_droplet_2017,Gomez_durotaxis_2020} to full-scale computational studies \citep{Bueno_tensotaxis_2017,Bueno_wettability_2018}.
\cite{Kumar_dewetting_2004} and \cite{matar_nonlinear_2005} first developed a long-wave approach to study the non-linear evolution of thin liquid films dewetting near soft elastomeric layers. \cite{charitatos_soft_2020} followed Matar's work for droplet spreading on soft solid substrates, while \cite{charitatos_droplet_2021}, extended their own work developing a one-sided model to examine droplet evaporation on viscoelastic substrates.

Building on the latter model, the present paper presents a detailed and comprehensive theoretical model for the investigation of the dynamics of a system of two evaporating droplets residing on a compliant solid substrate. The droplets may interact through the developed elastic stresses in the viscoelastic substrate which is modelled using the Kelvin-Voigt model. When the droplets are exposed in atmosphere, a further question is raised, concerning the effect of the local variations in vapour concentration on the evaporation rate of the droplets. To this end, we develop a two-sided evaporation model following a similar approach with earlier studies for rigid substrates \citep{sultan_evaporation_2005,Doumenc_coupling_2011}. Thus, apart from the viscoelastic substrate, our model unravels the potential of communication of the droplets through the atmosphere, while also taking fully into consideration the effects of evaporative cooling and induced thermocapillary phenomena.
To remove the stress singularity that
arises at the moving contact line, we assume the presence of sufficiently thin precursor film. The precursor film is stabilised and  evaporation therein is suppressed through the inclusion of a disjoining pressure.

The rest of the paper is organised as follows. The problem is formulated in \cref{model definition} and the equations governing the flow dynamics are discussed. The scaling and resulting evolution equations are presented in \cref{Scaling} and \cref{Evolution equations}, respectively. Results are presented and discussed in \cref{Results}, followed by concluding remarks in \cref{Conclusions}.

\section{Problem statement and model formulation}\label{model definition}
\subsection{Description of the problem}
\label{Description of the problem}
\begin{figure}
    \centering
    \includegraphics[width=\textwidth]{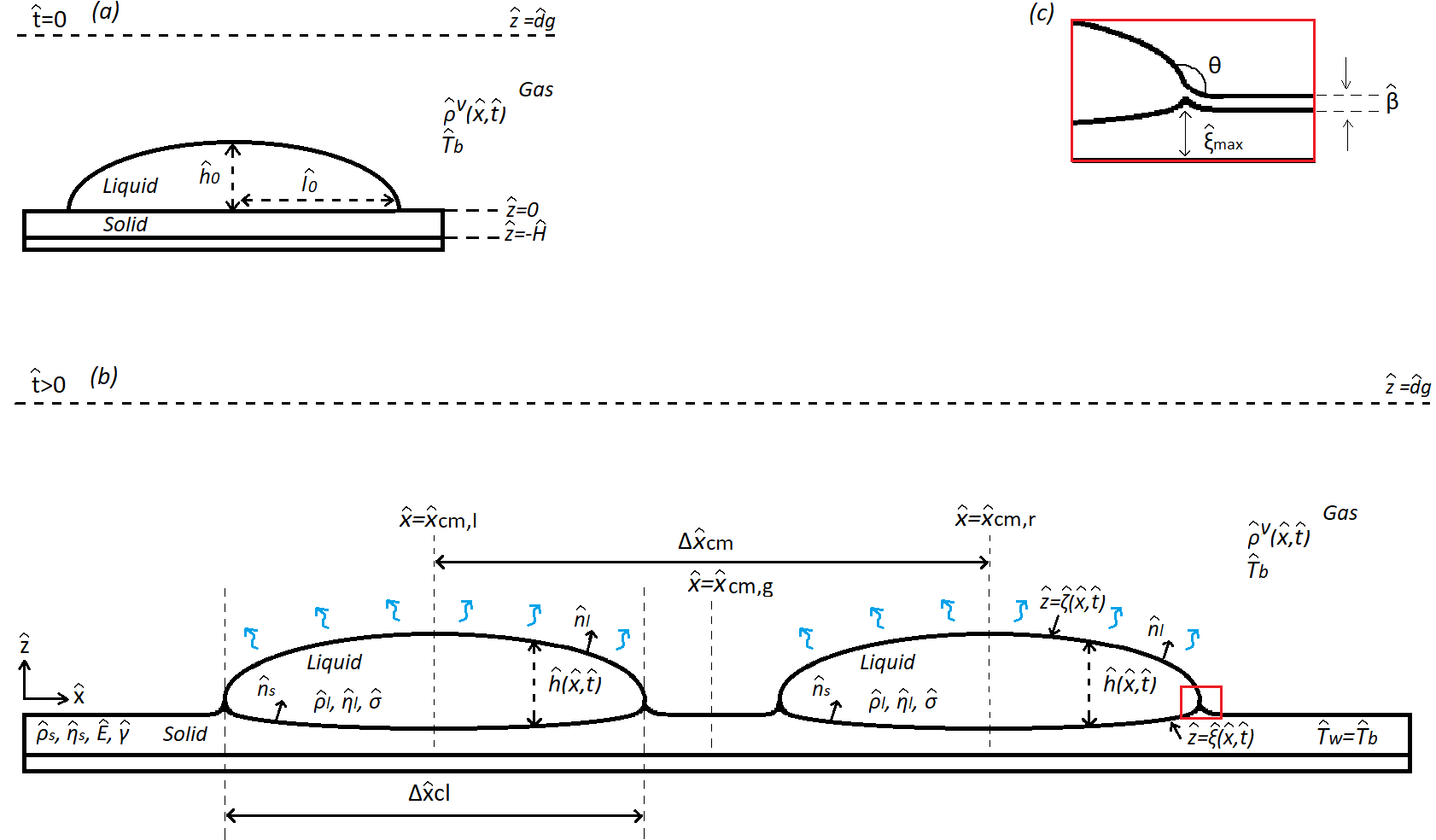}
    \caption{Schematic diagram of model geometry. (a) Initial configuration of a droplet with initial half-width $\hat{l}_0$ and initial height $\hat{h}_0$ resting on an undeformed compliant substrate at $\hat{z}=0$, which is attached to a rigid substrate at $\hat{z}=-\hat{H}$. (b) The soft solid deforms while the system of two droplets spreads and evaporates. (c) Magnified view of the contact line, where $\hat{\beta}$ is the precursor film thickness, $\hat{\theta}$ is the apparent contact angle and $\hat{\xi}_{max}$ denotes the maximum height of the wetting ridge. The local thickness of each droplet is given by $\hat{h}(\hat{x},\hat{t})=\hat{\zeta}(\hat{x},\hat{t})-\hat{\xi}(\hat{x},\hat{t})$.}
    \label{fig:1}
\end{figure}

We consider the behaviour of a single or a system of two two-dimensional sessile evaporating droplets, with initial 
cross-sectional area $\hat{V}$, placed on an incompressible linear viscoelastic solid substrate, which is also referred to as soft substrate. At $\hat{t}=0$, the droplet is resting on the soft substrate and has an initial footprint half-width $\hat{l}_0$ and an initial height $\hat{h}_0$ (Fig.\ref{fig:1}a). The liquid-solid interface is originally located at $\hat{z}=0$. The soft substrate is originally undistorted and attached to a rigid substrate at $\hat{z}=-\hat{H}$; the rigid substrate is highly conductive with constant temperature $\hat{T}_b$. The soft substrate is characterized by density $\hat{\rho}_s$, viscosity $\hat{\eta}_s$, thermal conductivity $\hat{\lambda}_w$, shear modulus $\hat{E}$ and constant liquid-solid interfacial tension $\hat{\gamma}$, which is independent of strain; the presence of an immiscible liquid solvent in the soft substrate is assumed. The droplet is assumed to have constant density $\hat{\rho}_l$, viscosity $\hat{\eta}_l$, thermal conductivity $\hat{\lambda}$, specific heat capacity $\hat{c}_p$ and saturation temperature $\hat{T}_{sat}$. The liquid-gas interfacial tension, $\hat{\sigma}$, is assumed to be a linear function of temperature
\begin{equation}
\hat{\sigma} = \hat{\sigma}_0 - \frac{\partial \hat{\sigma}}{\partial \hat T}(\hat{T}_s - \hat{T}_{ref}),
\label{surface tension lg}
\end{equation}
where $\hat{\sigma}_0$ is the surface tension at the reference temperature $\hat{T}_{ref}$, and $\hat{T}_s$ is the local temperature at the liquid-gas interface. The reference temperature was considered to be equal to the bulk temperature of the gas $\hat{T}_{b}$.

At $\hat{t}>0$, the droplet, being in an unsaturated environment, evaporates causing the deformation of the soft solid substrate.
The liquid-solid interface is located at $\hat{z}=\hat{\xi}(\hat{x},\hat{t})$, with an outward normal unit vector of 
$\hat{n}_s=\left(-\frac{\partial \hat{\xi}}{\partial \hat{x}},1\right) \bigg / \sqrt{1+\left(\frac{\partial \hat{\xi}}{\partial \hat{x}}\right)^2}$ 
while the liquid-air interface is located at $\hat{z}=\hat{\zeta}(\hat{x},\hat{t})$, with an outward normal unit vector of 
$\hat{n}_l=\left(-\frac{\partial \hat{\zeta}}{\partial \hat{x}},1\right) \bigg / \sqrt{1+\left(\frac{\partial \hat{\zeta}}{\partial \hat{x}}\right)^2}$. 
The tangential unit vectors are 
$\hat{t}_s=\left(1, \frac{\partial \hat{\xi}}{\partial \hat{x}}\right)\bigg / \sqrt{1+\left(\frac{\partial \hat{\xi}}{\partial \hat{x}}\right)^2}$ and 
$\hat{t}_l=\left(1, \frac{\partial \hat{\zeta}}{\partial \hat{x}}\right)\bigg / \sqrt{1+\left(\frac{\partial \hat{\zeta}}{\partial \hat{x}}\right)^2}$ for the liquid-solid and the liquid-air interface respectively. 

We assume that the droplet is released into a thin precursor film; evaporation in the film is stabilised by the disjoining pressure which accounts for intermolecular van der Waals interactions. The inclusion of the precursor film removes the stress singularity that can arise at the moving contact line
(see Fig. \ref{fig:1}c). This approach also allows us to easily account for the evaporation of multiple droplets as well as their interactions; in Fig. \ref{fig:1}b we depict a system of two evaporating droplets, of the same initial radius and height. Ahead of the contact line the dimensional precursor film thickness is denoted with $\hat{\beta}$ (see Fig. \ref{fig:1}c) and the apparent contact angle is $\hat{\theta}$. Concerning the wetting ridge, its maximum height is denoted with $\hat{\xi}_{max}$. Moreover, the length of each droplet, that is the distance between the two contact lines, is noted as $\Delta \hat{x}_{cl}$, while the length of the computational domain is noted as $L_x$ (i.e. $0<\hat{x}<\hat{L}_x$). In this domain, the global center of mass of the system is located at $\hat{x}=\hat{x}_{cm,g}$ and the center of mass of each droplet is located at $\hat{x}=\hat{x}_{cm,l}$ and $\hat{x}=\hat{x}_{cm,r}$, respectively. Consequently, the distance between the two centers of mass is denoted as $\Delta \hat{x}_{cm}=\hat{x}_{cm,r}-\hat{x}_{cm,l}$. The presented model geometry in Fig. \ref{fig:1} constitutes the reference layout for the rest of this paper.

In the present work, the droplets are assumed to be so thin that the droplet aspect ratio $\mathbf{\epsilon}=\hat{h}_0/\hat{R}_0$ is considered to be much smaller than unity; $\hat{R}_0$ is a characteristic length scale, defined as $\hat{R}_0=2\hat{l}_0/3$. Under this assumption, we will employ the lubrication approximation to derive a reduced set of governing evolution equations. Furthermore, gravitational forces are neglected, since the solid and liquid Bond numbers $Bo_s=\hat{\rho}_s\hat{g}{\hat{R}_0}^2/\hat{\sigma}$ and $Bo_l=\hat{\rho}_l\hat{g}{\hat{R}_0}^2/\hat{\sigma}$ are assumed to be less than unity; this condition typically holds for small droplets. A 2-D Cartesian coordinate system $(\hat{x},\hat{z})$ is used to model the velocity field, which is described by a function of $\boldsymbol{\mathbf{\hat{v}}}=(\hat{v}_x, \hat{v}_z)$, whereas the solid displacement is given by $\boldsymbol{\mathbf{\hat{u}}}=(\hat{u}_x, \hat{u}_z)$. Our model can describe a typical system of water droplets drying on polydimethylsiloxane (PDMS) substrates and the physical properties of such a system are given in Table \ref{Table:1}. 

%
\begin{table}
\def~{\hphantom{0}}
\setlength{\tabcolsep}{8pt}
\begin{tabular}{ll|ll|ll}
{\textbf{Property}}                                &{\textbf{Notation}}   &{\textbf{Value}}\\[3pt]
Density of the liquid phase $(kg \cdot m^{-3})$    &$\hat{\rho}_l$     &999 \\
Viscosity of the liquid phase $(mPa\cdot s)$       &$\hat{\eta}_l$	 &1.001   \\
Thermal conductivity of the liquid phase $(W\cdot m^{-1}\cdot K^{-1})$ &$\hat{\lambda}$ &0.603 	\\
Thermal conductivity of the solid phase $(W\cdot m^{-1}\cdot K^{-1})$  &$\hat{\lambda}_w$  &0.15  \\
Surface tension of the liquid gas interface $(mN\cdot m^{-1})$ &$\hat{\sigma}_{0}$  &72.8 \\
Surface tension of the liquid solid interface $(mN \cdot m^{-1})$ &$\hat{\gamma}$ &42.5 \\
Latent heat of evaporation $(KJ \cdot kg^{-1})$    &$\hat{L}_v$	&2454  	\\
Diffusion coefficient of the vapour in the gas phase $(m^2 \cdot s^{-1})$ &$\mathcal{\hat{D}}_v$  &$2.42 \times 10^{-5}$ 	\\
Saturation pressure of the liquid phase $(Pa)$ &$\hat{P}_{sat}$  &2313.35 	\\
Vapour concentration far from the droplet $(kg \cdot m^{-3})$ &$\hat{\rho}^{v}_{ref}$  &$17.099 \times 10^{-3}$	\\
Temperature derivative of surface tension $(N \cdot m^{-1} \cdot K)$ &$\frac{\partial \hat{\sigma}}{\partial \hat{T}}$  &$2 \times 10^{-4}$ 	\\
Initial droplet radius $(m)$ &$\hat{R}_{0}$  &0.001 \\
Universal gas constant $(J\cdot mole^{-1} \cdot K^{-1})$   &$\hat{R}$	&8.314 \\
Molecular weight of water $(kg \cdot mol^{-1})$            &$\hat{M}$      &$1.8 \times 10^{-2}$ \\
Accommodation coefficient    &$\alpha$      &1  \\
\end{tabular}
\caption{Properties of water and PDMS at 20$^\circ$C and 1 atm.}
\label{Table:1}
\end{table}

\subsection{Liquid phase}
\label{Liquid phase}
The mass, momentum and energy conservation equations for the liquid are given by:
\begin{equation}
\hat{\nabla} \cdot \mathbf{\hat{v}} = 0,
\label{l-continuity}
\end{equation}
\begin{equation}
\hat{\rho}_l \bigg( \frac{\partial \mathbf{\hat{v}}}{\partial \hat{t}}  + \mathbf{\hat{v}} \cdot \hat{\nabla} \mathbf{\hat{v}} \bigg ) = - \hat{\nabla} \hat{p}_l  + \hat{\eta}_l \hat{\nabla}^2 \mathbf{\hat{v}},
\label{l-mom}
\end{equation}
\begin{equation}
\frac{\partial \hat{T}}{\partial \hat{t}} + \mathbf{\hat{v}} \cdot \hat{\nabla} \hat{T}  - \hat{\alpha}_l \hat{\nabla}^2 \hat{T} = 0,
\label{l-energy}
\end{equation}
where $\hat{p_l}$ is the liquid pressure, $\hat{\nabla}=(\partial_{\hat{x}},\partial_{\hat{y}})$ is the gradient operator, $\hat{T}$ is the temperature and $\hat{\alpha}_l=\frac{\hat{\lambda}}{\hat{\rho}_l \hat{c}_p}$ is the thermal diffusivity of the liquid.
Along the free interface $\hat{z}=\hat{\zeta}(\hat{x},\hat{t})$, the liquid velocity $\mathbf{\hat{v}}=(\hat{v}_x, \hat{v}_z)$ differs from the velocity of the interface $\mathbf{\hat{v}_s}=(\hat{v}_{xs}, \hat{v}_{zs})$. If the evaporative flux is denoted by $\hat{J}$, then
\begin{equation}
\hat{J}=\hat{\rho}_l (\boldsymbol{\hat{\mathbf{v}}}-\boldsymbol{\hat{\mathbf{v}}_s}) \cdot \boldsymbol{\hat{n_l}}.
\label{evap flux}
\end{equation}

Furthermore, along the free interface, the local mass, energy and force balances are given by:
\begin{equation}
\hat{J}=\hat{\rho}_l (\boldsymbol{\hat{\mathbf{v}}}-\boldsymbol{\hat{\mathbf{v}}_s}) \cdot \boldsymbol{\hat{n}}_l=\hat{\rho}_g (\boldsymbol{\hat{\mathbf{v}}_g}-\boldsymbol{\hat{\mathbf{v}}_s}) \cdot \boldsymbol{\hat{n}_l},
\label{mass evap flux}
\end{equation}
\begin{equation}
\hat{J} \hat{L}_v + \hat{\lambda} \hat{\nabla} \hat{T} \cdot \boldsymbol{\hat{n}_l} - \hat{\lambda}_g \hat{\nabla} \hat{T}_g \cdot \boldsymbol{\hat{n}_l}=0,
\label{energy bal}
\end{equation}
\begin{equation}
\hat{J} (\boldsymbol{\hat{\mathbf{v}}} - \boldsymbol{\hat{\mathbf{v}}_g}) - \boldsymbol{\hat{n}_l} \cdot (-\hat{p}_l \mathrm{I} +\hat{\eta}_l (\hat{\nabla} \boldsymbol{\mathbf{\hat{v}}} +(\hat{\nabla} \boldsymbol{\mathbf{\hat{v}}})^T) -\hat{p}_g \cdot \boldsymbol{\hat{n}_l}+ \hat{\Pi} \cdot \boldsymbol{\hat{n}_l}+ 2\hat{\kappa}_l \hat{\sigma} \boldsymbol{\hat{n}_l} +\hat{\nabla}_s \hat{\sigma} = 0,
\label{force bal}
\end{equation}
where $\hat{\rho}_g$, $\hat{\lambda}_g$, $\mathbf{\hat{v}_g}$ and $\hat{T}_g$ denote the density, the thermal conductivity, the velocity and the temperature of the gas phase respectively. $\hat{L}_v$ is the specific latent internal heat of vaporization, $\mathrm{I}$ is the identity tensor, $\hat{\kappa}_l$ is the mean curvature of the free interface, while $\hat{\nabla}_s$ is the surface gradient operator. In the above equations, $\hat{\kappa}_l=\hat{\nabla}_s \cdot \boldsymbol{\hat{n}_l}$ and $\hat{\nabla}_s=(I-\boldsymbol{\hat{n}_l} \boldsymbol{\hat{n}_l}) \cdot \hat{\nabla}$. Finally, $\hat{\Pi}$ stands for the disjoining pressure, which, taking into consideration the van der Waals interaction, equals to 
\begin{equation}
\hat{\Pi}=\hat{A}_1 \bigg[\bigg(\frac{\hat{A}_2}{\hat{h}}\bigg)^n - \bigg(\frac{\hat{A}_2}{\hat{h}}\bigg)^c \bigg],
\label{dsj p}
\end{equation}
where $\hat{A_1}=\hat{A}_{Ham}/\hat{A}_2^3$, is a constant that describes the intermolecular interactions between the liquid-gas and the liquid-solid interfaces, $\hat{A}_{Ham}$ the Hamaker constant, $\hat{A}_2$ is a constant of the same order of magnitude as the precursor film thickness $\hat{\beta}$. $\hat{h}$ denotes the droplet thickness and $n>c>1$.
Moreover, the kinematic boundary condition along the moving interface $\hat{z}=\hat{\zeta}(\hat{x},\hat{t})$, is described as:
\begin{equation}
\frac{\partial \hat{\zeta}}{\partial \hat{t}} + \hat{v}_{xs} \frac{\partial \hat{\zeta}}{\partial \hat{x}}= \hat{v}_{zs}.
\label{kinematic}
\end{equation}
\subsection{Gas phase}
\label{Gas phase}
The gas phase is assumed to comprise air and vapour, but it is not saturated by vapour. Typically, we may consider that $\Lambda_g=\frac{\hat{\lambda}}{\hat{\lambda}_g}\ll 1$, and under this assumption the bulk temperature of the gas can be assumed to be constant and equal to $\hat{T}_b$. Moreover, we assume that the viscosity of the gas, $\hat{\eta}_g$, is much smaller than the viscosity of the liquid phase, i.e. $\hat{\eta}_g/\hat{\eta}_l \ll 1$ and therefore the gas can be considered as inviscid and passive with respect to the fluid.

The droplet evaporation is approached using the generalised diffusion-limited model developed by \cite{sultan_evaporation_2005},
in which evaporation is considered limited by the solvent vapour diffusion in the air; this model is able to capture the
transition between the diffusion-limited and the one-sided model. The vapour concentration $\hat{\rho}^v$ in the gas phase is described by the Laplace's equation, due to the fact that the gas phase is considered to be at rest and the characteristic evaporation time is much larger than the respective diffusion time. As a result, we get
\begin{equation}
\hat{\nabla}^2 \hat{\rho}^v=0.
\label{vap laplace}
\end{equation}

The vapour mass flux $\hat{J}$ is assumed to be limited by the rate of diffusion and thus

\begin{equation}
\hat{J}=-\mathcal{\hat{D}}_v(\boldsymbol{\hat{n}_l} \cdot \hat{\nabla} \hat{\rho}^v)|_{\hat{\zeta}},
\label{diffusion}
\end{equation}
where $\mathcal{\hat{D}}_v$ the diffusion vapour coefficient. Considering also that the vapour mass flux $\hat{J}$ is proportional to the departure from equilibrium at the liquid-gas interface 
\citep{schrage_theoretical_1953,plesset_flow_1976, moosman_evaporating_1980}, the following linear constitutive equation, most commonly known as Hertz-Knudsen equation, for $\hat{J}$ can be used
\begin{equation}
\hat{J}=\alpha \bigg(\frac{\hat{R} \hat{T}_{s}}{2\pi \hat{M}}\bigg)^{\frac{1}{2}} (\hat{\rho}^{ve}-\hat{\rho}^v|_{\hat{\zeta}}),
\label{Hertz_Knudsen} 
\end{equation}
where $\alpha$ is the accommodation coefficient, usually considered equal to unity near equilibrium. $\hat{R}$ denotes the universal gas constant, $\hat{T}_{s}$ denotes the temperature of the liquid-gas interface, $\hat{\rho}^v$ is the local vapour concentration in the gas phase and $\hat{\rho}^{ve}$ is the equilibrium vapour concentration. 

In order to get a boundary condition for the vapour concentration $\hat{\rho}^v$ at $\hat{z}=\hat{\zeta}$, we can combine Eqs. (\ref{diffusion}) and (\ref{Hertz_Knudsen}), which leads to
\begin{equation}
-\mathcal{\hat{D}}_v(\boldsymbol{\hat{n}_l} \cdot \hat{\nabla} \hat{\rho}^v)|_{\hat{\zeta}}=\alpha \bigg(\frac{\hat{R} \hat{T}_{s}}{2\pi \hat{M}}\bigg)^{\frac{1}{2}} (\hat{\rho}^{ve}-\hat{\rho}^v|_{\hat{\zeta}}).
\label{vap BC for conc}
\end{equation}

Finally, following a similar procedure as described by \citep{moosman_evaporating_1980}, the following equation for the equilibrium vapour concentration can be derived
\begin{equation}
\hat{\rho}^{ve}=\hat{\rho}^v_{ref}+\frac{\hat{M} \hat{\rho}^v_{ref}}{\hat{\rho}_l \hat{R} \hat{T}_{ref}}(-2\hat{H}_l \hat{\sigma} -\hat{\Pi})+\frac{\hat{L}_v \hat{M} \hat{\rho}^v_{ref}}{\hat{R} \hat{T}^2_{ref}}(\hat{T}_s-\hat{T}_{ref}), 
\label{vap bal chem pot}
\end{equation}
where $\hat{\rho}^v_{ref}$ is the equilibrium vapour concentration at the reference temperature.

At the far-field boundary, the most natural choice would be to impose a constant vapour concentration at infinity. However, since it is known that for the diffusion-limited model there is no analytical solution in two-dimensional half-space \citep{Schofield_2020}, we follow a similar approach to \cite{Schofield_2020} considering a finite domain of the gas phase and the far-field condition is replaced by a similar Dirichlet condition at a distant, but finite, boundary. Thus, far from the droplet ($\hat{z}=\hat{d}_g$), the vapour concentration is maintained at a constant initial vapour concentration $\hat{\rho}_{vi}$:
\begin{equation}
\hat{\rho}^v\vert_{\hat{z}=\hat{d}_g}=\hat{\rho}^{vi}.
\label{vap humidity}
\end{equation}

\subsection{Soft solid substrate}
\label{Soft solid substrate}
The mass, momentum and energy conservation equations for the soft solid substrate are given by:
\begin{equation}
\hat{\nabla} \cdot \mathbf{\hat{u}} = 0,
\label{s-continuity}
\end{equation}
\begin{equation}
\hat{\rho}_s \frac{\partial^2 \mathbf{\hat{u}}}{\partial \hat{t}^2}  = \hat{\nabla} \cdot  \mathrm{\hat{T}_s},
\label{s-mom}
\end{equation}
\begin{equation}
\frac{\partial \hat{T}_w}{\partial \hat{t}} -\hat{\alpha}_w \hat{\nabla}^2 \hat{T}_w = 0,
\label{s-energy}
\end{equation}
where $\hat{p}_s$ is the pressure in the solid, $\hat{\alpha}_w$ is the thermal diffusivity of the solid, $\hat{T}_w$ is the temperature of the solid surface and $\mathrm{\hat{T}_s}$ is the solid stress tensor. Following the work of \cite{Kumar_dewetting_2004}, \cite{matar_nonlinear_2005} and \cite{charitatos_soft_2020,charitatos_droplet_2021}, who modeled the soft elastomer layer as a linear viscoelastic material, we consider that the viscoelastic solid is described by the Kelvin-Voigt model and therefore the solid stress tensor is defined as
\begin{equation}
\mathrm{\hat{T}_s}= -\hat{p}_s \mathrm{I} + \hat{E}[\hat{\nabla} \boldsymbol{\hat{\mathbf{u}}}+ (\hat{\nabla} \boldsymbol{\hat{\mathbf{u}}})^T] + \hat{\eta}_s (\partial /\partial \hat{t}) [\hat{\nabla} \boldsymbol{\hat{\mathbf{u}}}+ (\hat{\nabla} \boldsymbol{\hat{\mathbf{u}}})^T].
\end{equation}

Finally, Eq. (\ref{s-mom}) gives
\begin{equation}
\hat{\rho}_s \frac{\partial^2 \mathbf{\hat{u}}}{\partial \hat{t}^2}  =  - \hat{\nabla} \hat{p}_s + \hat{E} \hat{\nabla}^2 \mathbf{\hat{u}} + \hat{\eta}_s \hat{\nabla}^2 \frac{\partial \mathbf{\hat{u}}}{\partial \hat{t}}.
\label{s-mom2}
\end{equation}

At $\hat{z}=-\hat{H}$, the application of the no-slip and no-displacement boundary condition yields: $\hat{v}_x=\hat{v}_z=0$ and $\hat{u}_x=\hat{u}_z=0$, while the temperature at the bottom of the solid substrate is considered to be equal to $\hat{T}_b$, i.e. $\hat{T}_w|_{\hat{z}=-\hat{H}}=\hat{T}_b$. 

Along the liquid-solid interface, at $\hat{z}=\hat{\xi}(\hat{x},\hat{t})$, we consider thermal equilibrium $\hat{T}_w|_{\hat{\xi}}=\hat{T}|_{\hat{\xi}}$ and continuity of thermal flux:
\begin{equation}
\hat{\lambda}_w \left( \boldsymbol{\hat{n}_l} \cdot \hat{\nabla}\hat{T}_w \right)\bigg \vert_{\hat{\xi}}=\hat{\lambda} \left( \boldsymbol{\hat{n}_l} \cdot \hat{\nabla}\hat{T} \right)\bigg \vert_{\hat{\xi}}.
\label{cont thermal flux}
\end{equation}

In addition, the combination of the no-slip and the no-penetration boundary conditions at $\hat{z}=\hat{\xi}(\hat{x},\hat{t})$, that is the liquid-solid interface, form the continuity-of-velocity boundary condition:
\begin{equation}
{\frac{\partial \boldsymbol{\hat{\mathbf{u}}}}{\partial \hat{t}}}\bigg \vert_{\hat{z}=0} =\boldsymbol{\hat{\mathbf{v}}}|_{\hat{\xi}}.  
\label{cont velocity}
\end{equation}

The normal and tangential force balances on the liquid-solid interface lead to:
\begin{equation}
\boldsymbol{\hat{n}_s} \cdot \mathrm{\hat{T}_l} \cdot \boldsymbol{\hat{n}_s} - \boldsymbol{\hat{n}_s} \cdot \mathrm{\hat{T}_s} \cdot \boldsymbol{\hat{n}_s} + 2 \hat{\gamma} \hat{\kappa}_s = 0,
\label{s-normal force bal}
\end{equation}
\begin{equation}
\boldsymbol{\hat{n}_s} \cdot \mathrm{\hat{T}_l} \cdot \boldsymbol{\hat{t}_s} - \boldsymbol{\hat{n}_s} \cdot \mathrm{\hat{T}_s} \cdot \boldsymbol{\hat{t}_s} = 0,
\label{s-tang force bal}
\end{equation}
where $\hat{\kappa}_s=\hat{\nabla}_s \cdot \boldsymbol{\hat{n}_s}$ and $\mathrm{\hat{T}_l}$ stands for the liquid stress tensor, defined as $\mathrm{\hat{T}_l}= -\hat{p}_l \mathrm{I}+\hat{\eta}_l[\hat{\nabla} \boldsymbol{\hat{\mathbf{v}}}+ (\hat{\nabla} \boldsymbol{\hat{\mathbf{v}}})^T]$.


\section{Scaling}\label{Scaling}
In order to render the aforementioned equations and boundary conditions non-dimensional, we use the scalings shown below: 
\begin{equation}
\left. \begin{array}{l}
\displaystyle
(\hat{x}, \hat{z}, \hat{\xi}, \hat{\zeta}) = \hat{R}_0 (x, \epsilon z, \epsilon \xi, \epsilon \zeta), \quad  
\hat{t} = \frac{\hat{R}_0}{{\hat{U}}} t; \\[16pt]
\displaystyle 
\hat{\sigma} = \hat{\sigma}_0 \sigma, \quad
(\hat{p}_l, \hat{p}_s, \hat{\Pi})  = \frac{\hat{\eta}_l {\hat{U}}}{\epsilon^2 \hat{R}_0} 
 (p_l, p_s, \Pi); \\[16pt] 
\displaystyle
(\hat{u}_x, \hat{u}_z)=\hat{R}_0 (u_x, \epsilon u_z), \quad
(\hat{v}_x, \hat{v}_z)=\hat{U} (v_x,\epsilon v_z); \\[16pt]
\displaystyle
\hat{T} = \hat{T}_{ref} + T\Delta\hat{T},\quad 
\hat{J} = \frac{\hat{\lambda} \Delta \hat{T}}{\hat{L}_v \hat{h}_0} J, \quad
\hat{\rho^v} = \hat{\rho}^{v}_{ref} \rho^v.
\end{array} \right\}
\label{eq scaling}
\end{equation}
where $\Delta \hat{T}=\epsilon^2 \hat{T}_{ref}$. As $\hat{T}_{ref}$ we consider the constant bulk temperature of the gas phase, $\hat{T}_b$. As characteristic velocity we use $\hat{U}=\epsilon^3 \hat{\sigma}_0 / \hat{\eta}_l$. Note that henceforth all the variables in the following equations are dimensionless. 

\subsection{Liquid phase}
\label{Liquid_phase_scaled}

By substituting these scalings and taking into consideration that $\epsilon \ll 1$, the leading order equations for the liquid are:
\begin{equation}
\frac{\partial v_x}{\partial x} + \frac{\partial v_z}{\partial z} = 0,
\label{l-continuity d}
\end{equation}
\begin{equation}
-\frac{\partial p_l}{\partial x} + \frac{\partial^2 v_x}{\partial z^2}=0,
\label{l-momx d}
\end{equation}
\begin{equation}
\frac{\partial p_l}{\partial z}=0,
\label{l-momz d}
\end{equation}
\begin{equation}
\frac{\partial^2 T}{\partial z^2}=0.
\label{l-energy d}
\end{equation}

Along the free interface, i.e. $z=\zeta(x,t)$, we get for the mass, energy and force balances in the normal and tangential coordinate:
\begin{equation}
EJ =-\frac{\partial \zeta}{\partial x} (v_x-v_{xs}) +(v_z-v_{zs}),
\label{evap flux d}
\end{equation}
\begin{equation}
\frac{\partial T}{\partial z}\bigg\vert_{\zeta}=- J,
\label{energy bal d}
\end{equation}
\begin{equation}
p_l|_{\zeta}=p_g - \Pi- C_l^{-1} \sigma \frac{\partial^2 \zeta}{\partial x^2},
\label{l normal force bal d}
\end{equation}
\begin{equation}
\frac{\partial v_x}{\partial z}\bigg\vert_{\zeta}= \left( \epsilon^2 C_l \right)^{-1} \frac{\partial \sigma}{\partial x},
\label{l tang force bal d}
\end{equation}
where $\sigma=1-Ma T_s$, $C_l=\frac{\hat{\eta}_l \hat{U}}{\epsilon^3 \hat{\sigma}_0}$, $Ma=\frac{\partial \hat{\sigma}}{\partial \hat{T}}\frac{\Delta \hat{T}}{\hat{\sigma}_0}$ and $E=\frac{\hat{\lambda} \Delta \hat{T}}{\hat{L}_v \hat{h}_0 \hat{\rho}_l \hat{U} \epsilon}$ the evaporation number, which represents the ratio between the capillary time $t_c=\hat{R}_0/\hat{U}$ and the evaporation time $t_e=\frac{\hat{h}^{2}_0 \hat{\rho}_l \hat{L}_v}{\hat{\lambda} \Delta \hat{T}}$, as it is derived from the scaling. In Eq.(\ref{l normal force bal d}) the gas pressure has been set equal to zero (datum pressure) without loss of generality. 

The kinematic equation, i.e. Eq. (\ref{kinematic}) in combination with Eq. (\ref{evap flux d}) gives the evolution equation for the liquid-gas interface $z=\zeta(x,t)$:
\begin{equation}
\frac{\partial \zeta}{\partial t} + v_x\vert_{\zeta} \frac{\partial \zeta}{\partial x} - v_z\vert_{\zeta}+ EJ= 0.
\label{kinematic d}
\end{equation}

Finally, the scaled disjoining pressure is given by the following expression:
\begin{equation}
\Pi=\mathcal{A} \bigg[\bigg(\frac{B}{h}\bigg)^n - \bigg(\frac{B}{h}\bigg)^c \bigg],
\label{dsj P d}
\end{equation}
where $B=\frac{\hat{A}_2}{\hat{h}_0}$ and $\mathcal{A}=\frac{\hat{A}_{Ham}}{\hat{A}_2^3} \frac{\epsilon \hat{h}_0}{\hat{\eta}_l \hat{U}}$ the dimensionless Hamaker constant.
\subsection{Soft solid substrate}
\label{Soft_solid_substrate_scaled}

Using the same scaling, the leading equations for the soft solid become:
\begin{equation}
\frac{\partial u_x}{\partial x} + \frac{\partial u_z}{\partial z} = 0,
\label{s-continuity d}
\end{equation}
\begin{equation}
-\frac{\partial p_s}{\partial x} + G \frac{\partial^2 {u_x}}{\partial z^2} + m \frac{\partial}{\partial t} \left( \frac{\partial^2 u_x}{\partial z^2} \right)= 0,
\label{s-momx d}
\end{equation}
\begin{equation}
\frac{\partial p_s}{\partial z}=0,
\label{s-momz d}
\end{equation}
\begin{equation}
\frac{\partial^2 T_w}{\partial z^2}=0,
\label{s-energy d}
\end{equation}
where 
$m=\hat{\eta}_s/\hat{\eta}_l$. 
The ratio of elastic forces to interfacial tension forces is defined as $G=\epsilon^{-3} \hat{R}_0/\hat{l}_{ec}$, where $\hat{l}_{ec}=\hat{\sigma}_0/\hat{E}$ denotes the elastocapillary length  which sets the characteristic size of the deformation of the elastic substrate.

As far as the boundary conditions are concerned, at $z=\xi(x,t)$ we get: $T_w \vert_{\xi}=T\vert_{\xi}$, whereas at $z=-H$ we get $v_x=v_z=0$, $u_x=u_z=0$ and we set $T_w\vert_{-H}=0$. The dimensionless continuity of thermal flux along $z=\xi(x,t)$ is given by:
\begin{equation}
\frac{\partial T_w}{\partial z}\bigg \vert_{\xi}=\frac{1}{\Lambda_w} \frac{\partial T}{\partial z}\bigg \vert_{\xi},
\label{cont thermal flux d}
\end{equation}
where $\Lambda_w=\hat{\lambda}_w/\hat{\lambda}$ denotes the ratio of thermal conductivity between the solid and the liquid.

At the liquid-solid interface, i.e. $z=\xi(x,t)$, both the continuity-of-velocity and the normal and tangential force balances render to the following form:
\begin{equation}
{\frac{\partial u_x}{\partial t}}\bigg \vert_{0} =v_x|_{\xi},  
\label{cont velocity x d}
\end{equation}
\begin{equation}
{\frac{\partial u_z}{\partial t}}\bigg \vert_{0} =v_z|_{\xi},  
\label{cont velocity z d}
\end{equation}
\begin{equation}
p_s=p_{cap,l}+p_{cap,s},
\label{normal force bal d}
\end{equation}
\begin{equation}
\frac{\partial v_x}{\partial z} - G \frac{\partial u_x}{\partial z} -m \frac{\partial}{\partial t} \left( \frac{\partial u_x}{\partial z} \right) =0,
\label{tang force bal d}
\end{equation}
where $p_{cap,l}=-{C_l}^{-1}\sigma\partial^2 \zeta/\partial x^2$ and $p_{cap,s}=-{C_s}^{-1} \partial^2 \xi/\partial x^2$ the capillary-like pressures in the liquid and the solid respectively, with $C_s=\frac{\hat{\eta}_l \hat{U}}{\epsilon^3 \hat{\gamma}}$ denoting the capillary number at the liquid-solid interface. 

\subsection{Gas phase}
\label{Gas_phase_scaled}
Since the gas phase in the atmosphere may extend to large distances above the liquid phase, the scaling in the $z$-direction shown in Eq. (\ref{eq scaling}) is not appropriate and therefore we employ the same scaling with the $x$-direction (i.e. $\hat{z} = \hat{R}_0 z'$ and $\hat{\zeta}= \hat{R}_0 \zeta'$). The dimensionless conservation equation for the vapour concentration is then given by:
\begin{equation}
\frac{\partial^2 \rho^v}{\partial x^2} + \frac{\partial^2 \rho^v}{\partial z'^2}=0.
\label{vap conc d}
\end{equation}

The above equation is subjected to the following boundary conditions far from the droplet ($z'=d_g$) and along the liquid-gas interface ($z'=\zeta'(x,t)$):
\begin{equation}
\rho^v\vert_{d_g}=\mathcal{H}, 
\label{vap BC humidity d}
\end{equation}
\begin{equation}
Pe_v J=-(\boldsymbol{n_l} \cdot \nabla \rho^v)|_{\zeta'},
\label{vap BC d}
\end{equation}
where $\mathcal{H}=\hat{\rho}^{vi}/\hat{\rho}^{v}_{ref}$ denotes the relative humidity and $Pe_v=\frac{\hat{\lambda} \Delta \hat{T} \hat{R}_0}{\hat{h}_0\mathcal{\hat{D}}_v \hat{\rho}^{v}_{ref}\hat{L}_v}$.

Solving Eqs. (\ref{vap conc d})-(\ref{vap BC d}) we evaluate the vapour concentration in the gas phase, therefore making it possible to compute the vapour mass flux using the following dimensionless constitutive equation
\begin{equation}
K J=\rho^{ve} - \rho^v \vert_{\zeta'}.
\label{evol J}
\end{equation}
where $K=\frac{\hat{\lambda} \Delta \hat{T}}{\alpha \hat{\rho}^v_{ref} \hat{L}_v \hat{h}_0} \sqrt{\frac{2 \pi \hat{M}}{\hat{R} T_s}}$. In the above equation the equilibrium vapour concentration is given by:
\begin{equation}
\rho^{ve} = 1 +\delta p_l +\psi T_s.
\label{evol rho_ve}
\end{equation}
where $\delta=\frac{\hat{M} \hat{\eta}_l \hat{U}}{\hat{\rho}_l \hat{R} \hat{T}_{ref} \epsilon^2 \hat{R}_0}$ and $\psi=\frac{\hat{L}_v \hat{M} \Delta \hat{T}}{\hat{R} \hat{T}^2_{ref}}$.

In order to evaluate the precursor film thickness $\beta$, we can combine Eq. (\ref{evol J}) with Eq. (\ref{evol rho_ve}). 
Taking into account that far from the droplets the film is flat and at equilibrium with the environment, the following equation is derived
\begin{equation}
\delta \mathcal{A} \bigg[\bigg(\frac{B}{\beta}\bigg)^n - \bigg(\frac{B}{\beta}\bigg)^c \bigg]=1-\mathcal{H}.
\label{evol beta}
\end{equation}
%

%
%
\noindent
An estimation of the order-of-magnitude of certain dimensionless parameters is depicted in Table 2.
\begin{table}
\begin{center}
\def~{\hphantom{0}}
\setlength{\tabcolsep}{10pt}
\begin{tabular}{ll|ll|ll}
{Parameter}          &{Definition}     &{Order-of-magnitude} \\
$\mathcal{A}$    &$\frac{\hat{A}_{Ham}}{\hat{A}_2^3} \frac{\epsilon \hat{h}_0}{\hat{\eta}_l \hat{U}}$   &\num{200}-\num{500}	\\
$B$    &$\frac{\hat{A}_2}{\hat{h}_0}$   &\num{e-3}-\num{e-4}	\\ 
$K$   &$\frac{\hat{\lambda} \Delta \hat{T}}{\alpha \hat{\rho}^v_{ref} \hat{L}_v \hat{h}_0} \sqrt{\frac{2 \pi \hat{M}}{\hat{R}_g T_s}}$ &\num{e-5}-\num{e-1}\\ 
$Pe_v$   &$\frac{\hat{\lambda} \Delta \hat{T} \hat{R}_0}{\hat{h}_0\mathcal{\hat{D}}_v \hat{\rho}^{v}_{ref}\hat{L}_v}$	 &\num{e-2}-\num{1}   \\ 
$\psi$  &$\frac{\hat{L}_v \hat{M} \Delta \hat{T}}{\hat{R}_g \hat{T}^2_{ref}}$ &\num{e-1}-\num{1}	\\  
$\delta$  &$\frac{\hat{M} \hat{\eta}_l \hat{U}}{\hat{\rho}_l \hat{R}_g \hat{T}_{ref} \epsilon^2 \hat{R}_0}$ &\num{5e-4}-\num{e-3}	\\  
$E$  &$\frac{\hat{\lambda} \Delta \hat{T}}{\hat{L}_v \hat{h}_0 \hat{\rho}_l \hat{U} \epsilon}$ &\num{e-4}-\num{e-3}	\\  
$\Lambda_w$   &$\frac{\hat{\lambda}_w}{\hat{\lambda}}$   	&\num{1}  	\\  
$M_a$  &$\frac{\partial \hat{\sigma}}{\partial \hat{T}}\frac{\Delta \hat{T}}{\hat{\sigma}_{0}}$    &\num{e-4}-\num{5e-3} 	\\ 
${C_l}^{-1}$  &$\frac{\epsilon^3 \hat{\sigma}_{0}}{\hat{\eta}_l \hat{U}}$ 	&\num{1} \\  
${C_s}^{-1}$  &$\frac{\epsilon^3 \hat{\gamma}}{\hat{\eta}_l \hat{U}}$     &\num{0.5}  \\  
$G$  &$\frac{\hat{E} \hat{R}_0}{\hat{\sigma}_0 \epsilon^3}$  &\num{1}-\num{e5} \\  
$m$  &$\frac{\hat{\eta}_s}{\hat{\eta}_l}$   &\num{100}   \\
$\mathcal{H}$  &$\frac{\hat{\rho}^{vi}}{\hat{\rho}^{v}_{ref}}$ &\num{0}-\num{1} \\
\end{tabular}
\caption{Order-of-magnitude estimate for the dimensionless parameters assuming $\epsilon=0.1$, $\Delta \hat{T}=3 K$.}
\end{center}
\label{Table 2}
\end{table}

\section{Evolution equations}
\label{Evolution equations}

In order to derive the evolution equations, we make an approximation that the streamwise displacement follows a parabolic profile in $z$, since this is the simplest solution that satisfies Eq. (\ref{s-momx d}) \citep{matar_nonlinear_2005,ghosh_influence_2016,charitatos_soft_2020,charitatos_droplet_2021} for the soft solid:
\begin{equation}
u_x=b_1(x,t) z^2 + b_2(x,t) z +b_3(x,t),
\label{parab prof u}
\end{equation}
in which $b_1$, $b_2$ and $b_3$ are functions of both space and time that will be determined later;  a detailed derivation is given in the Appendix \ref{appendix:raw}.

Using the boundary condition at $z=-H$,  i.e. $u_x=0$, we get for $b_3(x,t)$:
\begin{equation}
b_3=b_2 H - b_1 H^2.
\label{b3}
\end{equation}

Regarding the coefficient $b_1(x,t)$, introducing Eq. (\ref{parab prof u}) into the $x$-component of the solid momentum, i.e. Eq. (\ref{s-momx d}), yields the following expression:
\begin{equation}
\frac{\partial b_1}{\partial t} = \frac{1}{m} \bigg(\frac{1}{2} \frac{\partial p_s}{\partial x} -G b_1 \bigg).
\label{b1t}
\end{equation}

From Eq. (\ref{normal force bal d}), using Eqs. (\ref{parab prof u}) and (\ref{s-momx d}), as well as the expression of the $x$-component of the liquid velocity (i.e. Eq. (\ref{vx2}) derived in the Appendix \ref{appendix:raw}), we conclude to an expression for the coefficient $b_2(x,t)$:
\begin{equation}
\frac{\partial b_2}{\partial t} = \frac{1}{m} \bigg( \frac{\partial p_l}{\partial x} (\xi-\zeta) - Gb_2 + \left( \epsilon^2 C_l \right)^{-1} \frac{\partial \sigma}{\partial x} \bigg). 
\label{b2t}
\end{equation}

In order to derive the evolution equation for $\zeta(x,t)$, we use the kinematic equation for the liquid, i.e. Eq. (\ref{kinematic d}). Using the expressions of $v_x$ and $v_z$ (i.e. Eqs. (\ref{vx2}) and (\ref{vz2}) respectively derived in the Appendix \ref{appendix:raw}), and setting $z=\zeta$, we get:
\begin{equation}
\begin{split}
\frac{\partial \zeta}{\partial t} = & \frac{\partial}{\partial x} \bigg[
\frac{1}{3} \frac{\partial p_l}{\partial x} (\zeta-\xi)^3 - \frac{1}{2} \left( \epsilon^2 C_l \right)^{-1} \frac{\partial \sigma}{\partial x} (\zeta-\xi)^2 - H \zeta \bigg( \frac{\partial b_2}{\partial t} - H \frac{\partial b_1}{\partial t}\bigg) \bigg] \\ & + \xi \frac{\partial}{\partial x} \bigg(H \frac{\partial b_2}{\partial t} -H^2 \frac{\partial b_1}{\partial t}\bigg) + \frac{2 H^3}{3} \frac{\partial^2 b_1}{\partial x \partial t} - \frac{H^2}{2} \frac{\partial^2 b_2}{\partial x \partial t}- EJ. 
\end{split}
\label{evol zeta}
\end{equation}

Using the expressions of the material derivatives of $\xi$ and $u_z(x,0,t)$, (i.e. Eqs. (\ref{mat der ksi}) and (\ref{mat der uz}) respectively derived in the Appendix \ref{appendix:raw}), leads to an evolution equation for $\xi(x,t)$:
\begin{equation}
\frac{\partial \xi}{\partial t}+\frac{\partial \xi}{\partial x} \bigg( H \frac{\partial b_2}{\partial t}- H^2 \frac{\partial b_1}{\partial t}\bigg)=\frac{\partial}{\partial t} \bigg(\frac{2H^3}{3} \frac{\partial b_1}{\partial x}- \frac{H^2}{2} \frac{\partial b_2}{\partial x}\bigg).
\label{evol ksi}
\end{equation}

Furthermore, we can easily result in an evolution equation for the droplet thickness $h(x,t)=\zeta(x,t)-\xi(x,t)$ by subtracting Eq. (\ref{evol ksi}) from Eq. (\ref{evol zeta}):
\begin{equation}
\frac{\partial h}{\partial t} + \frac{\partial q}{\partial x} = -EJ,
\label{evol h}
\end{equation}
where the liquid flowrate, $q$, is given by:
\begin{equation}
q=-\frac{1}{3}\frac{\partial p_l}{\partial x} (\zeta-\xi)^3 +\frac{1}{2} \left( \epsilon^2 C_l \right)^{-1} \frac{\partial \sigma}{\partial x} (\zeta-\xi)^2+ H (\zeta-\xi) \bigg(\frac{\partial b_2}{\partial t}-H \frac{\partial b_1}{\partial t}\bigg).
\label{q}
\end{equation}

By integrating the energy equation, i.e. Eq. (\ref{s-energy d}), with respect to $z$, and using the continuity of thermal flux at $z=\xi(x,t)$, i.e. Eq. (\ref{cont thermal flux d}) and the boundary condition, $T_w\vert_{-H}=0$, the following evolution equation for the temperature in the soft solid substrate can be derived:
\begin{equation}
T_w=-\frac{J}{\Lambda_w}(z+H).
\label{Tw}
\end{equation}

Similarly, by integrating the energy equation, Eq. (\ref{l-energy d}), with respect to z, and using the energy balance, Eq. (\ref{energy bal d}), and the fact that $T |_\xi=T_w\vert_{\xi}$, the following evolution equation for the temperature in the liquid phase can be derived:
\begin{equation}
T=-J (z-\xi)-\frac{J}{\Lambda_w}(\xi+H).
\label{T}
\end{equation}

In summary, we solve numerically the evolution equations Eqs. (\ref{b1t}), (\ref{b2t}), (\ref{evol ksi}), (\ref{evol h}) and (\ref{evol J}) on the domain $0<x<L_x$ and Eq. (\ref{vap conc d}) on the 2D domain $0<x<L_x$, $0<z<L_z$. The latter equation is subjected to boundary conditions Eq. (\ref{vap BC humidity d}) and (\ref{vap BC d}) in the $z$-direction and to the following condition in the $x$-direction:
\begin{equation}
\frac{\partial \rho_v}{\partial x} \bigg \vert_{x=0}= \frac{\partial \rho_v}{\partial x} \bigg \vert_{x=L_x}=0. 
\label{BC vap 2D}
\end{equation}

The numerical solution of the evolution equations Eqs. (\ref{b1t}), (\ref{b2t}), (\ref{evol ksi}) and (\ref{evol h}) is subjected to the following conditions at $x=0$ and $x=L_x$:
\begin{equation}
\left. \begin{array}{l}
\displaystyle
\frac{\partial \zeta}{\partial x} = \frac{\partial \xi}{\partial x} =  \frac{\partial^3 \zeta}{\partial x^3} = \frac{\partial^3 \xi}{\partial x^3} = \frac{\partial b_1}{\partial x} = \frac{\partial b_2}{\partial x} =0, \\[16pt]
\zeta-\xi = \beta,
\end{array} \right\}
\label{BC 1D}
\end{equation}
where $\beta=\hat{\beta}/\hat{h}_0$ is the dimensionless precursor film height. These conditions were concluded into, after the assumptions that both $\zeta$ and $\xi$ are horizontal at $x=0$ and $x=L$ and that the dimensionless precursor film thickness equals to the distance between the two interfaces at these positions. Furthermore, the liquid flow rate and the solid displacement in the $z$-direction were considered equal to zero at all ends.

Concerning the initial conditions, we assumed a flat liquid-solid interface at $t=0$:
\begin{equation}
b_1(x,0)=b_2(x,0)=\xi(x,0)=J(x,0)=0. 
\label{initial conditions}
\end{equation}

As far as the initial shape of the droplet thickness is concerned, we use a fourth order polynomial which satisfies $\frac{\partial h}{\partial x} = \frac{\partial^3 h}{\partial x^3} =0$ at the droplet center ($x=x_{cm,l}$ or $x_{cm,r}$) and $\frac{\partial h}{\partial x} = 0$ as well as $h = \beta$ at distance $l_0$ from the droplet center, respectively.

The length and height of the computational domain was taken equal to $L_x=\hat{L}_x/\hat{R}_0=16$ and $L_z=\hat{L}_z/\hat{R}_0=5$, respectively. The dimensionless initial droplet footprint half-width, $l_0$, was defined equal to $l_0=1.5$ and the dimensionless initial droplet cross-sectional area was considered to be equal to $V=2/3$. Moreover, the initial center of mass for each droplet was taken at $x=x_{cm,l}=4$ and at $x=x_{cm,r}=12$, hence the initial distance between the two centers of mass was given by $\Delta x_{cm}=x_{cm,r}-x_{cm,l}=8$, while the center of mass of the system was initially located at $x=x_{cm,g}=8$. 

The above set of equations is solved using the Finite Element Method and it has been implemented in COMSOL Multiphysics commercial software. We applied a fully implicit finite difference scheme (BDF) to solve the system of the evolution equations and we selected the PARDISO iterative solver for the intermediate time-stepping. Typically we use 10000 elements for the discretization of the system geometry and the moving mesh of the surrounding atmosphere was appropriately refined using free triangular cells; numerical checks showed that increasing the number of elements further led to negligible changes. The simulations stop when the system mass has decreased by 80\%.

\section{Results and discussion} \label{Results}

Droplet evaporation on compliant substrates is a parametrically rich problem. We begin our study by examining the case of the evaporation of a single droplet on a soft substrate in \cref{sec:single_droplet}, while in \cref{sec:multiple_droplets} we proceed with simulations for a system of two interacting droplets. Numerical solutions were obtained over a wide range of parameter values. The ‘base’ case, however, has broadly typical values of $\epsilon=0.1$, $l_0=1.5$, $H=0.1$, $A=500$, $B=0.005$, $n=3$, $c=2$, $E=10^{-4}$, $\mathcal{H}=0.5$, $K=0.2$, $\psi=0.1$, $Pe_v=0.1$, $\delta=10^{-3}$, $m=100$, $C_l^{-1}=1$, $C_s^{-1}=0.5$,  unless noted otherwise in the text. In the figures that follow, we define a scaled time $t'=t/t_{ev}$ where $t_{ev}$ is defined as the time that the system mass has decreased by 80\%.

\subsection{Evaporation of a single droplet} \label{sec:single_droplet}

\subsubsection{Effect of thermocapillarity}
\label{Effect of thermocapillarity}

\begin{figure}
	\centering
	\vspace{1cm}
    (a) \hspace{0.47\textwidth} (b) \\
	\includegraphics[width=0.47\textwidth]{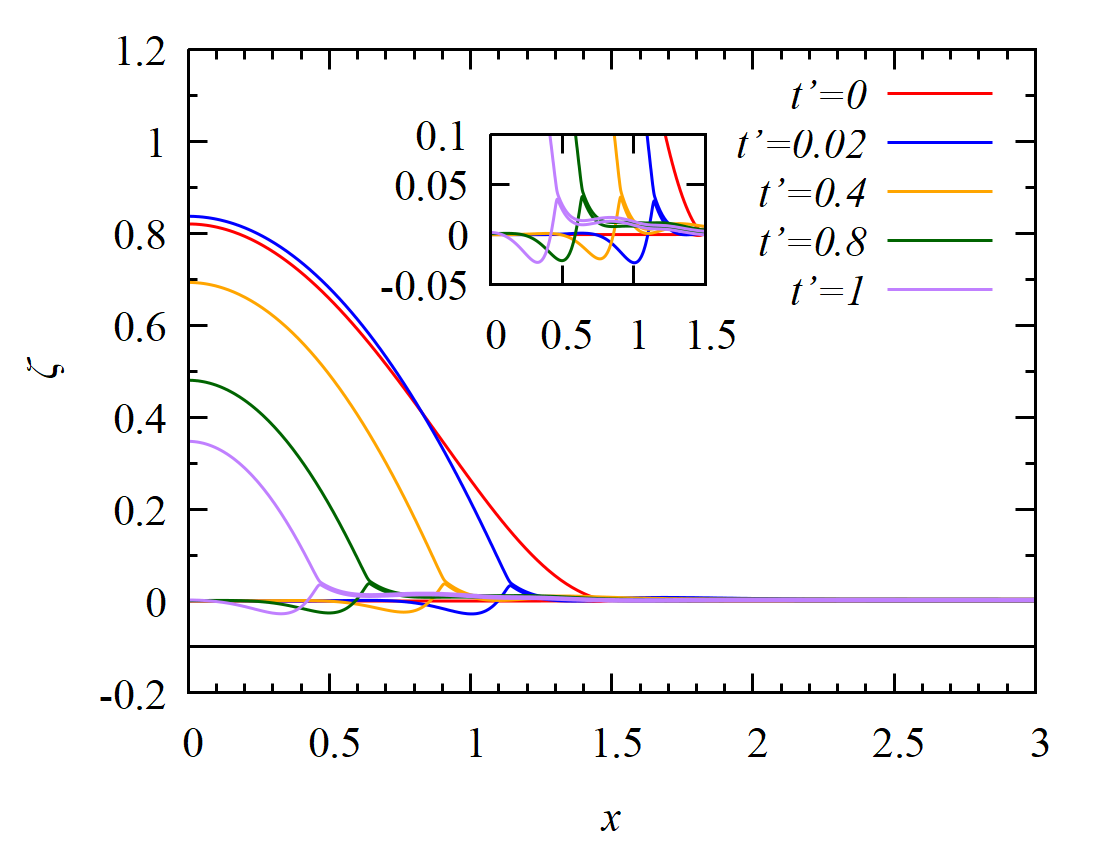} \hspace{0.5 cm}
        \includegraphics[width=0.47\textwidth]{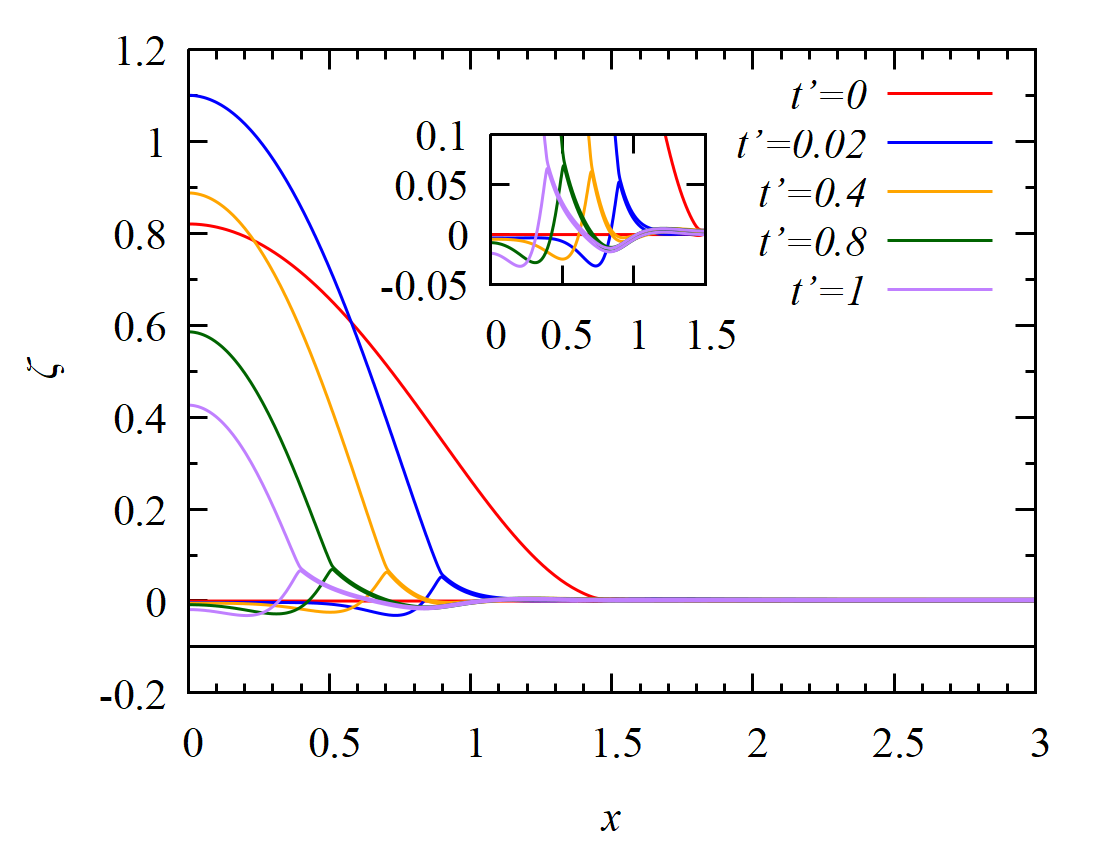}
    \caption{Time evolution of the liquid-air ($\zeta$) and the liquid-solid ($\xi$) interfaces for a single droplet for (a) $M_a=0$ ($t_{ev}=4242$) and (b) $M_a=5\cdot 10^{-3}$ ($t_{ev}=5366$) respectively, for $G=3$. The inset is an enlargement of the contact line region. The rest of the system parameters are the same with the 'base' case.}
    \label{fig:2}
\end{figure}

To set the stage, we begin with the simplest configuration, i.e. the evaporation of a single droplet on a soft substrate. Fig. \ref{fig:2} depicts the typical time evolution of the liquid-gas and the liquid-solid interfaces for a single sessile evaporating droplet, highlighting the contact line region in the inset of the same figure. \cite{charitatos_droplet_2021} considered a system similar to the present setup, albeit ignoring the effect of thermocapillarity and employing the one-sided model. In order to examine the effect of thermocapillary phenomena, we present in Fig. \ref{fig:2}a the evolution for $M_a=0$ and in Fig. \ref{fig:2}b for $M_a=0.005$. In the absence of thermocapillary stresses (Fig. \ref{fig:2}a), in line with \cite{charitatos_droplet_2021}, we notice a gradual decrease in the droplet footprint, which is accompanied by a small deformation of the soft substrate, due to the balance of the capillary forces along the liquid-solid interface and in the contact line region. Consequently, a wetting ridge is formed and as the droplet dries out, both the contact line and the wetting ridge retract as a result of the decrease in the droplet volume. 

\begin{figure}
	\centering
	\vspace{1cm}
    (a) \hspace{0.47\textwidth} (b) \\
	\includegraphics[width=0.47\textwidth]{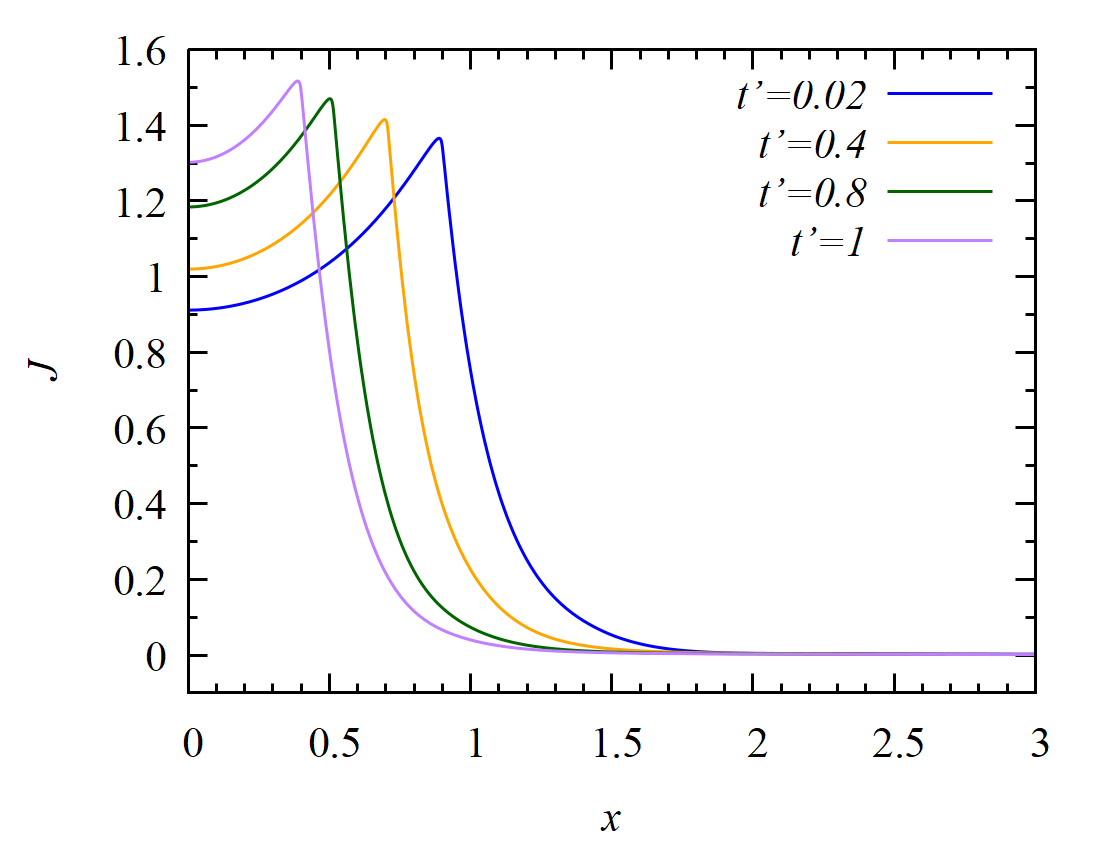} \hspace{0.5 cm}
	\includegraphics[width=0.47\textwidth]{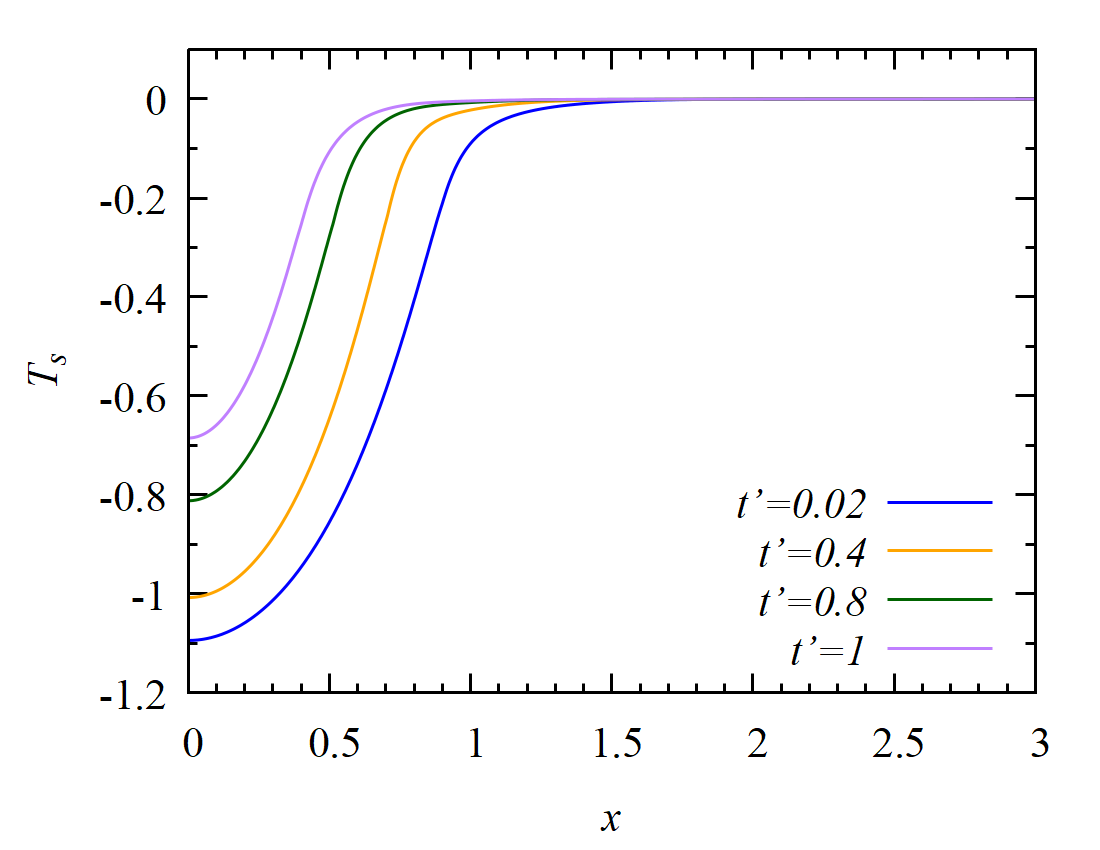}
    \caption{Time evolution of (a) the evaporation rate $J$ and (b) the interfacial temperature $T_s$, for a single droplet for  $M_a=5\cdot 10^{-3}$ and $G=3$ ($t_{ev}=5366$).The rest of the system parameters are the same with the 'base' case.}
    \label{fig:4n}
\end{figure}

\begin{figure}
	\centering
    \includegraphics[width=1.0\textwidth]{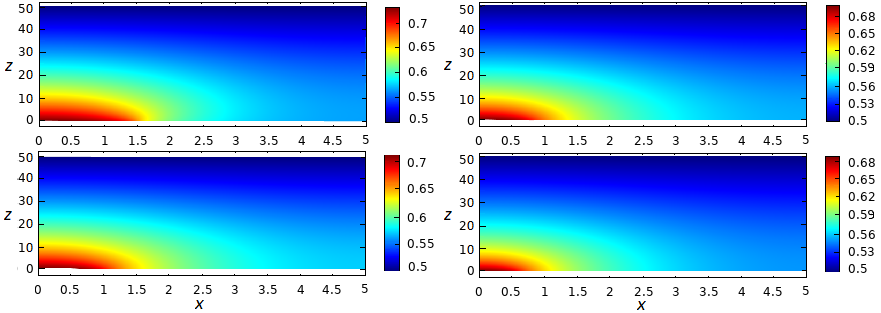}
    \caption{Gas phase concentration profiles at different time instants for $M_a=5\cdot 10^{-3}$ and $G=3$. From top to bottom: t'=0, t'=0.02 (left column),  t'=0.4, t'=0.8 (right column) respectively. The rest of the system parameters are the same with the 'base' case. }
    \label{fig:3}
\end{figure}

On the other hand, in the presence of thermocapillarity (see Fig. \ref{fig:2}b), the Marangoni stresses drive liquid towards the colder region (i.e. at droplet apex, see Fig.\ref{fig:4n}b) causing a faster retraction of the droplet. The faster motion of the contact line results in significantly larger substrate deformation, since for example at $t=100$ the maximum deformation of the wetting ridge (evaluated as the $z$-position of the contact line, see Fig. \ref{fig:1}c), is $\xi_{max}= 0.038$ and $\xi_{max}= 0.068$, in Figs. \ref{fig:2}a and \ref{fig:2}b, respectively.  Furthermore, it can be deduced from Fig. \ref{fig:2}b that for finite values of $M_a$ the loss of droplet mass is retarded; $t_{ev}$ is considerably larger in Fig \ref{fig:2}b as compared to Fig. \ref{fig:2}a. This is due to the fact that the action of thermocapillary stresses leads to a considerably smaller droplet footprint with larger distance of the droplet apex from the rigid solid (at $z=-H$). The increased droplet height inhibits the supply of heat from the substrate (maintained at a constant temperature) to the interface, which is continuously being cooled due to the effect of latent heat. This consequently leads to lower temperature along the liquid-gas interface and in turn results in the overall decrease of the evaporation rate; the evolution of the local evaporation flux is presented in Fig. \ref{fig:4n}a. 


To illustrate the vapour concentration field in the gas phase, we present the corresponding contour plot in Fig. \ref{fig:3}, for the case of $M_a=0.005$. It is noted that the far-field boundary is taken to be very far from the droplet (i.e. at $z=\epsilon^{-1} z'=50$) and as a result the droplet is difficult to be seen in this figure. We notice, though, that in the neighbourhood of the droplet the vapour concentration is high and decreases moving away from the droplet as expected. 


\subsubsection{Effect of substrate elasticity and thickness}
\label{Effect of substrate elasticity}

\begin{figure}
	\centering
	\vspace{1cm}
    (a) \hspace{0.47\textwidth} (b) \\
	\includegraphics[width=0.47\textwidth]{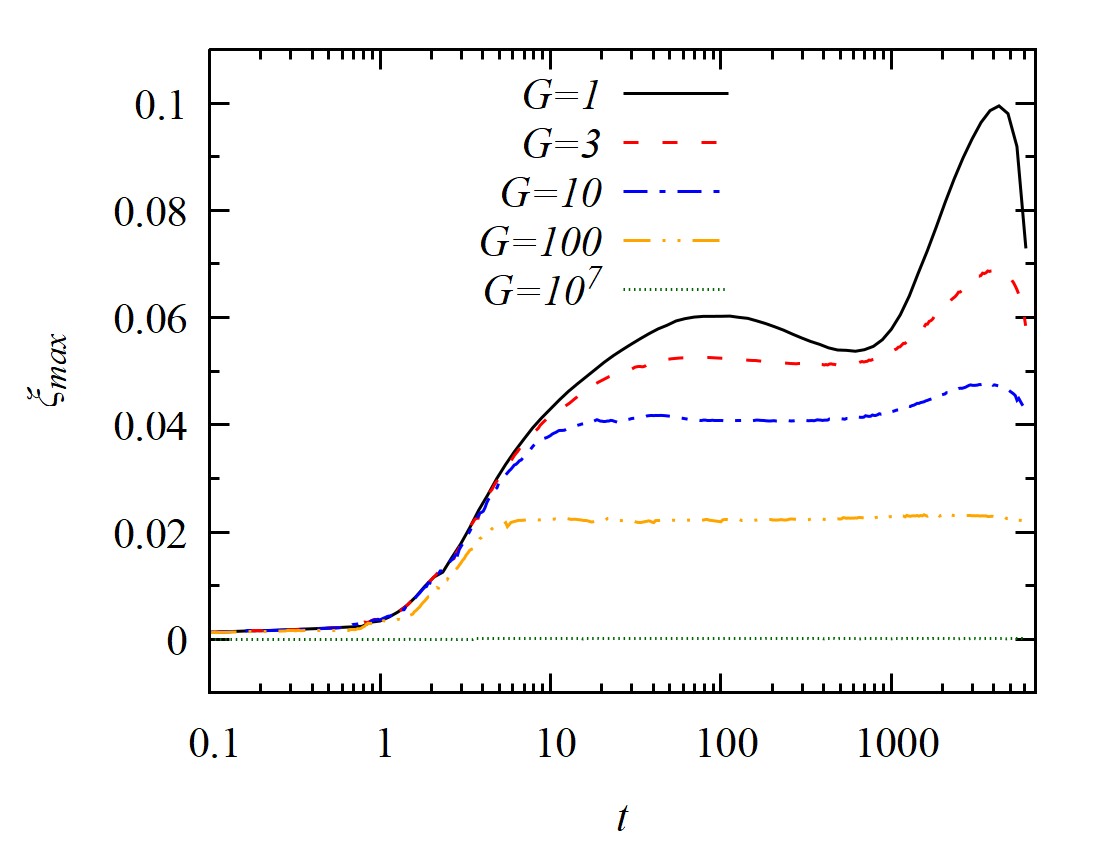} \hspace{0.5 cm}
	\includegraphics[width=0.47\textwidth]{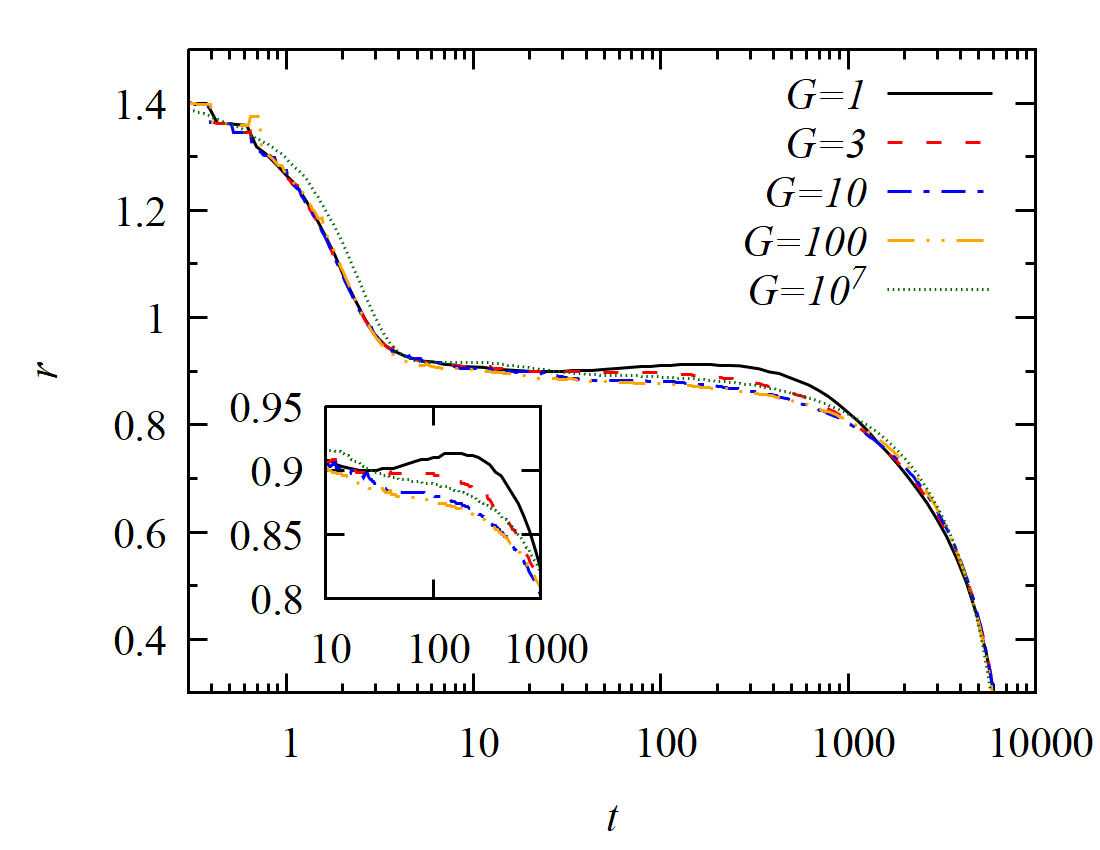} \\ 
    (c) \hspace{0.47\textwidth} (d) \\
	\includegraphics[width=0.47\textwidth]{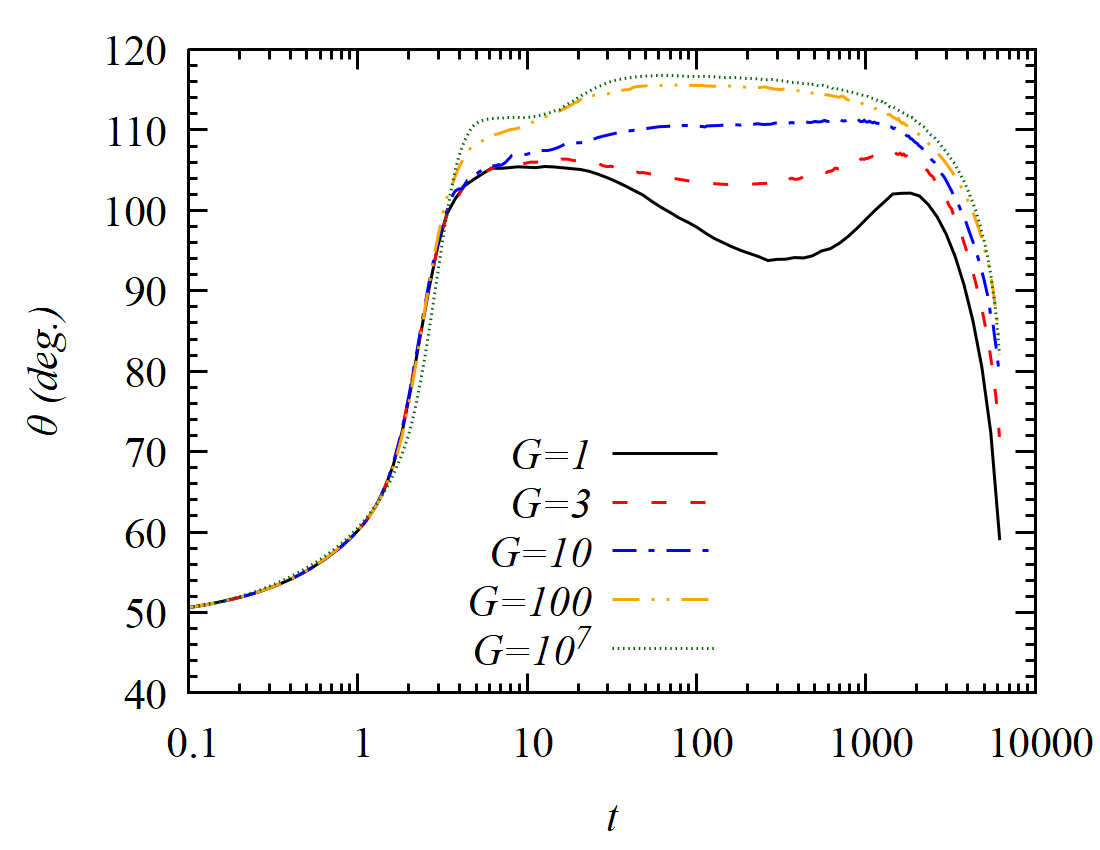}
    \hspace{0.5 cm}
	\includegraphics[width=0.47\textwidth]{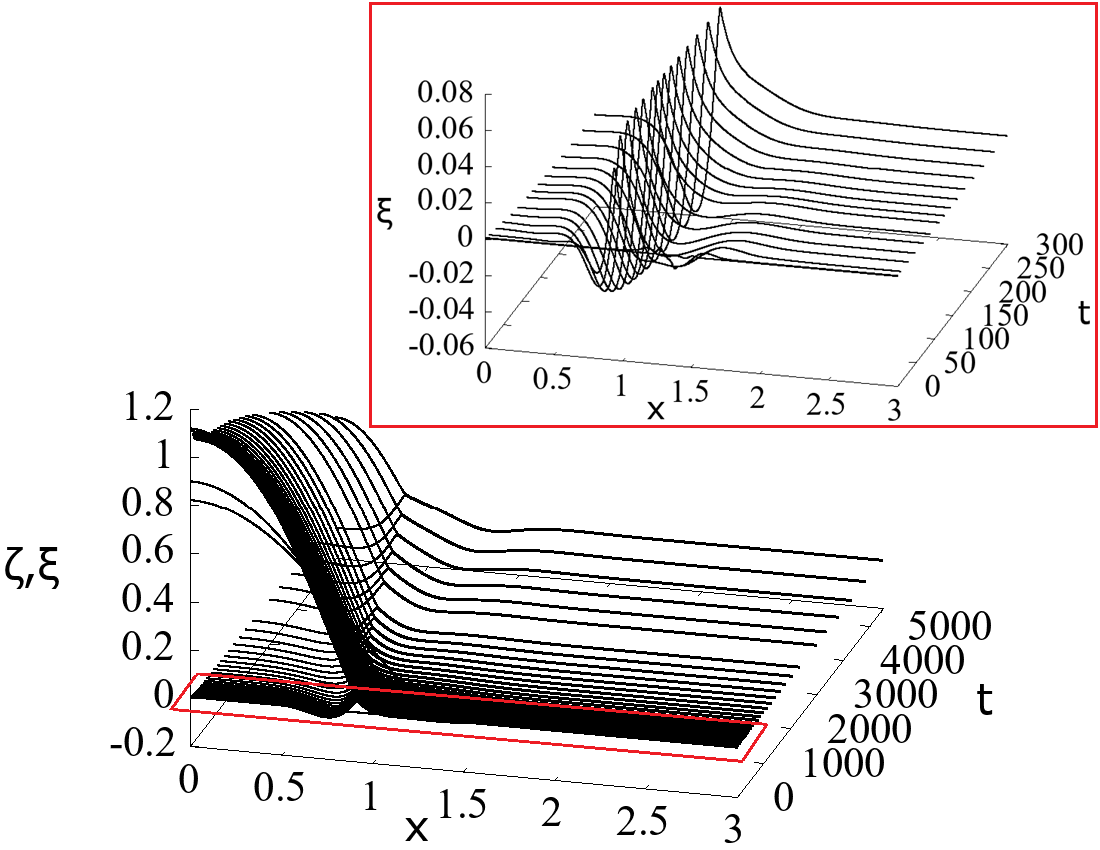}
    \caption{Time evolution of (a) the point of maximum deformation of the wetting ridge $\xi_{max}$, (b) the contact radius $r$ and (c) the apparent contact angle $\theta$ for a single droplet, varying substrate elasticity $G$ and for $M_a=0.005$. (d) Space-time plot of the droplet profiles at a soft substrate with $G=1$ and for $M_a=0.005$. The inset is a magnified view of the wetting ridge profiles during droplet spreading. The rest of the system parameters are the same with the 'base' case.}
    \label{fig:4}
\end{figure}

Here, we examine the effect of elasticity of the substrate by varying $G=\frac{\hat{E} \hat{R}_0}{\hat{\sigma}_0 \epsilon^3}$; this parameter measures the ratio of elastic to liquid-gas interfacial tension forces. $G$ is proportional to the shear modulus of the soft solid and therefore smaller values correspond to the case of softer substrates. By letting $G\rightarrow \infty$, the case of the rigid substrate can be recovered. In Fig. \ref{fig:4}a, we investigate the effect of substrate elasticity on the deformation of the soft solid, by plotting the evolution of the maximum deformation of the wetting ridge, $\xi_{max}$, with time. Naturally, it can be seen that the softer substrates deform more easily. 
Figs. \ref{fig:4}b and \ref{fig:4}c depict the time evolution of the contact radius and the apparent contact angle, respectively, of a single droplet evaporating on a rigid ($G=10^7$) and on soft solid substrates with $G=1, 3, 10, 100$. Following the work of \cite{charitatos_droplet_2021}, the apparent contact angle is defined as the largest angle between the tangent of the liquid-air interface $z=\zeta(x,t)$ and $z=0$. On the other hand, the contact radius is defined as the intersection point between the tangent of the largest angle and $z=0$.

At early times and for all examined cases, the droplet contact radius quickly decreases (Fig. \ref{fig:4}b), accompanied by an increase in the apparent contact angle (Fig. \ref{fig:4}c), while in parallel the size of the wetting ridge grows (Fig. \ref{fig:4}a). The initial droplet retraction, which is due to both droplet evaporation and the action of thermocapillary stresses, takes place faster for softer substrates. After the initial droplet retraction, the contact line remains apparently pinned for a significant amount of time ($t \approx 3-300$) with a relatively constant droplet footprint, indicating the stick-phase of the droplet spreading (see inset of Fig. \ref{fig:4}d). In fact, in the case of softer substrates (i.e. $G=1$) the constant contact radius is maintained throughout evaporation accompanied with a continuous decrease of the contact angle (see Fig. \ref{fig:4}c); the evaporation takes place in constant contact radius (CCR) mode. In contrast, for harder substrates (i.e. $G \geq 10$) the evaporation takes place in constant contact angle (CCA) mode, with a continuous slow decrease of the contact radius, in line with previous computational studies referring to droplet evaporation on rigid substrates \citep{pham_drying_2017}. This CCR mode observed in softer substrates has been previously reported in experimental studies concerning the evaporation of water droplets on compliant PDMS substrates \citep{lopes_evaporation_2012,gerber_wetting_2019}. After de-pinning, the contact line retracts continuously until the droplet fully evaporates. At the same time a non-monotonous behaviour of the contact angle is observed in line with previous experimental studies of \cite{lopes_evaporation_2012,lopes_influence_2013} and \cite{yu_experimental_2013}.

\begin{figure}
	\centering
	\vspace{1cm}
    (a) \hspace{0.47\textwidth} (b) \\
	\includegraphics[width=0.47\textwidth]{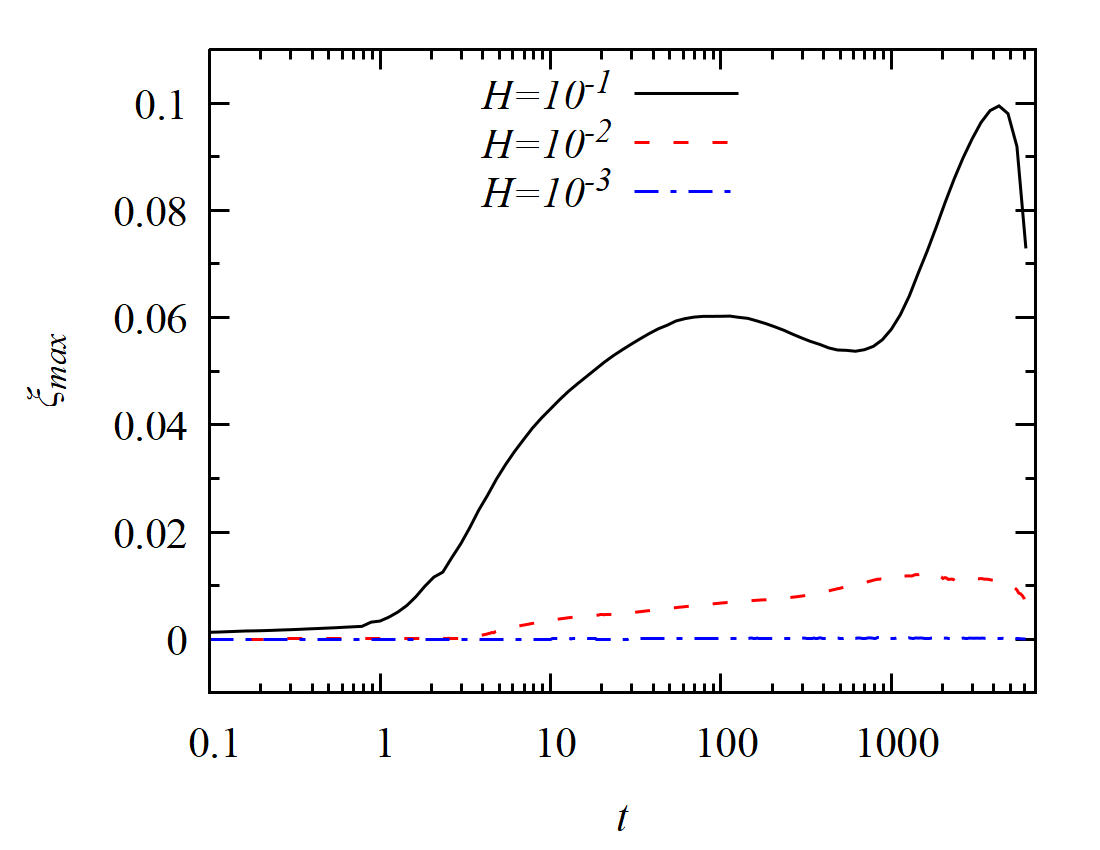} \hspace{0.5 cm}
	\includegraphics[width=0.47\textwidth]{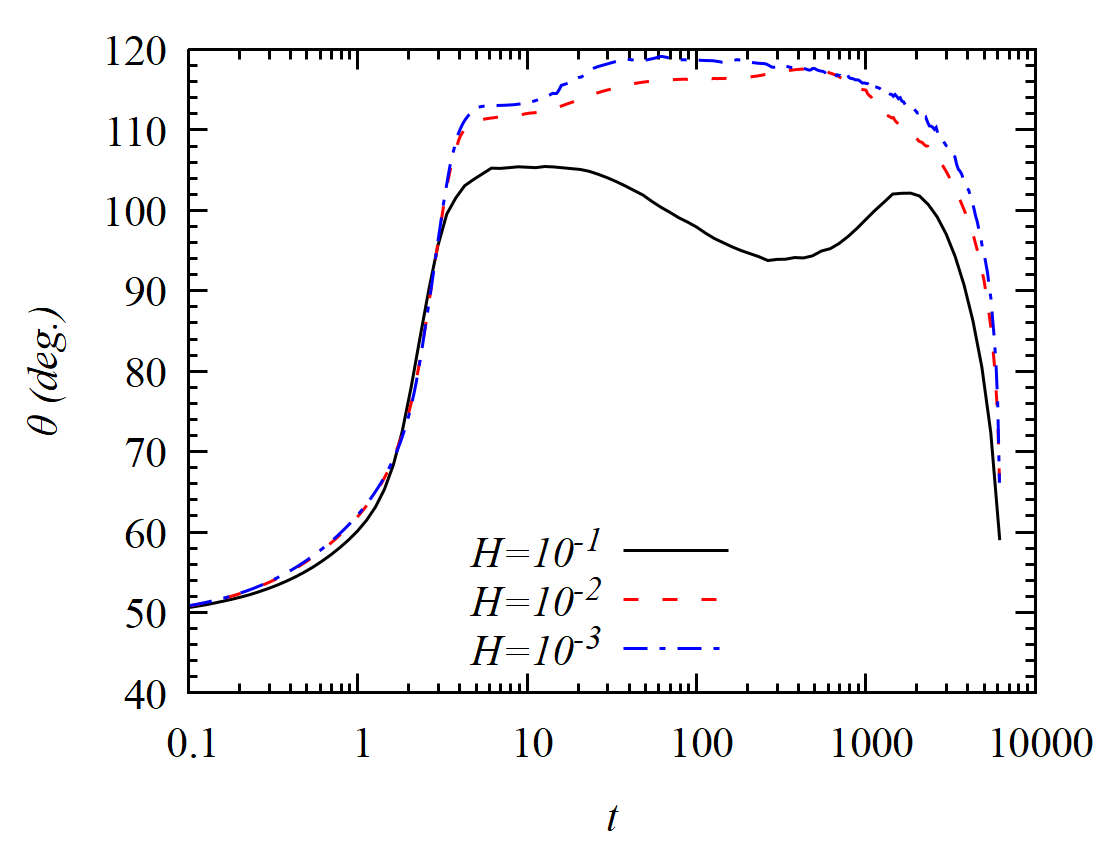}
    \caption{Time evolution of (a) the point of maximum deformation of the wetting ridge $\xi_{max}$ and (b) the apparent contact angle $\theta$ for a single droplet, varying substrate thickness $H$ and for $G=1$, $M_a=0.005$. The rest of the system parameters are the same with the 'base' case.}
    \label{fig:5}
\end{figure}


In Fig. \ref{fig:5}a, we investigate the effect of substrate thickness on the deformation of the soft solid, by plotting the maximum deformation of the wetting ridge, $\xi_{max}$, with time. It can be seen that the thicker substrates deform more easily than the thinner ones. Clearly, this is due to the fact that with decreasing thickness of the compliant substrate, less soft solid is available to deform, thereby increasing the resistance to the deformation of the substrate and leading to smaller wetting ridges. Consequently, making the substrate thinner can be seen as equivalent to making it more rigid, whereas thicker substrates behave similarly to softer ones. This effect is also reflected in the mode of evaporation. As it can be seen in Fig. \ref{fig:5}b, evaporation takes place in CCR mode for the thicker, hence softer, substrate ($H=10^{-1}$), and in CCA mode for the thinner, hence harder, substrate ($H=10^{-3}$), in line with the findings shown in Fig. \ref{fig:4}.

\subsection{Evaporation of a pair of droplets}
\label{sec:multiple_droplets}

Now that we have studied the basic characteristics of the flow for a single sessile evaporating droplet, we may proceed with the examination of a system of multiple volatile droplets. In particular, we will investigate the dynamics of a pair of droplets and focus on the effects of their interaction, either through the soft substrate or their atmosphere, on the dynamics of the drying process.

\begin{figure}
	\centering
	\vspace{1cm}
    (a) \hspace{0.47\textwidth} (b) \\
	\includegraphics[width=0.47\textwidth]{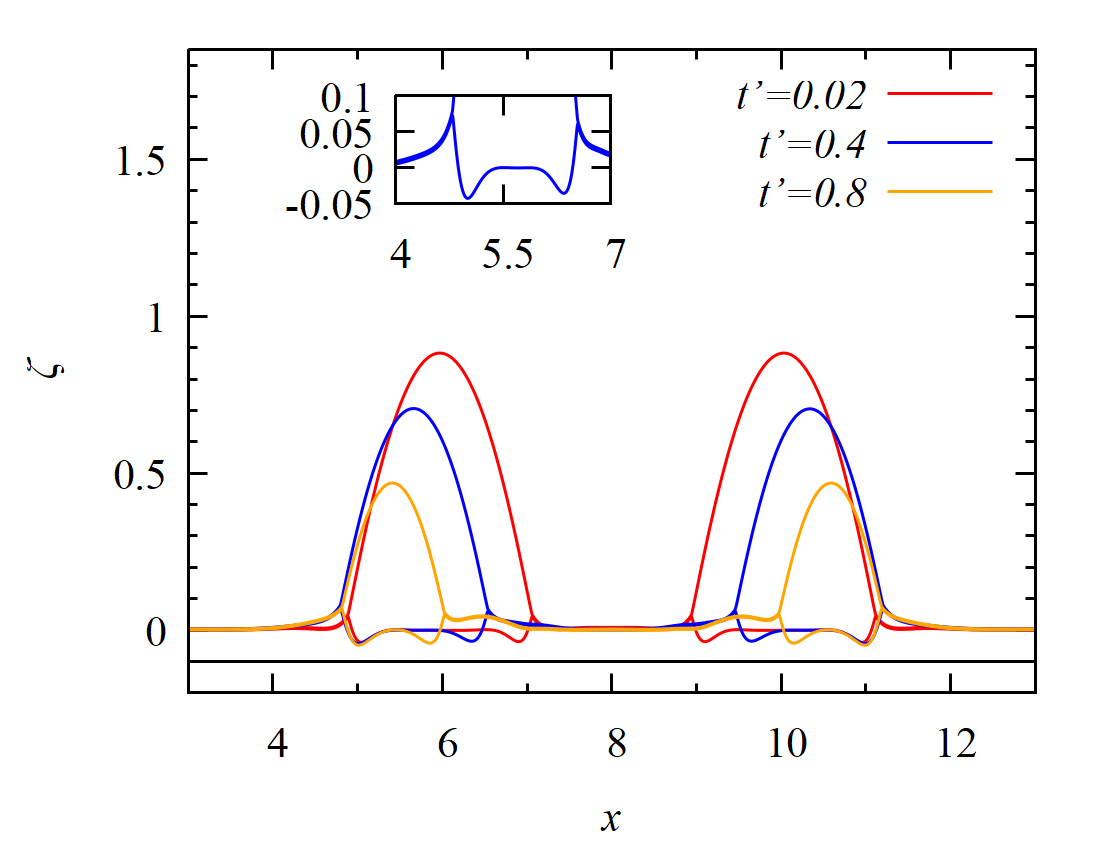} \hspace{0.5 cm}
	\includegraphics[width=0.47\textwidth]{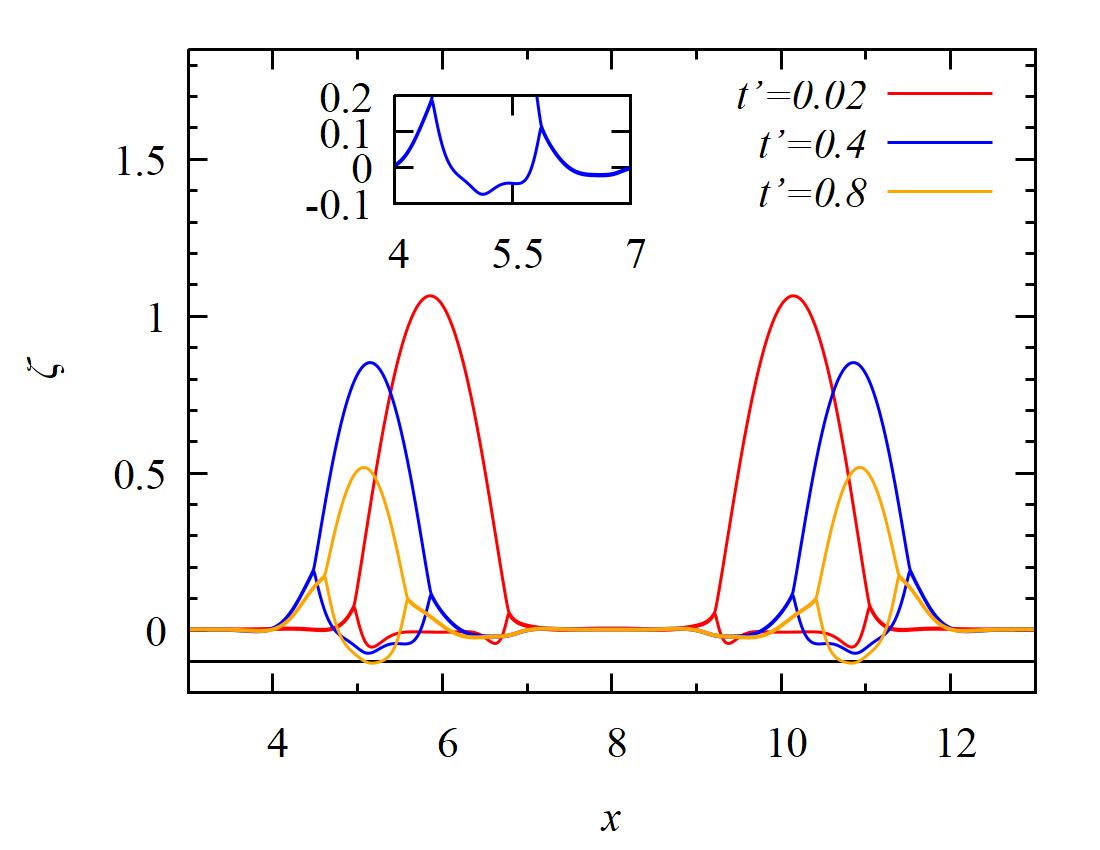} 
    \caption{Time evolution of the liquid-air ($\zeta$) and the liquid-solid ($\xi$) interfaces for 2 droplets drying on a soft substrate with $G=1$ and for (a) $M_a=0.001$ ($t_{ev}=4667$) and (b) $M_a=0.005$ ($t_{ev}=5889$), respectively. The inset is an enlargement of the height-range of the contact line region of the left drop at $t'=0.4$. The rest of the system parameters are the same with the 'base' case. }
    \label{fig:6}
\end{figure}

In Fig. \ref{fig:6}, we depict the time evolution of a pair of droplets evaporating on a compliant substrate with $G=1$. In these simulations, we fully take into account the effect of thermocapillarity and examine two cases with $M_a=0.001$ and $M_a=0.005$ in Figs. \ref{fig:6}a and \ref{fig:6}b, respectively.
An interesting observation is that in both cases the droplets appear to move away from each other as they dry out.  Regarding the deformation of the liquid-solid interface near the two contact lines, we observe that for low values of $M_a$ the height of the left and the right wetting ridge of each droplet is nearly symmetric (see inset of Fig. \ref{fig:6}a), whereas for higher values of $Ma$ the droplet deforms asymmetrically with the deformation of the soft solid in the inner region between the two droplets being somewhat smaller than the deformation in the outside region (see inset of Fig. \ref{fig:6}b). 
These observations provide a clear indication of the interaction of two droplets which may communicate either through the gas phase or through the developed stresses in the underlying viscoelastic substrate.

\subsubsection{Effect of the gas phase and thermocapillarity}
\label{Effect of the gas phase and thermocapillarity}

\begin{figure}
    \centering
    \includegraphics[width=1.0\textwidth]{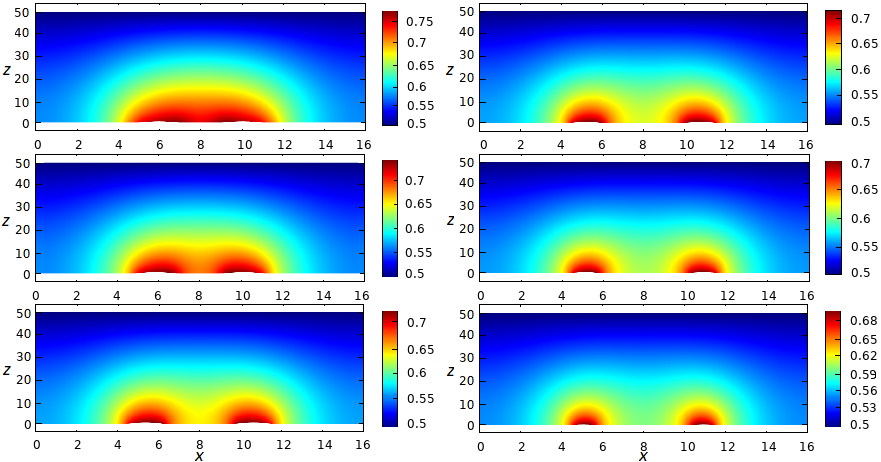}
    \caption{Gas phase concentration profiles at different time instants for  $G=1$ and $M_a=0.005$. From top to bottom: $t'=0.002$, $t'=0.02$, $t'=0.08$ (left column), $t'=0.2$, $t'=0.5$, $t'=0.8$ (right column) respectively ($t_{ev}=5889$). The rest of the system parameters are the same with the 'base' case.}
    \label{fig:7}
\end{figure}

\begin{figure}
	\centering
	\vspace{1cm}
    (a) \hspace{0.47\textwidth} (b) \\
	\includegraphics[width=0.47\textwidth]{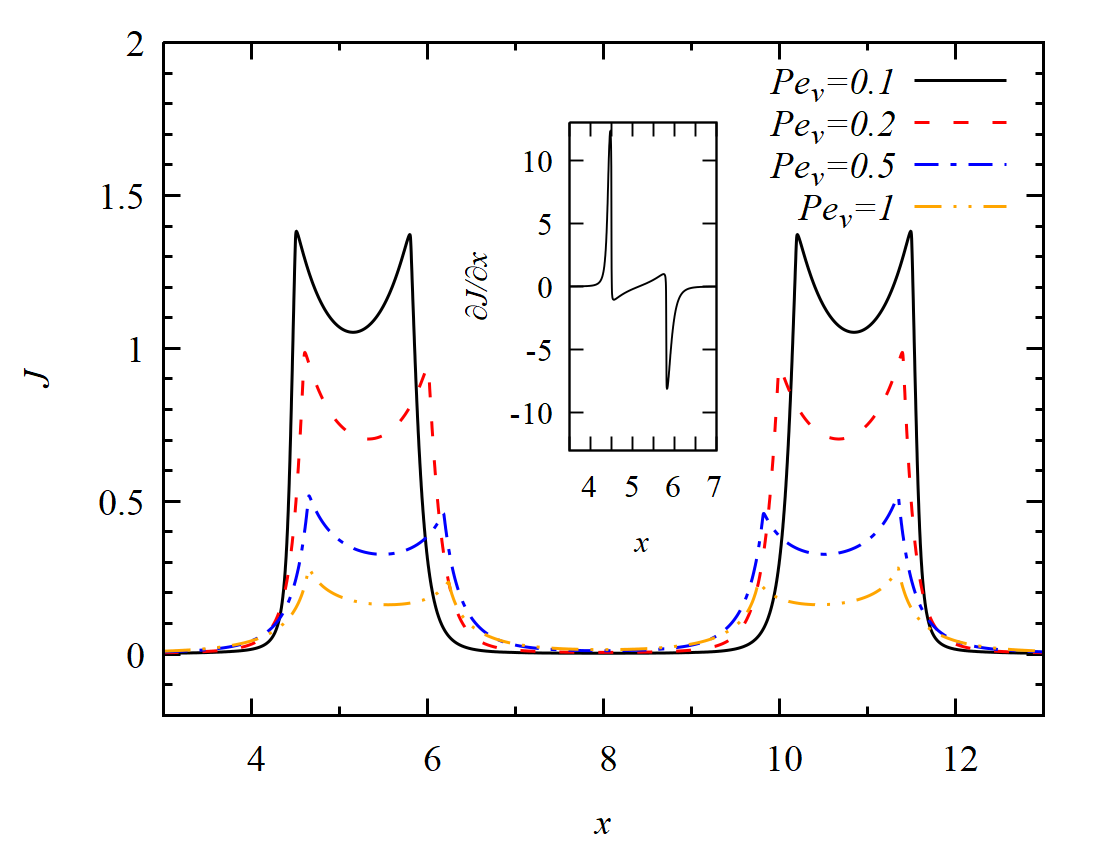} \hspace{0.5 cm}
	\includegraphics[width=0.47\textwidth]{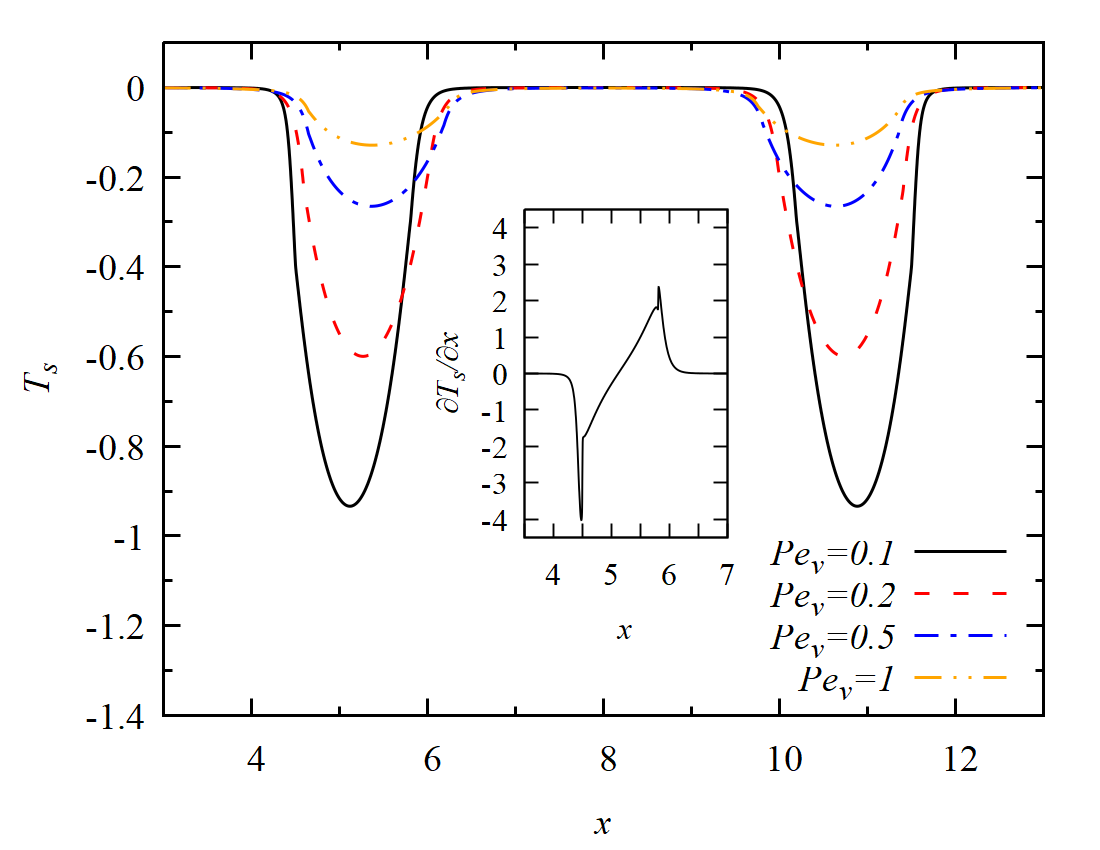} \\ 
    (c) \hspace{0.47\textwidth} (d) \\
	\includegraphics[width=0.47\textwidth]{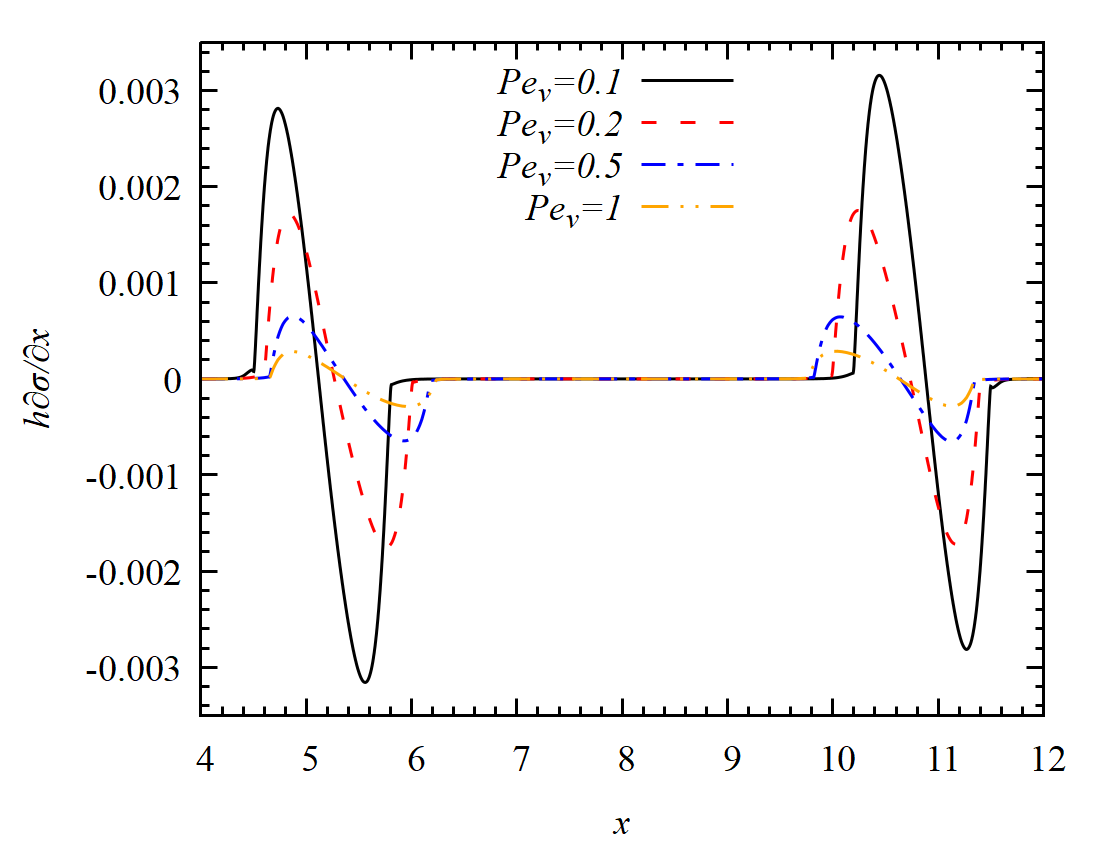} \hspace{0.5 cm}
	\includegraphics[width=0.47\textwidth]{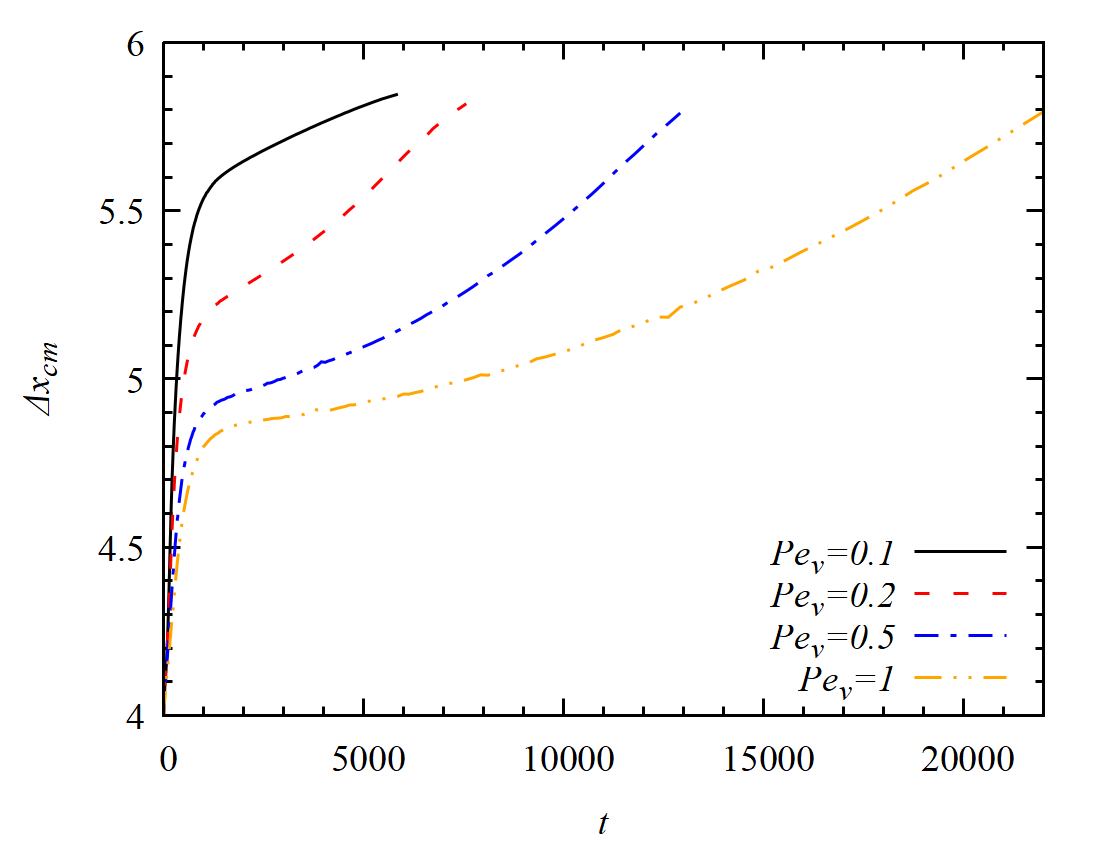}
    \caption{Effect of $Pe_v$ on the spatial profile of (a) the evaporation rate $J$, (b) the interfacial temperature $T_s$, (c) the Marangoni stresses, $h\frac{\partial \sigma}{\partial x}$, at $t'=0.5$. (d) The time evolution of the distance between the droplets' centers of mass, $\Delta x_{cm} = x_{cm,r}-x_{cm,l}$ for $G=1$, $M_a=0.005$ and the rest of the system parameters are the same with the 'base' case. The insets in panels (a) and (b) depict the spatial profiles for $Pe_v=0.1$ of $\partial J / \partial x$ and $\partial T_s / \partial x$, respectively.}
    \label{fig:8}
\end{figure}

In order to shed light on the physical mechanisms behind the observed dynamics, we will first focus on the gas phase and depict in Fig. \ref{fig:7} the vapour concentration in the atmosphere of the two droplets. As shown in this figure, the vapour concentration is higher between the two droplets than in their periphery. Since the evaporation flux is limited by diffusion (see Eq. \ref{vap BC d}), the higher saturation of the gas phase with vapour in the region between the two droplets results in weaker evaporation in that region; the spatial dependence of the evaporation flux $J$ is plotted in Fig. \ref{fig:8}a.
For all values of $Pe_v$ that we have examined, $J$ acquires an asymmetric profile along the liquid-gas interface of each droplet; this can be seen more clearly by plotting $\partial J / \partial x$ in the inset of the same figure for $Pe_v=0.1$. 

\begin{figure}
	\centering
	\vspace{1cm}
    (a) \hspace{0.47\textwidth} (b) \\
	\includegraphics[width=0.47\textwidth]{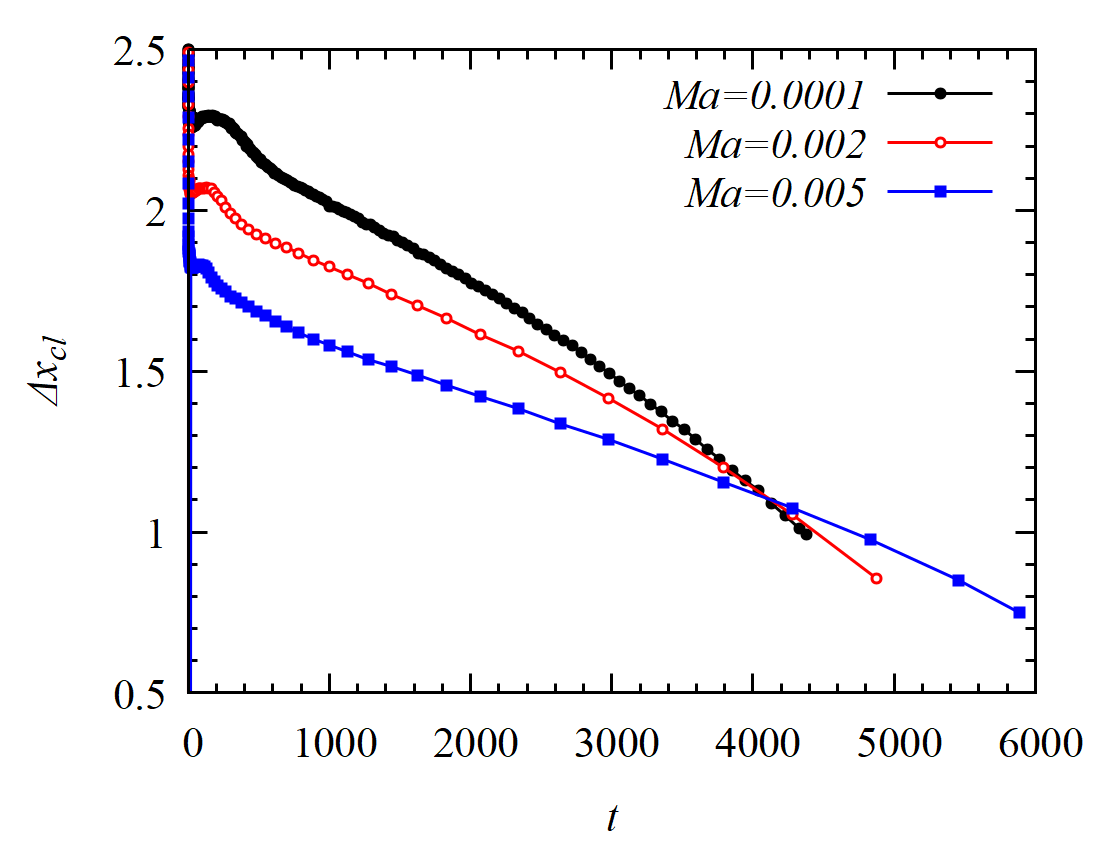} \hspace{0.5 cm}
	\includegraphics[width=0.47\textwidth]{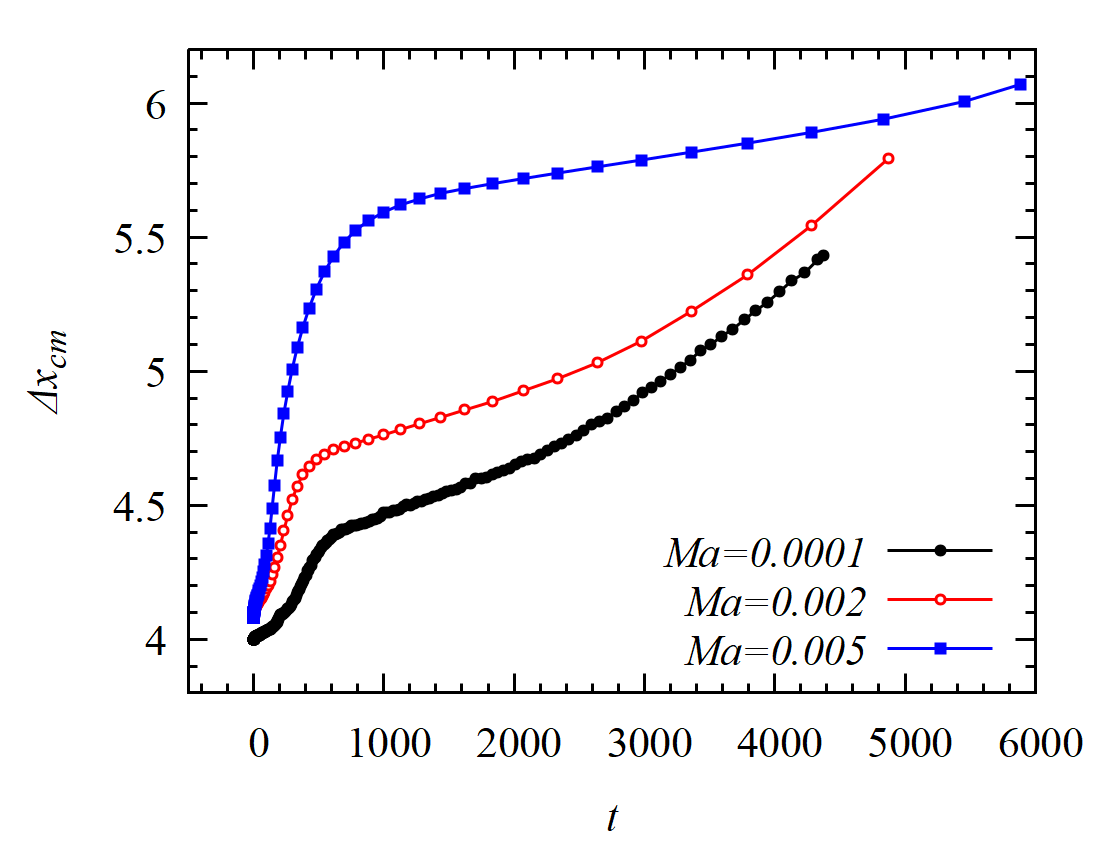} \\ 
    (c) \\
	\includegraphics[width=0.47\textwidth]{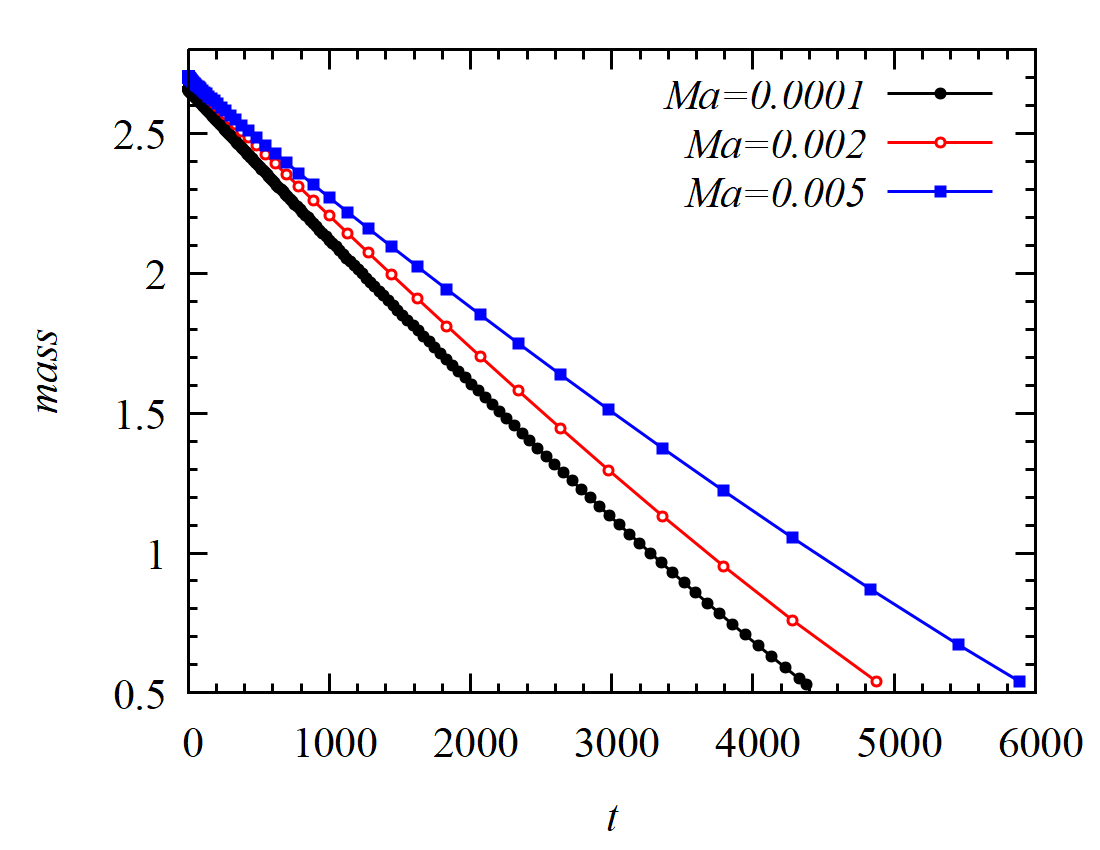}
    \caption{Time evolution of (a) the length of footprint of the left drop, $\Delta x_{cl}$, (b) the distance between the two centers of mass, $\Delta x_{cm}$, and (c) the system mass, for different values of $M_a$ and for $G=1$. The rest of the system parameters are the same with the 'base' case.} 
    \label{fig:9}
\end{figure}

Due to the effect of the latent heat, the evaporation flux affects the local interfacial temperature, which is depicted in Fig. \ref{fig:8}b; the liquid-gas interface is cooler than the rest of the drop and the interfacial temperature is lowest at the droplet apex.
The presence of temperature gradients affects in turn the flow field inside the droplet due to the action of Marangoni stresses, the spatial dependence of which, is plotted in Fig. \ref{fig:8}c; the Marangoni stress is proportional to $h \partial \sigma / \partial x$. Focusing first on each droplet, we notice that the Marangoni stresses, exhibiting opposite signs in the regions left and right from the droplet apex, act as a compressive force reducing the footprint of the droplet. The effect of thermocapillarity on the droplet footprint is shown very clearly in Fig. \ref{fig:9}a where we plot the length of footprint of the left drop, $\Delta x_{cl}$ (see also Fig. \ref{fig:1}c), defined as the distance between the maxima of left and right wetting ridge. Additionally, the asymmetric profile of the evaporation flux along the liquid-gas interface also induces a symmetry breaking in the interfacial temperature profile; this is clearly shown in the inset of Fig. \ref{fig:8}b where we plot the spatial dependence of $\partial T_s / \partial x$ for the droplet on the left side of the domain for $Pe_v=0.1$.

As a result, the Marangoni stresses not only compress the droplet but also contribute to their repulsion. This is demonstrated in Fig. \ref{fig:9}b where we plot the evolution of the distance between the centers of mass of the droplets on the left and the right side of the domain, $\Delta x_{cm}=x_{cm,r}-x_{cm,l}$ (see also Fig. \ref{fig:1}c), with time for different values of $M_a$. A similar effect is also shown in Fig. \ref{fig:8}d where enhanced repulsion is found for lower values of $Pe_v$; increase of $Pe_v$ corresponds to slower vapour diffusion enhancing the difference in the evaporation flux between the two sides of the droplets as shown in Fig. \ref{fig:8}a. Regarding the droplet lifetime, thermocapillarity plays a dual role; on one hand enhancing the evaporation rate in the region between the two droplets, as they move away from each other, but at the same time reducing the overall evaporation due to the compressive action of Marangoni stresses on the droplets. As it can be seen in Fig. \ref{fig:9}c, where we plot the total mass of the system, the droplet lifetime increases considerably with $M_a$, thus indicating that the latter effect is more significant.



\subsubsection{Effect of substrate elasticity in the absence of thermocapillarity}
\label{Effect of substrate elasticity in the absence of thermocapillarity}

\begin{figure}
	\centering
	\vspace{1cm}
    (a) \hspace{0.47\textwidth} (b) \\
	\includegraphics[width=0.47\textwidth]{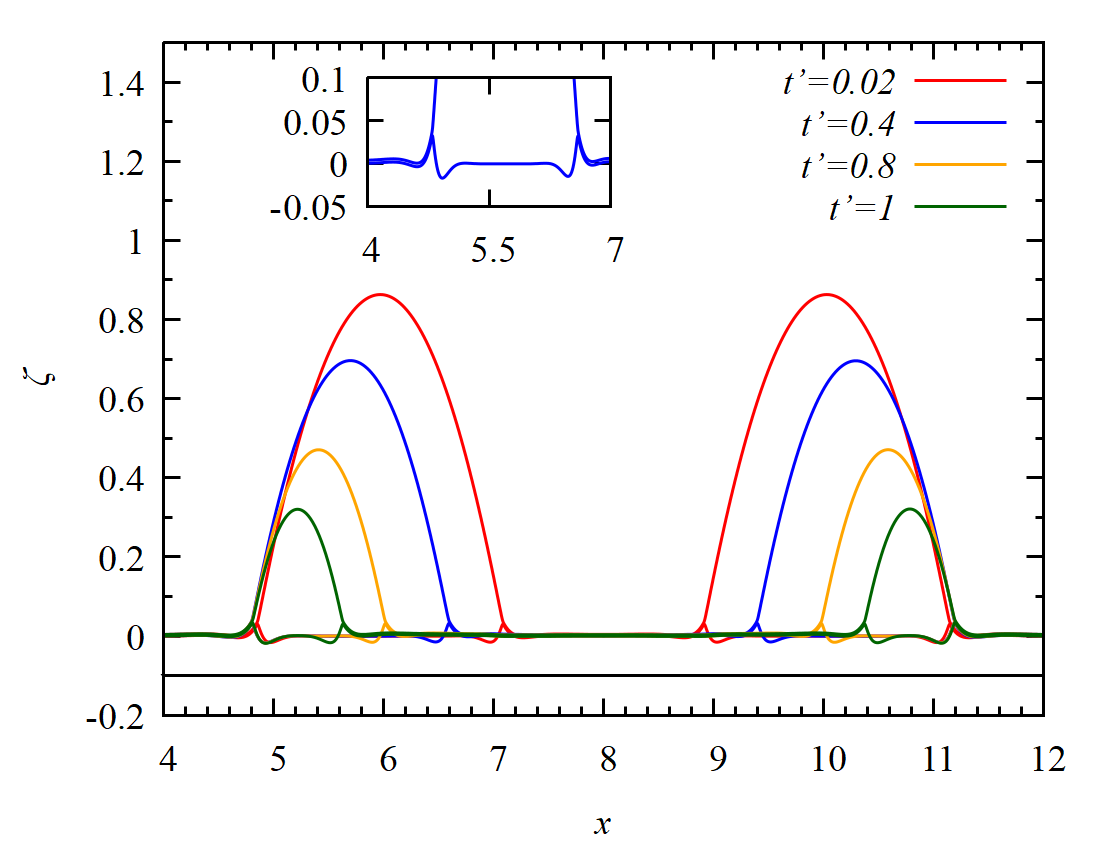} \hspace{0.5 cm}
	\includegraphics[width=0.47\textwidth]{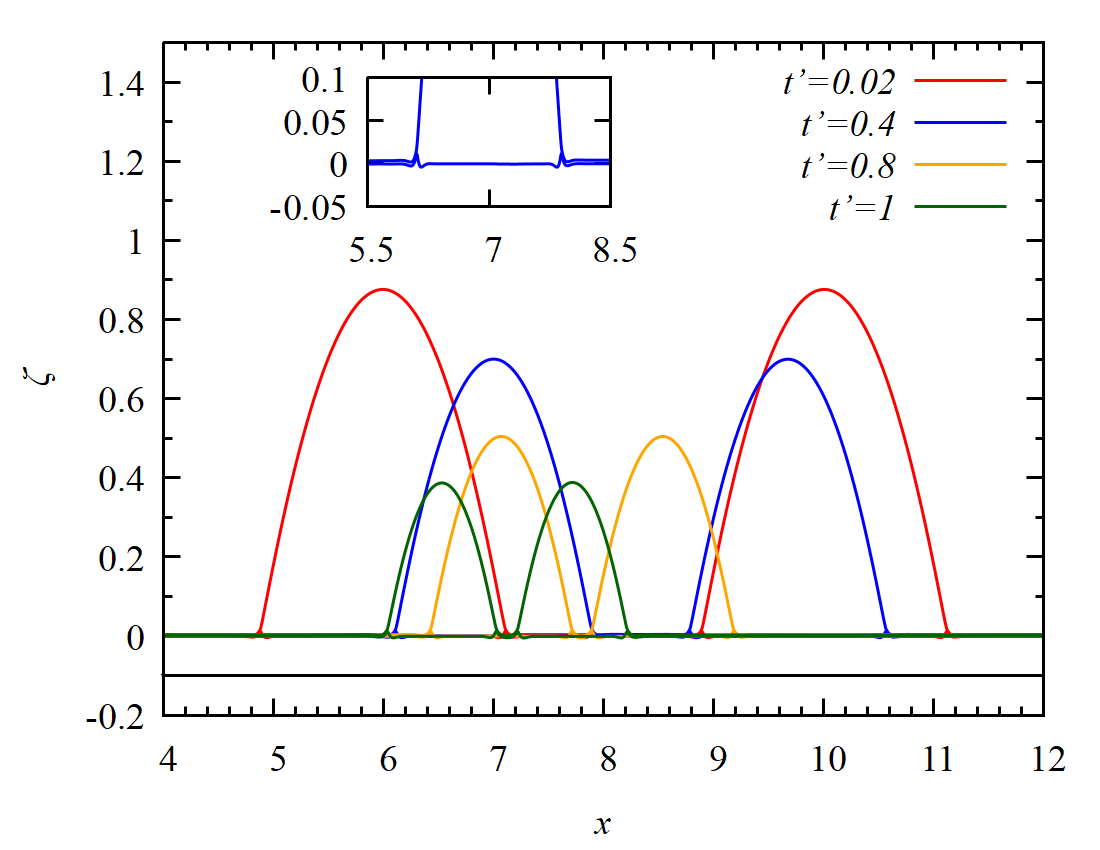} \\ 
    (c) \hspace{0.47\textwidth} (d) \\
	\includegraphics[width=0.47\textwidth]{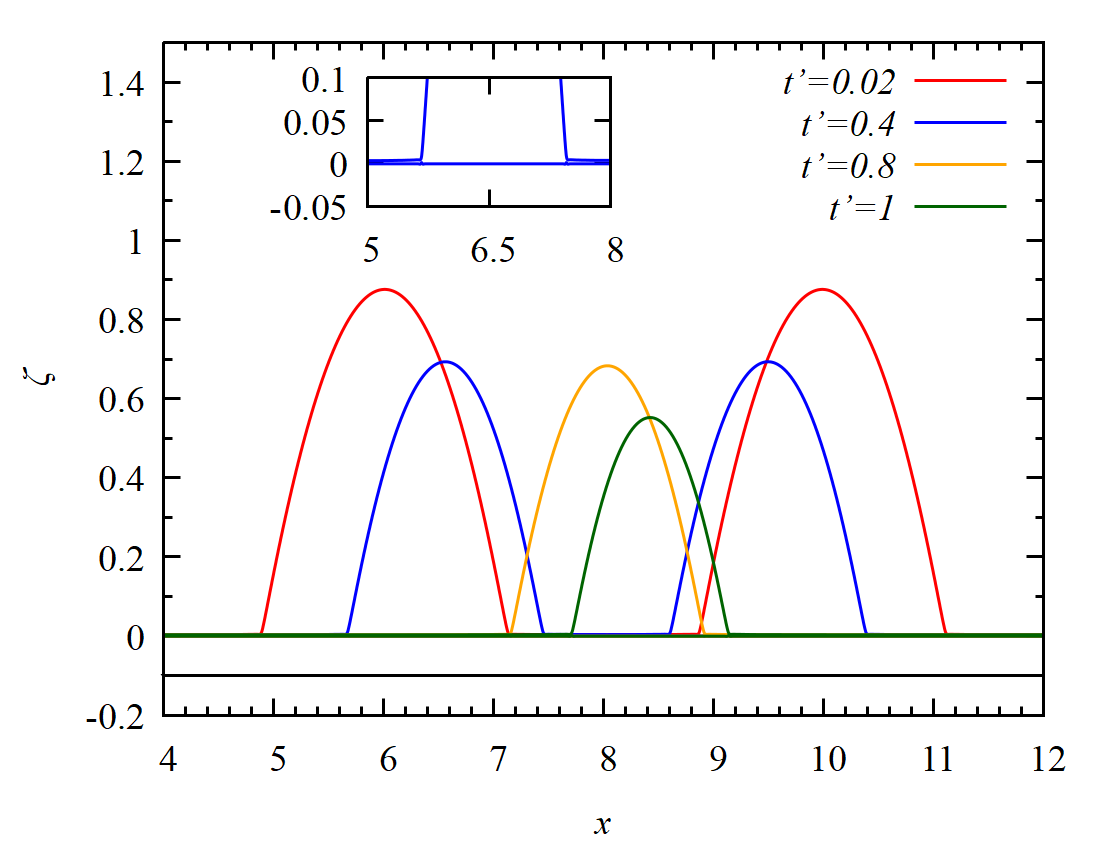} \hspace{0.5 cm}
 	\includegraphics[width=0.47\textwidth]{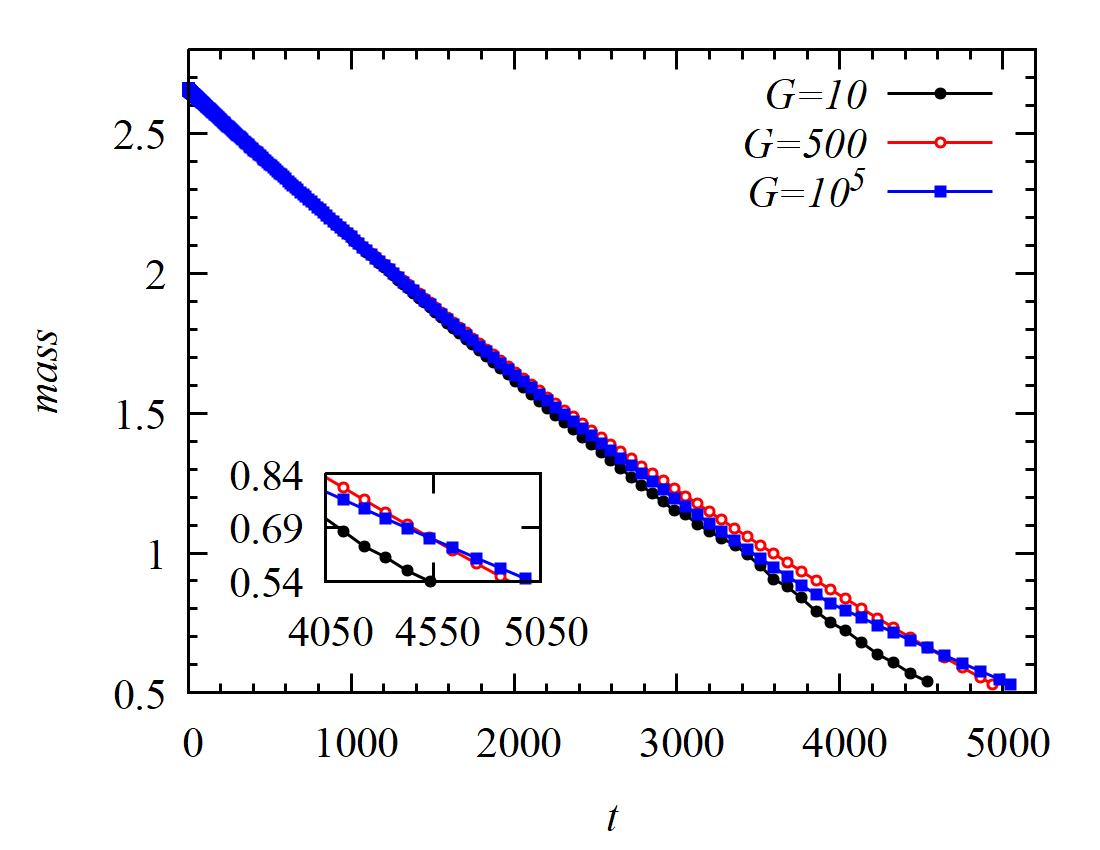}
    \caption{Time evolution of the liquid-air ($\zeta$) and the liquid-solid ($\xi$) interfaces for 2 droplets with (a) $G=10$ ($t_{ev}=4537$), (b) $G=500$ ($t_{ev}=4934$) and (c) $G=10^5$ ($t_{ev}=5048$) respectively, for $M_a=0$. The inset is an enlargement of the height-range of the contact line region of the left drop at $t'=0.4$. (d) Time evolution of the system mass, varying substrate elasticity $G$. The rest of the system parameters are the same with the 'base' case. }
    \label{fig:10}
\end{figure}

\begin{figure}
	\centering
	\vspace{1cm}
    (a) \hspace{0.47\textwidth} (b) \\
	\includegraphics[width=0.47\textwidth]{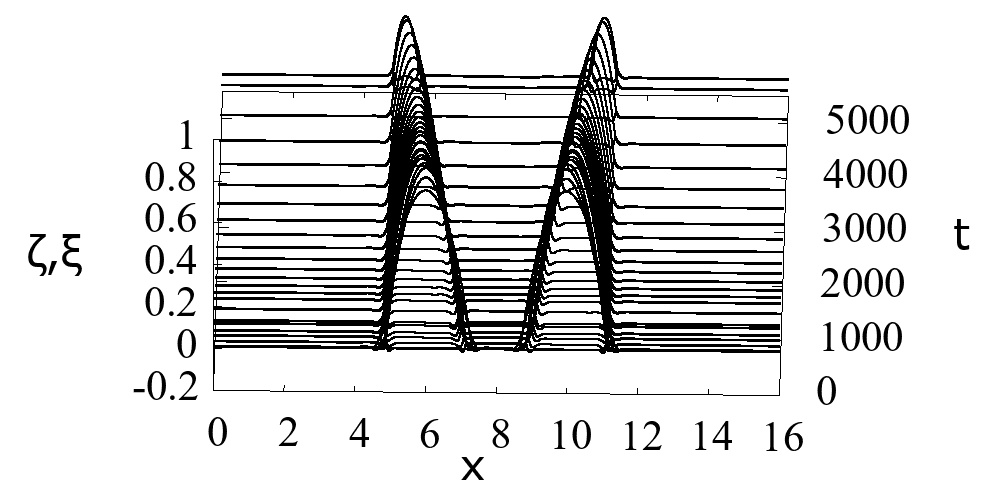} \hspace{0.5 cm}
	\includegraphics[width=0.47\textwidth]{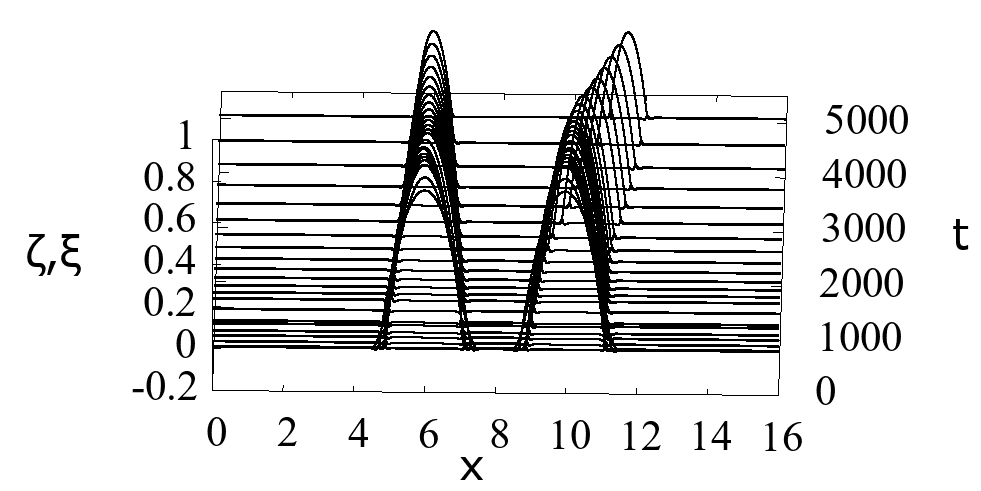}\\ 
    (c) \hspace{0.47\textwidth} (d) \\
	\includegraphics[width=0.47\textwidth]{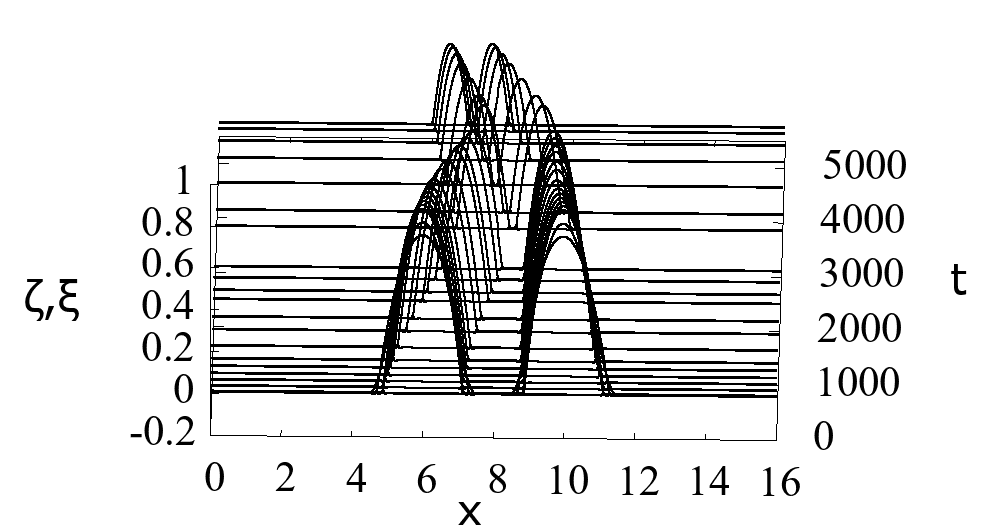}
    \hspace{0.5 cm}
        \includegraphics[width=0.47\textwidth]{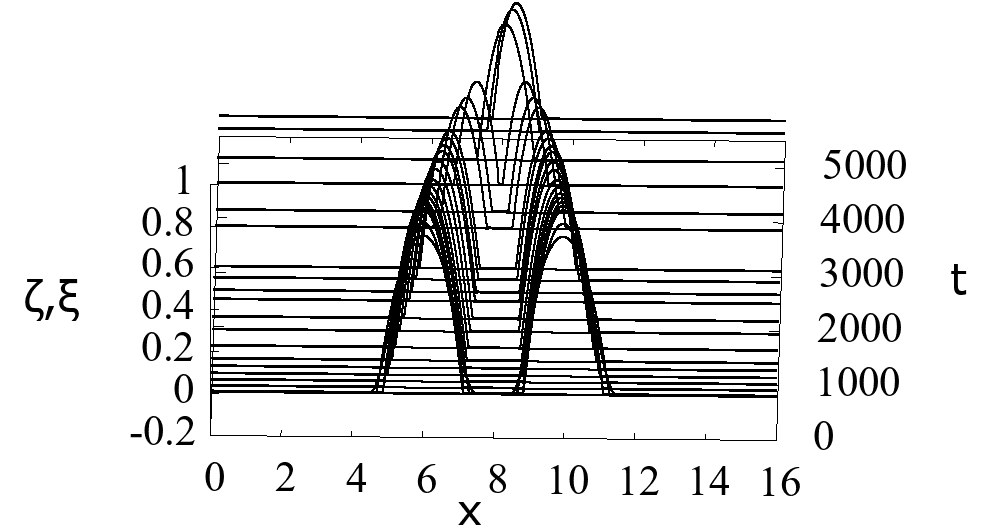}\\
    (e) \hspace{0.47\textwidth} (f) \\
        \includegraphics[width=0.47\textwidth]{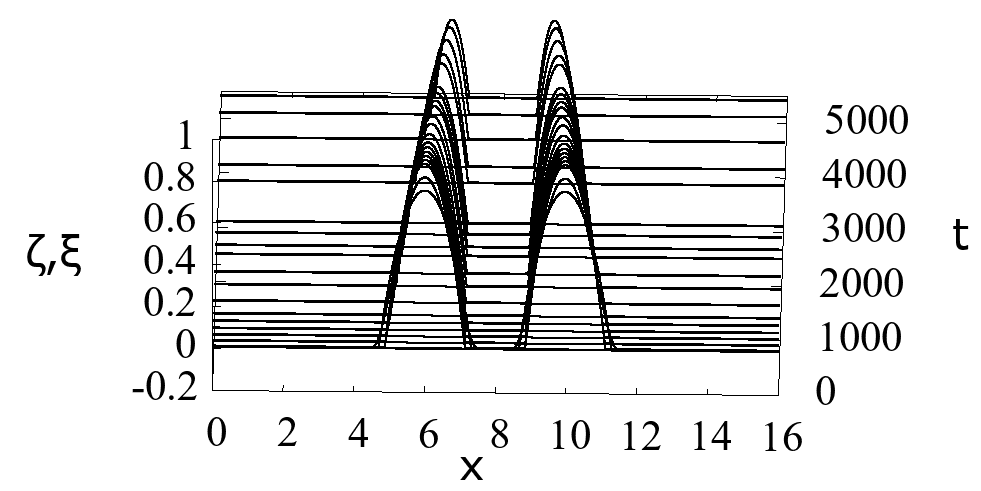}
    \hspace{0.5 cm}
        \includegraphics[width=0.47\textwidth]{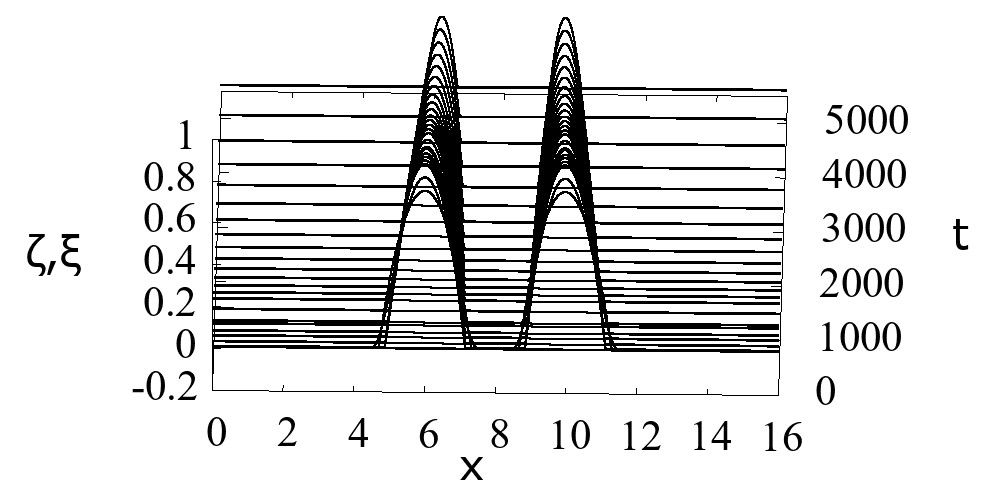}
     \caption{Space-time plots of the droplet profiles for (a) $G=10$, (b) $G=100$, (c) $G=500$, (d) $G=10^5$, (e) $G=10^6$ and (f) $G=10^7$, for $M_a=0$. The rest of the system parameters are the same with the 'base' case.}
    \label{fig:12}
\end{figure}

In order to investigate the effect of the substrate elasticity, we plot in Fig. \ref{fig:10} the time evolution of a pair of droplets evaporating on solid substrates with $G=10,500,10^5$; here, we neglect the effect of Marangoni stresses, i.e. $M_a=0$. Interestingly, we find that in the case of soft substrates the droplets repulse as they dry out (see Fig. \ref{fig:10}a), whereas in the case of stiffer substrates the droplets are attracted to each other (see Fig. \ref{fig:10}b). We note that in the latter case the droplets approach each other but do not coalesce; this behaviour is found in a well-define range of $G$ (i.e. $300 \leq G \leq 2 \times 10^4$ for $Ma=0$ and $2 \times 10^3 \leq G \leq 5 \times 10^4$ for $Ma=10^{-4}$). For very stiff substrates (see Fig. \ref{fig:10}c for $G=10^5$), the two droplets eventually coalesce, and the drying process continues as for a single droplet. The dynamics for these three cases are also presented in the form of space-time plots in Fig. \ref{fig:12} (see panels \ref{fig:12}a, \ref{fig:12}c and \ref{fig:12}d for $G=10,500,10^5$, respectively).

\begin{figure}
	\centering
	\vspace{1cm}
    (a) \hspace{0.47\textwidth} (b) \\
	\includegraphics[width=0.47\textwidth]{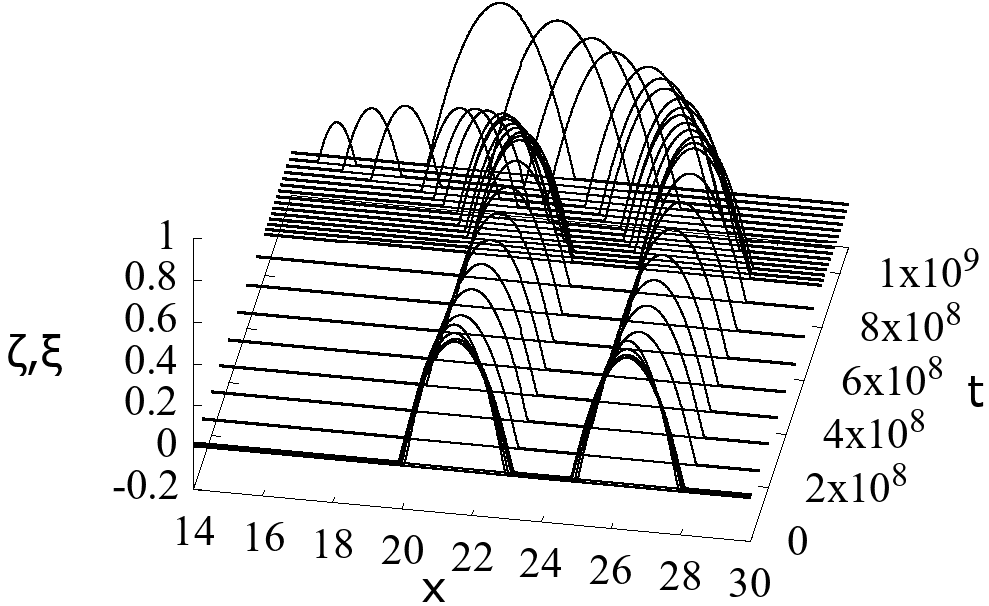} 
	\includegraphics[width=0.47\textwidth]{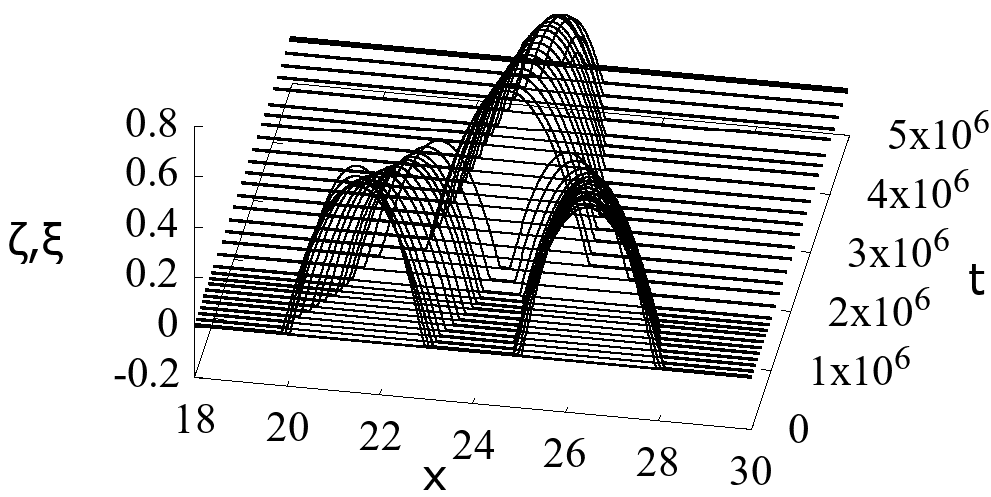} \hspace{0.5 cm}
     \caption{Space-time plots of the droplet profiles for (a) $E=10^{-8}$ and (b) $E=10^{-4}$, for $M_a=0$ and $G=10^7$. The domain length is $L=48$, $\mathcal{A}=120$ and the rest of the system parameters are the same with the 'base' case.}
    \label{fig:12b}
\end{figure}

\begin{figure}
	\centering
	\vspace{1cm}
    (a) \hspace{0.47\textwidth} (b) \\
	\includegraphics[width=0.47\textwidth]{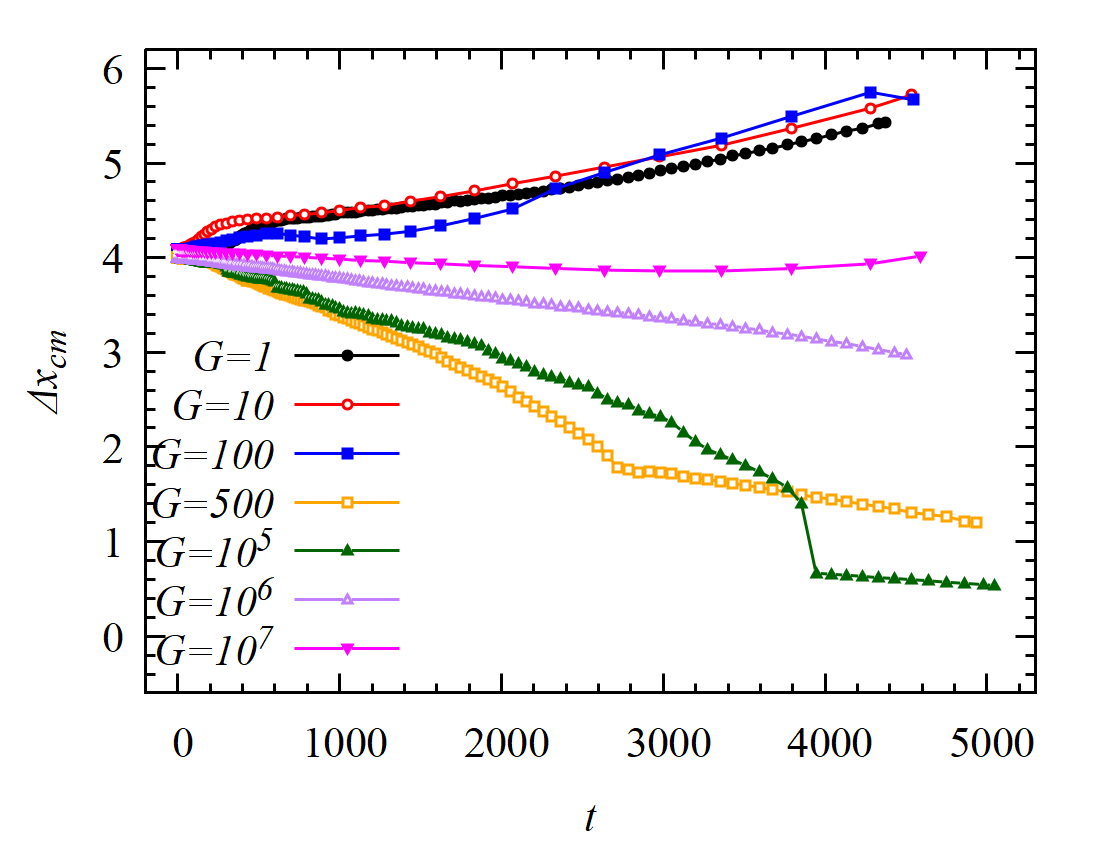} \hspace{0.5 cm}
	\includegraphics[width=0.47\textwidth]{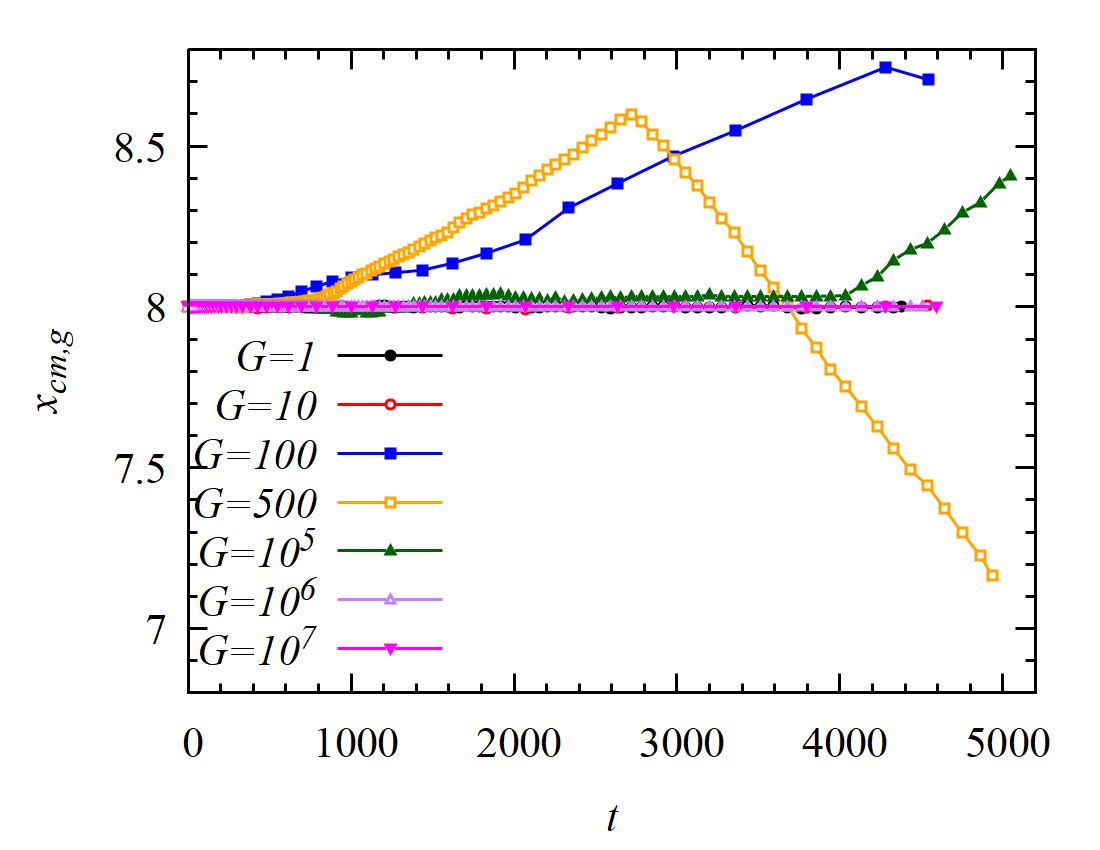} 
     \caption{Time evolution of (a) the distance between the two centers of mass $\Delta x_{cm}$ and (b) the center of mass of the system $x_{cm,g}$, for $M_a=0$. The rest of the system parameters are the same with the 'base' case.}
    \label{fig:11}
\end{figure}

Inspecting the space-time plots presented in Fig. \ref{fig:12} and comparing against the work of \cite{henkel_gradient-dynamics_2021} for non-volatile droplets, we notice a significant difference. For all the cases of volatile droplets that we have examined, varying the softness of the substrate, we find that mass transfer from one drop to the other does not take place and hence there is no droplet coarsening through Ostwald ripening mode; the latter mode was found to be dominant in the study of \cite{henkel_gradient-dynamics_2021}. For substrates with $G=10^5$ where droplets coalesce, the coarsening process takes place with a typical translation mode as shown in Fig. \ref{fig:12}d. After the two contact lines of the neighboring droplets come in contact, there is fast droplet coalescence. This is in contrast to the case shown in Fig. \ref{fig:12}c ($G=500$) where after the two adjacent contact lines touch each other, the droplets do not coalesce. Furthermore, we note that in parallel to the results of \cite{henkel_gradient-dynamics_2021}, the translation mode which defines exclusively the coarsening behaviour in Fig. \ref{fig:12}d leads to a symmetric movement of the droplets towards each other, without displacing the system center of mass (see also Fig. \ref{fig:10}c and Fig. \ref{fig:11}b). 
To check whether our model can capture the emergence of Ostwald ripening in the case of non-volatile droplets, we further examine in Fig. \ref{fig:12b}a the limit of negligible evaporation ($E=10^{-8}$) for a substrate with high rigidity ($G=10^7$); the equilibrium contact angle is taken to be approximately equal to $44^o$ ($A=120$) in order to consider a system with similar wetting characteristics as in \cite{henkel_gradient-dynamics_2021}. As it is clearly shown in this figure, in the limit of non-volatile droplets the Ostwald ripening emerges at late times, as expected, with mass transferring from one drop to the other until the smaller droplet vanishes, and all the mass is contained in the larger remaining droplet. On the other hand, in Fig. \ref{fig:12b}b we examine the same system with two volatile droplets ($E=10^{-4}$). As shown, evaporation takes place on much faster time scales and coalescence eventually occurs with the translation mode.

As explained by \cite{karpitschka_liquid_2016}, interaction of non-volatile droplets on the surface of elastic solids is determined by the balance of elasticity and capillary forces and the resulting local deformation of the soft solid in the contact line region. Depending on the stiffness (or thickness) of the substrate, the elastic meniscus in the contact line region between the two droplets rotates by an angle as compared to the meniscus of an isolated drop and the direction of the rotation determines whether the drop–drop interaction is attractive or repulsive.
In the case of drying droplets, though, the shape of the wetting ridge is not determined merely by elastocapillary phenomena but can also be significantly affected by the local evaporation rate (see relevant discussion in \cite{charitatos_droplet_2021}). Given the fact that the evaporation mass flux between the two contact lines of each droplet differs, this will also contribute to the imbalance of forces between the inner and outer contact lines, thereby affecting the mode of droplet interaction.
To examine in more detail the complex droplet dynamics of our system, we plot in Fig. \ref{fig:11}a the evolution of the distance between the center of mass of the droplets, $\Delta x_{cm}$, with time. It can be seen that in the case of soft substrates ($G \leq 100$) the droplets repulse, since $\Delta x_{cm}$ continuously increases throughout evaporation. For harder substrates ($G>100$), though, the imbalance acts in the opposite direction pushing the droplets towards each other. In the case of nearly rigid substrates ($G \geq 10^6$), the deformation of the viscoelastic solid is so small that this imbalance does not play an important role and thus the distance between the two droplets does not change significantly during evaporation.

The different modes of the drying process affect also the lifetime of the droplets. As shown in Fig. \ref{fig:10}d, the evaporation is faster in softer substrates, due to the increased distance between the two droplets and the fact that less amount of vapor is trapped amidst the repulsing droplets leading to enhanced evaporation fluxes. In contrast, the greater amount of vapour trapped amidst the droplets when they attract in stiffer substrates, retards the evaporation significantly. 
Moreover, we notice that although for soft substrates the symmetry of the system is preserved throughout the drying process, this is not the case for substrates with intermediate stiffness. In fact, as it can be seen in Fig. \ref{fig:10}b (and Fig. \ref{fig:12}c), the pair of droplets at late stages of evaporation starts moving to the left exhibiting a clear symmetry breaking; the mechanisms for this behaviour will be investigated in detail below. Similarly, as shown in Fig. \ref{fig:10}c (and Fig. \ref{fig:12}d) for $G=10^5$, the droplet that has emerged after the coalescence of the two droplets appears to move slightly to the right, also indicating a symmetry breaking of the system, albeit with a somewhat smaller droplet displacement from the system center of mass.

As noted above, the elasticity of the substrate affects not only the relative distance between the droplets but may also lead under conditions to a symmetry breaking with the center of mass of the system, $x_{cm,g}$ being displaced from its initial position, i.e. the droplets appear to be 'walking' along the viscoelastic substrate. In Fig. \ref{fig:11}b, we depict the effect of $G$ on the evolution of the position of the system center of mass, $x_{cm,g}$. As shown in this figure, for very soft and very hard substrates (i.e. $G=10$ and $G \geq 10^6$) the center of mass of the system remains at $x_{cm,g}=8$ and the symmetry is preserved throughout the drying process. This is not the case, though, for substrates with intermediate stiffness where  symmetry breaking is found; we note that the system symmetry is considered broken when the center of mass of the system has moved  $\pm \SI{10}{\percent}$ of its initial maximum height (i.e. 0.1 dimensionless distance) from its initial position. It should be pointed that these asymmetric solutions are spontaneous and emerge due to disturbances of the numerical finite element scheme, while they appear to be stable with the increase in mesh resolution. To make sure that the symmetry breaking is not artificially introduced by the imposed boundary conditions, we varied the size of the domain or even applied periodic boundary conditions in the $x$-direction; these efforts are presented in detail in the Appendix \ref{appendix:raw2}. As discussed therein, neither the type of imposed boundary conditions or the domain size qualitatively affect the observed droplet behaviour. It is important to note that a spontaneous symmetry breaking has been also a matter of interest in earlier experimental and computational studies \citep{hernandez-sanchez_symmetric_2012, leong_droplet_2020} of non-volatile droplets; \cite{leong_droplet_2020} examined the growth of an inflating droplet on viscoelastic soft substrate and also observed asymmetric solutions for substrates with intermediate stiffness.

\subsubsection{Effect of substrate elasticity in the presence of thermocapillarity}
\label{Effect of substrate elasticity in the presence of thermocapillarity}

\begin{figure}
	\centering
	\vspace{1cm}
    (a) \hspace{0.47\textwidth} (b) \\
	\includegraphics[width=0.47\textwidth]{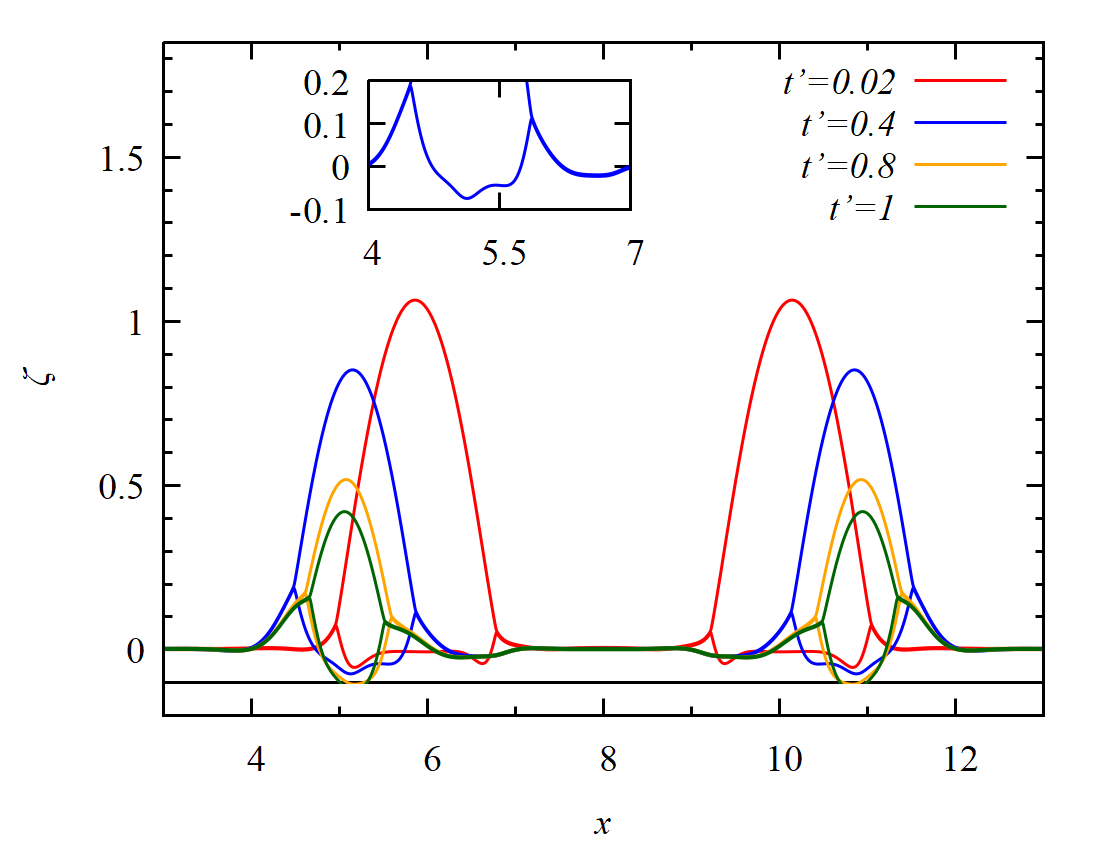} \hspace{0.5 cm}
	\includegraphics[width=0.47\textwidth]{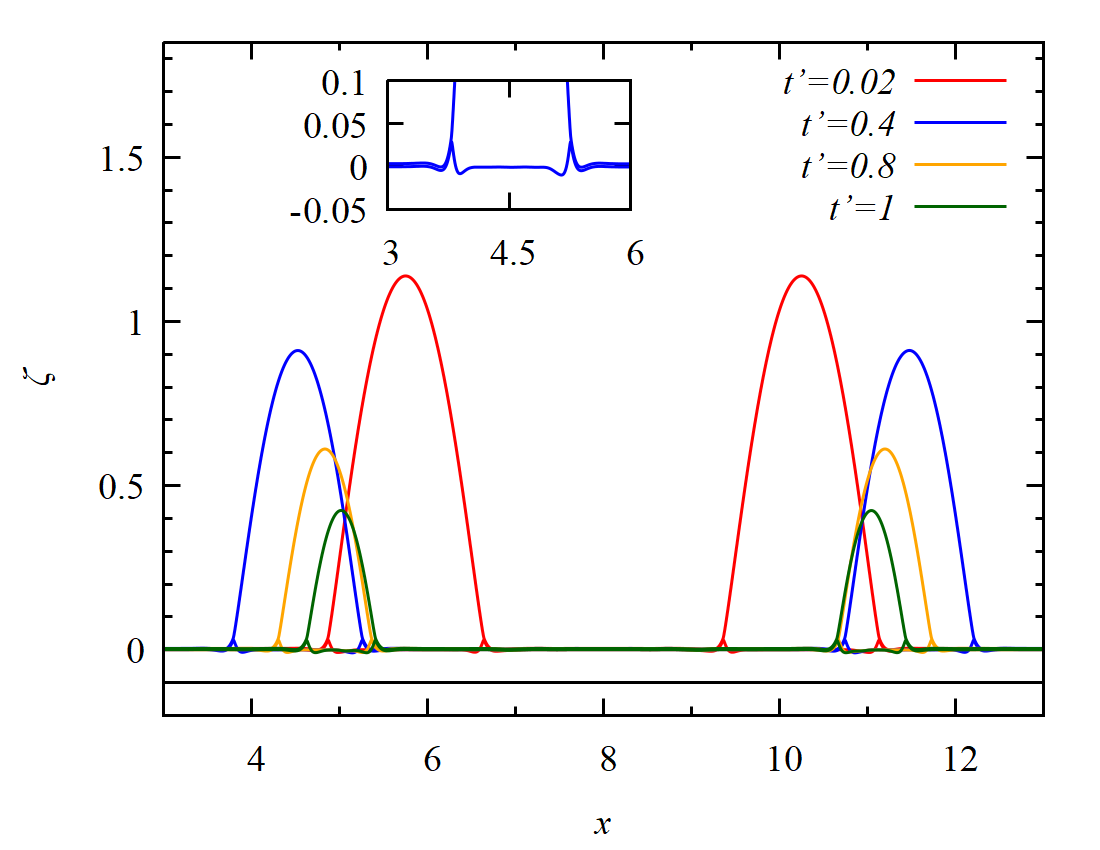} \\ 
    (c) \hspace{0.47\textwidth} (d) \\
	\includegraphics[width=0.47\textwidth]{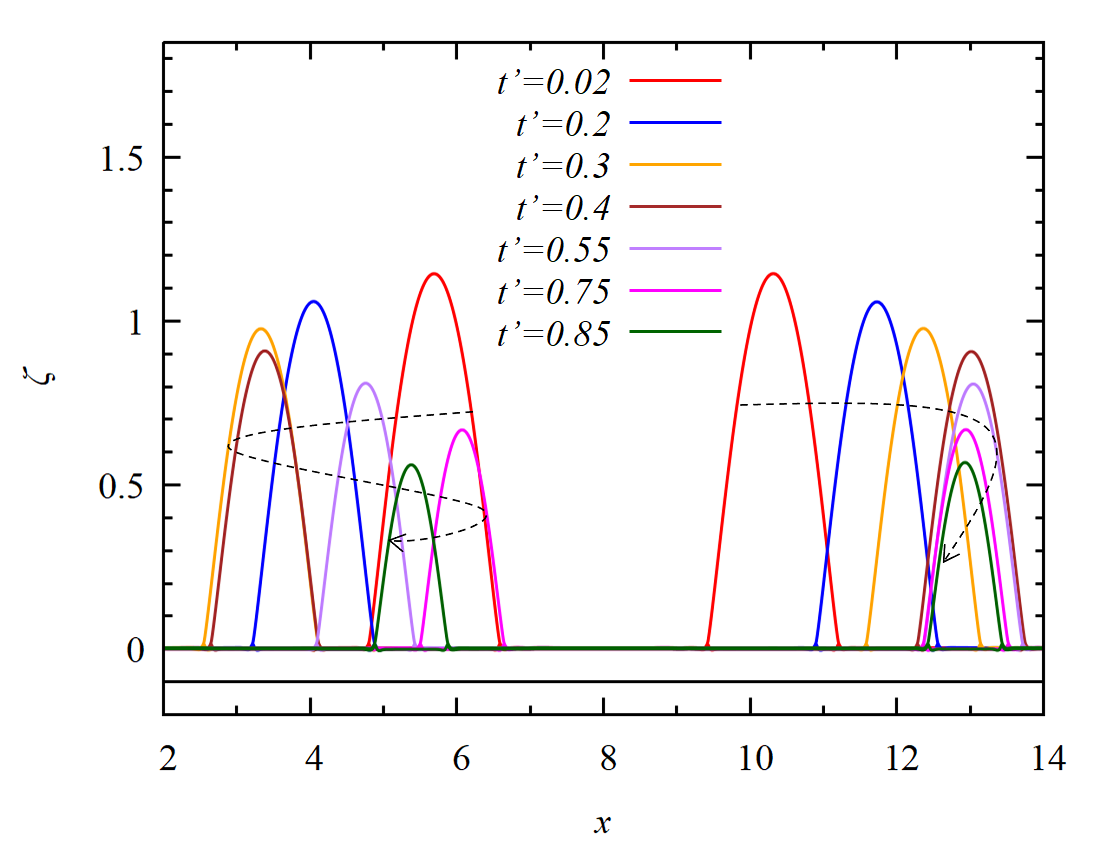} \hspace{0.5 cm}
 	\includegraphics[width=0.47\textwidth]{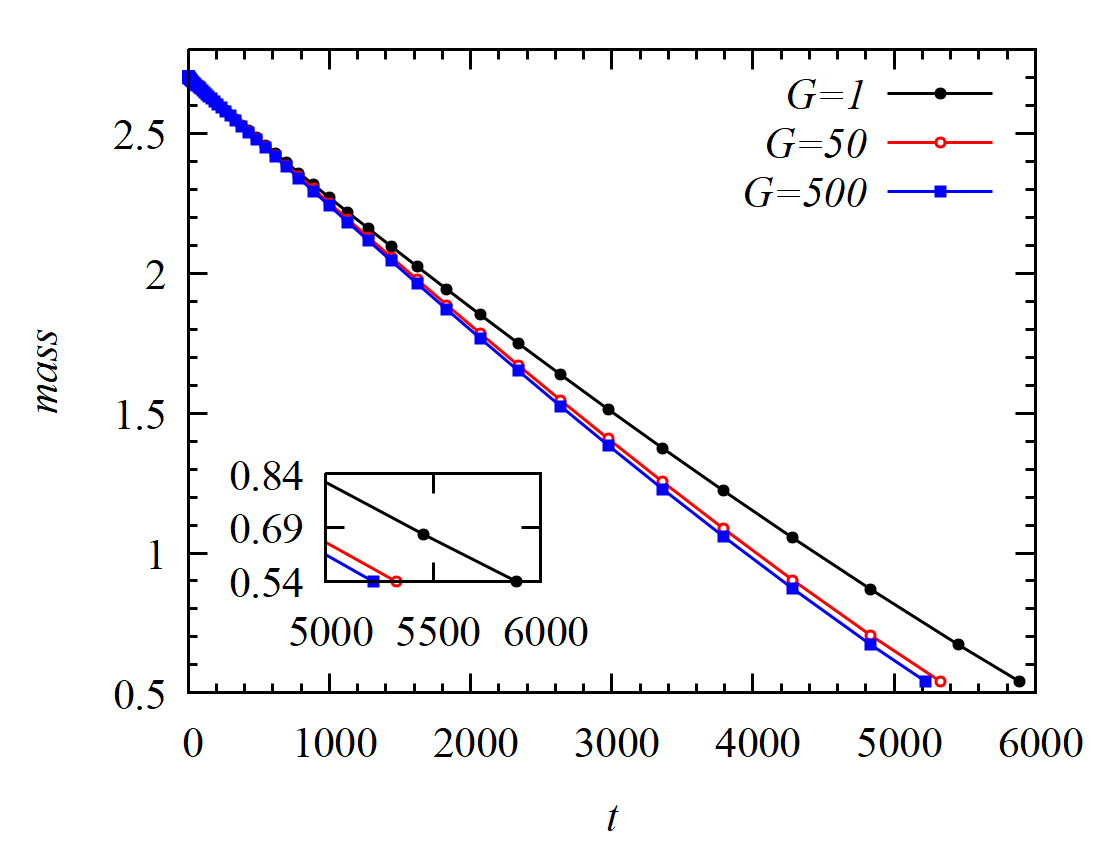}
    \caption{Time evolution of the liquid-air ($\zeta$) and the liquid-solid ($\xi$) interfaces for 2 droplets with (a) $G=1$ ($t_{ev}=5889$), (b) $G=50$ ($t_{ev}=5329$) and (c) $G=500$ ($t_{ev}=5222$) respectively, for $M_a=0.005$. The inset is an enlargement of the height-range of the contact line region of the left drop at $t'=0.4$. (d) Time evolution of the system mass varying substrate elasticity $G$. The rest of the system parameters are the same with the 'base' case.}
    \label{fig:13}
\end{figure}

\begin{figure}
	\centering
	\vspace{1cm}
    (a) \hspace{0.47\textwidth} (b) \\
	\includegraphics[width=0.47\textwidth]{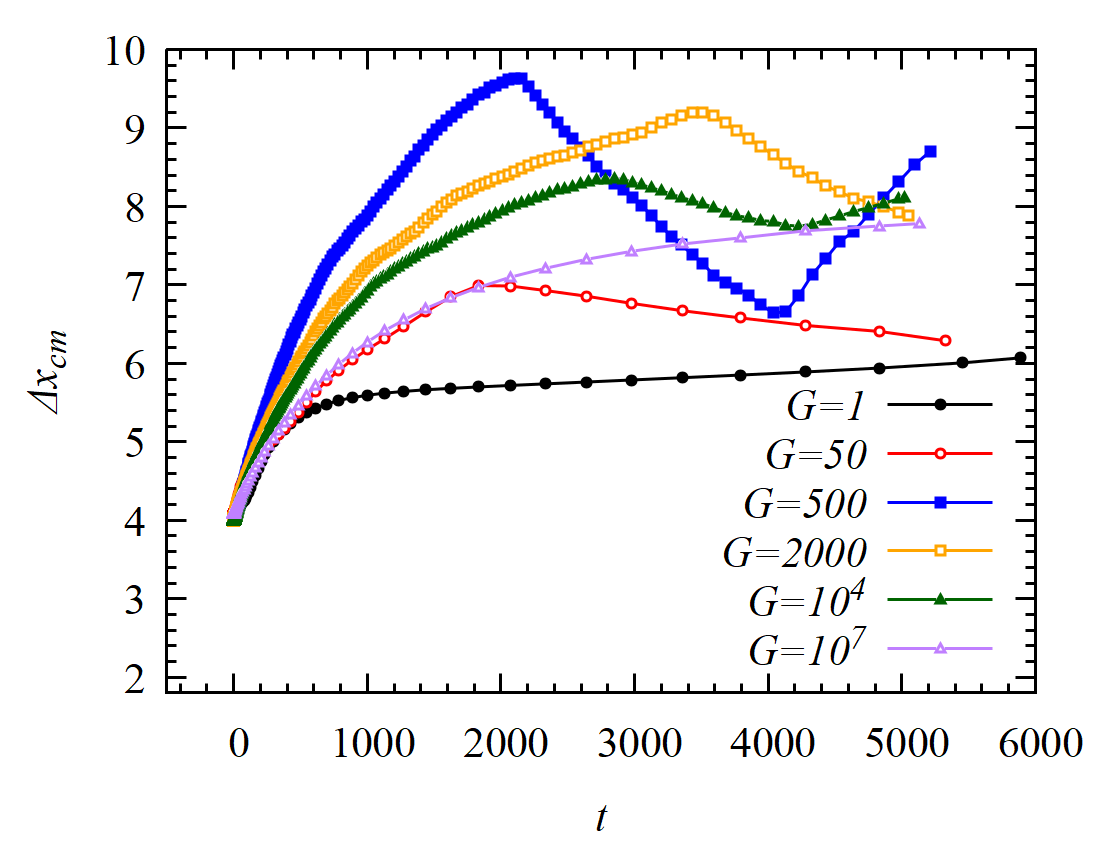} \hspace{0.5 cm}
	\includegraphics[width=0.47\textwidth]{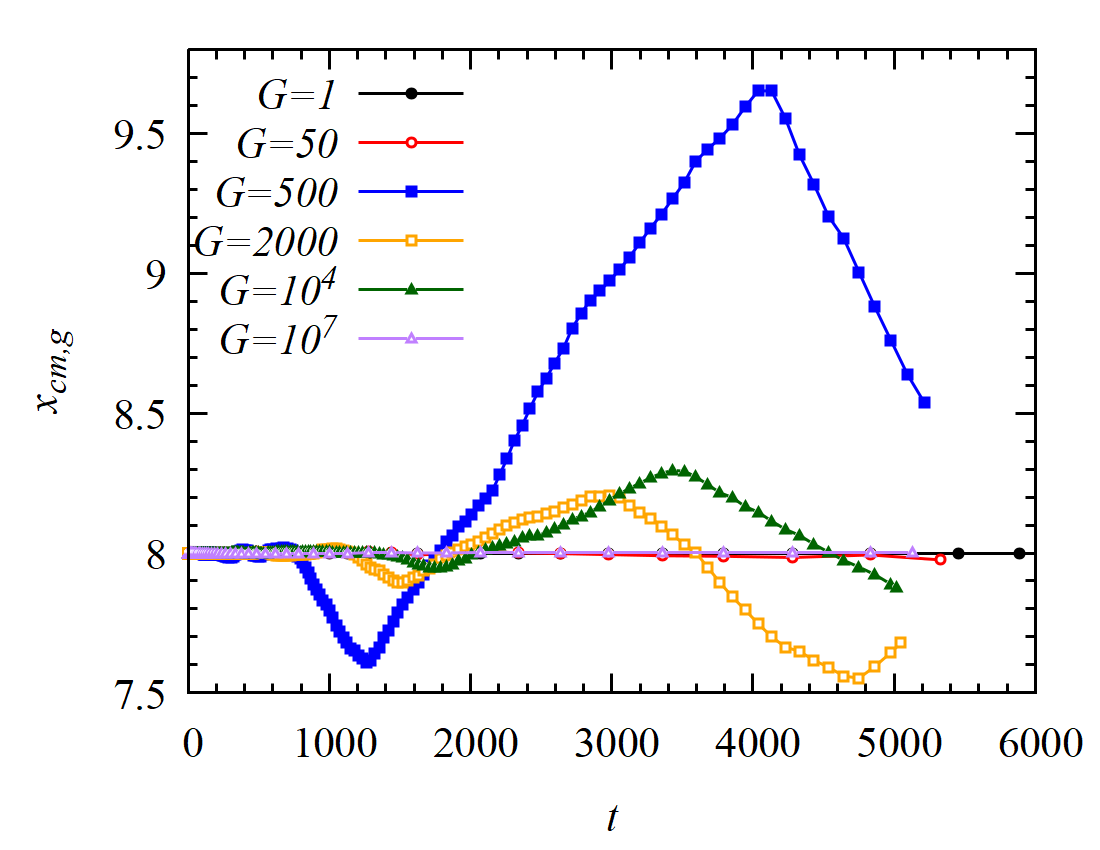} \\
    (c) \\
	\includegraphics[width=0.47\textwidth]{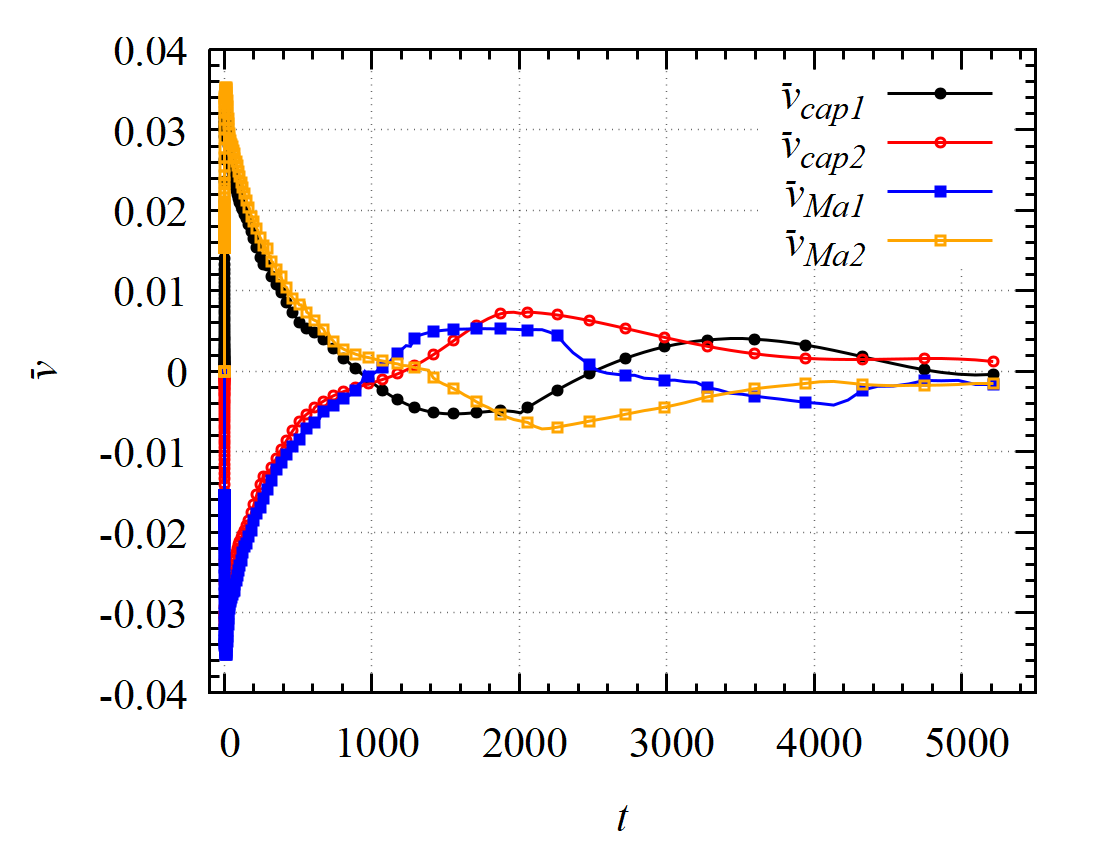}
    \caption{Time evolution of (a) the distance between the two centers of mass $\Delta x_{cm}$ and (b) the center of mass of the system $x_{cm,g}$, for $M_a=0.005$. (c) The contribution of the Marangoni stresses and capillary forces in the average $x$-velocity of droplet 1 and 2 for $G=500$ and $M_a=0.005$. The rest of the system parameters are the same with the 'base' case.}
    \label{fig:14}
\end{figure}

Next, we take into account the effect of thermocapillary stresses. In Fig. \ref{fig:13}, we depict the droplet dynamics for $M_a=0.005$ and three different values of $G=1,50,500$. Regardless of the elasticity strength of the substrate, it is shown that in all cases the droplets repulse from each other. This behaviour is markedly different from the one discussed previously in the absence of thermocapillary effects where the droplets are attracted to each other for substrates with intermediate or high values of $G$. As discussed in Fig. \ref{fig:9}, the Marangoni stresses play a dual role both acting as to compress the droplet footprint and also contributing to droplet repulsion. As shown in Fig. \ref{fig:14}a where we plot the distance $\Delta x_{cm}$ between the two droplets for a wide range of $G$ values for $M_a=0.005$, the latter contribution is dominant and thus always leading the droplets to repulse from each other at the early stages of the drying process. Nevertheless, we notice that at later stages and for substrates with intermediate values of $G$ (i.e. for $ G=50, 500, 2000, 10^4$) the droplet distance eventually starts decreasing indicating that the droplets are attracted to each other. This can be attributed to the fact that, at these late stages of the drying process, the droplet distance has increased considerably allowing the vapour concentration to acquire a more uniform profile along the interface of each droplet, leading to a more uniform evaporation flux and in turn to smaller temperature gradients. As a result, the thermal Marangoni stresses are significantly reduced, while the capillary forces induced by the substrate elasticity become dominant and drive the droplets closer to each other. For harder substrates, the capillary forces, as explained above, are weaker due to the fact that the substrate is less susceptible to elastic deformations and therefore the droplets continue to repulse due to the action of Marangoni stresses throughout the drying process. 

Interestingly, we also notice in Fig. \ref{fig:13}c, i.e. for a substrate with intermediate stiffness ($G=500$), that the droplets initially repulse, then they are attracted and eventually symmetry breaking takes place; in Fig. \ref{fig:13}c, a dashed arrow is drawn to indicate the motion of each droplet. As shown in Fig. \ref{fig:14}b where we plot the evolution of the global center of mass $x_{cm,g}$ with time, we find that the symmetry is preserved only for extremely soft and extremely stiff substrates, while for intermediate values of $G$ the droplets appear to 'walk' along the substrate. It should be noted that the emergence of this symmetry breaking at late stages of evaporation takes place spontaneously (i.e. at no specific time instant) and there is no preferred direction; it is triggered by numerical disturbances and appears to be stable and persistent with the increase in mesh resolution.

In Fig. \ref{fig:14}c, we make an effort to rationalize and elucidate the mechanisms responsible for the symmetry breaking, shown in Fig. \ref{fig:13}c (i.e. for $M_a=0.005$ and $G=500$). In order to examine the contribution of various forces, i.e. capillary, Marangoni and elastic, on the motion of the droplets, we evaluate their contributions to the mean velocity in the $x$-direction of each droplet, as follows 
\begin{equation}
\overline{v} =\frac{\int_{x_{cl}}^{x_{cr}} \int_0^h v_x  \,dz\,dx} {\int_{x_{cl}}^{x_{cr}} h \,dx} = \overline{v}_{cap,l}+\overline{v}_{cap,s}+\overline{v}_{Ma}+\overline{v}_{el}.
\end{equation}
The terms $\overline{v}_{cap,l}$ and $\overline{v}_{cap,s}$ denote the contributions from the capillary forces along the liquid-gas and liquid-solid interfaces, respectively, while the term $\overline{v}_{el}$ corresponds to the contribution of the elastic stresses and  $\overline{v}_{Ma}$ corresponds to the contribution of the Marangoni stresses; the analytical expressions for the various contributions are given in the Appendix \ref{appendix:raw3}. We note that for a number of different cases that we have examined the dominant contributions come from the capillary forces and the Marangoni stresses along the liquid-gas interface, evaluated by the terms $\overline{v}_{cap,l}$ and $\overline{v}_{Ma}$ respectively; $\overline{v}_{cap,s}$ and $\overline{v}_{el}$ were typically found to be two orders of magnitude smaller than $\overline{v}_{cap,l}$ and $\overline{v}_{Ma}$, and thus neglected here. Nevertheless, it is important to note that despite the fact that $\overline{v}_{el}$ is typically very small, substrate elasticity implicitly contributes to the effect of capillary forces of the liquid-gas interface through the induced deformation of the contact line region. The time evolution of $\overline{v}_{cap,l}$ and $\overline{v}_{Ma}$ is depicted in Fig. \ref{fig:14}c for both droplets; indexes 1 and 2 correspond to the droplet on the left and right, respectively. 


At early times (i.e. approximately for $t<800$), the contribution of the capillary and Marangoni stresses have similar magnitudes in both droplets and the symmetry is preserved (see Fig. \ref{fig:14}b). The capillary forces act antagonistically with the Marangoni stresses pushing the droplets in opposite directions. The droplets, however, repulse due to the slightly higher magnitude of the Marangoni contribution. 
At later times (i.e. for $t>800$), a disturbance in the local deformation of the solid causes an imbalance between the two droplets (see Fig. \ref{fig:14}c) leading to symmetry breaking and driving the system center of mass away from its initial position.
%
%
From Fig. \ref{fig:14}b, it becomes evident that whether some disturbance will lead to a symmetry breaking or not, is a matter of the substrate elasticity. On the one hand, when the substrate is soft, it is very flexible and its deformation is very large. The size of an arising disturbance is insignificant compared to the size of the total substrate deformation and, as a result, the system center of mass will remain constant and the symmetry will be preserved. On the other hand, in extremely stiff substrates, the deformation of the liquid-solid interface is very small, quickly damping any possible disturbance that could lead to an imbalance between the two droplets. 
However, at intermediate values of substrate elasticity, there can be a competition between this disturbance and the substrate deformation, which might eventually lead to an imbalance in the induced capillary and Marangoni stresses between the two droplets and thus to a symmetry breaking, if the size of the disturbance grows considerably as compared to the substrate deformation.

\section{Conclusions}\label{Conclusions}

In this paper we have studied the two-dimensional dynamics of a system of one or two droplets evaporating on a viscoelastic solid substrate. Lubrication theory is used to simplify the equations of mass, momentum, energy and the force balances applied in the liquid and the solid phases, considering the Kelvin-Voigt model to account for substrate viscoelasticity. Our model takes into consideration the effect of thermal Marangoni stresses, as well as the droplet interaction through both the compliant substrate and the surrounding vapour. The model accounts for the presence of the vapour employing a two-sided approach and considering the diffusion-limited model. The contact line is modelled assuming a precursor film ahead of the droplet. 

We have carried out a parametric study to investigate how the evaporation process, the flow dynamics and the interaction of droplets are affected by the physical properties of the compliant substrate (e.g. thickness, shear modulus) and vapour diffusion in the atmosphere affecting the local evaporation rate. In the case of a single droplet, it was found that for thinner substrates the elastic effects become decreasingly important and thus making the substrate thinner can be seen as equivalent to making it more rigid. Moreover, it is shown that on softer (or thicker) substrates the solid deforms affecting the wetting of the droplet and promoting evaporation in CCR mode, in line with experimental observations in the literature \citep{lopes_evaporation_2012, lopes_influence_2013,yu_experimental_2013,gerber_wetting_2019}; the CCA mode is observed for harder (or thinner) substrates. Lastly, the effect of evaporative cooling and the action of thermocapillary stresses lead to smaller droplet footprints, resulting in an overall decrease of the evaporation rate, capturing the trend observed in earlier studies \citep{Talbot_evaporation_2012,Schofield_lifetimes_2018,
Dunn_conductivity_2009} in the case of rigid substrates. On the other hand, in the case of a system of a pair of volatile droplets, it is shown that the droplets may communicate both through the viscoelastic substrate and the induced deformations of the liquid-solid interface and also through the vapour that diffuses in the atmosphere of the droplets. The delicate interplay between the elastic stresses in the substrate, the capillary pressure and the thermal Marangoni stresses determine the mode of droplet interaction. 

\begin{figure}
    \centering
    \includegraphics[width=1\textwidth]{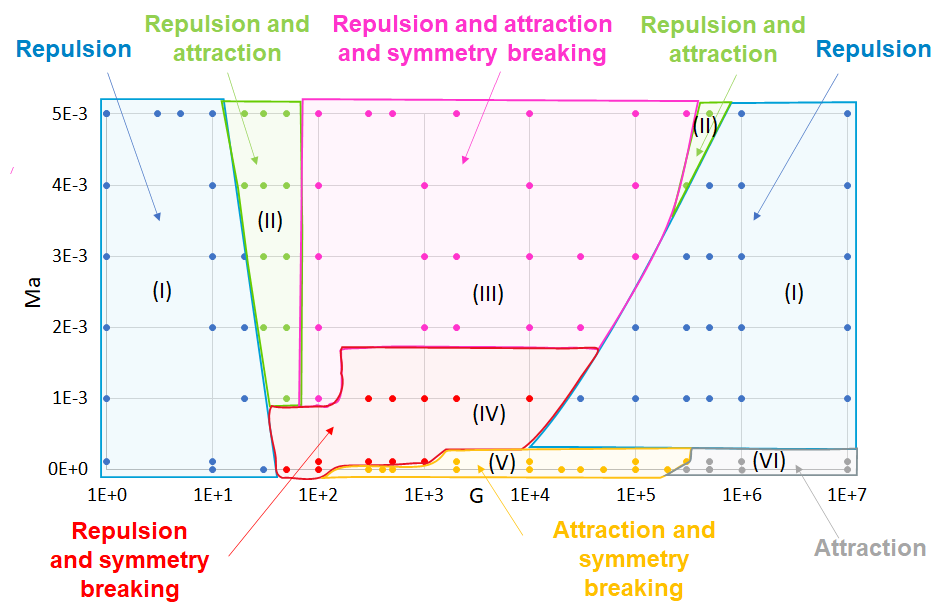}
    \caption{
    Map of the dynamic regimes depending on the value of Marangoni number, $M_a$, and substrate elasticity $G$. The rest of the system parameters are the same with the 'base' case. We note that the borders in this map have been added as a visual guide and are not precise.}
    \label{fig:15}
\end{figure}

To summarize the rich dynamics of this complex system, we produced the parametric map of the dynamic regimes depicted in Fig. \ref{fig:15}, varying the values of Marangoni number, $M_a$, and substrate elasticity $G$. In order to characterize the different regimes of the map, we consider that the system symmetry is broken when the center of mass of the system has moved  $\pm \SI{10}{\percent}$ of the initial maximum droplet height (i.e. 0.1 dimensionless distance) from its initial position. In the absence of thermocapillary stresses (i.e. $M_a=0$), the droplets repulse on soft substrates (region I) whereas they are attracted to each other (and even coalesce) on stiff substrates (region VI) with symmetry breaking arising in the case of substrates with intermediate stiffness (region V). 
Droplet coalescence that takes place for substrates with intermediate stiffness closely resemble the translation mode of droplet coarsening observed for intermediate elasticity in \cite{henkel_gradient-dynamics_2021}. The translation mode is mediated by elastic deformation recovering the inverted Cheerios effect \citep{karpitschka_liquid_2016}. 
On very stiff and very soft substrates, however, the dominance of the Ostwald ripening effect found by \cite{henkel_gradient-dynamics_2021} in the case of non-volatile droplets is significantly suppressed, due to the effect of evaporation that suppresses mass transfer between the droplets. 
Themocapillarity, on the other hand, apart from being responsible for the asymmetry of the wetting ridges on the two sides of each droplet between the two contact lines of each droplet, has also a drastic effect on the dynamics of the two droplets causing the repulsion of the droplets at the early stages of the drying process, irrespective of the stiffness of the viscoelastic solid (regions I, II and III). For substrates with intermediate stiffness, though, the droplet repulsion may also be followed by a phase of droplet attraction at later stages of the evaporating process (regions II and III), and under conditions to a symmetry breaking (region III); the symmetry is preserved either for very soft or very stiff substrates (regions I, II).
Spontaneous symmetry breaking was also found in the study of \cite{leong_droplet_2020} who examined the coalescence of inflating droplets on viscoelastic substrates, indicating that these systems are prone to instability as a result of the delicate interplay amongst elastic, capillary and Marangoni forces.

Our findings clearly indicate that the flow dynamics can be very interesting with important implications for the optimal design of soft substrates for controlled evaporation of droplets. Our comprehensive model can be easily extended to more realistic setups such as the simulation of multiple 3D droplets or more complex systems such as the evaporation of particle-laden droplets. We believe that the present work should be complemented in the future with detailed experimental studies. To the best of our knowledge, such studies are lacking, despite the vast experimental work that exists on evaporating droplets on rigid substrates. 

\section{Acknowledgements}
The authors gratefully acknowledge the financial support received from Hellenic Foundation for Research and Innovation (HFRI) and the General Secretariat for Research and Technology (GSRT), under Grant Agreement No. 792. We kindly thank the anonymous reviewers for their constructive comments.
\section{Declaration of interests}
The authors report no conflict of interest.

\section{Data availability statement}
The data that support the findings of this study are available from the corresponding author, upon reasonable request.

\appendix 
\section{Detailed derivation of the evolution equations for the soft substrate}
\label{appendix:raw}

The displacement in the $z$-direction, i.e. $u_z(x,t)$, can be evaluated by integrating the continuity equation for the solid, i.e. Eq. (\ref{s-continuity d}): 
\begin{equation}
u_z=b_4(x,t) - \bigg(\frac{\partial b_1}{\partial x} \frac{z^3}{3}+ \frac{\partial b_2}{\partial x} \frac{z^2}{2} + \frac{\partial b_3}{\partial x} z \bigg).
\label{uz1}
\end{equation}

In the above equation, $b_4(x,t)$ can be determined by using the fact that the displacement of the soft substrate at the interface with the rigid solid at $z=-H$ is zero and also using Eq. (\ref{b3}):
\begin{equation}
b_4 = \frac{\partial b_1}{\partial x} \bigg(\frac{2H^3}{3}\bigg)+ \frac{\partial b_2}{\partial x} \bigg(\frac{-H^2}{2}\bigg).
\label{b4}
\end{equation}

Introducing Eq. (\ref{b4}) into Eq. (\ref{uz1}) we get the following expression for $u_z$:
\begin{equation}
u_z = \frac{\partial b_1}{\partial x} \frac{1}{3} (2H^3 +3zH^2 -z^3) - \frac{\partial b_2}{\partial x} \frac{1}{2} {(z+H)}^2.
\label{uz2}
\end{equation}

At any time instant, the position of the liquid-solid interface, i.e. $\xi(x,t)$, is equal to the soft solid deformation at $z=0$ (i.e. the position of the undeformed liquid-solid interface) and therefore from Eq.(\ref{uz2}) we get:
\begin{equation}
\xi(x,t)=u_z(x,0,t)= \frac{\partial b_1}{\partial x} \frac{2H^3}{3} - \frac{\partial b_2}{\partial x} \frac{H^2}{2}.
\label{uz0}
\end{equation}

Turning our attention to the liquid phase, by integrating the $x$-component of the momentum, i.e. Eq. (\ref{l-momx d}), we get the following expression for $v_x$:
\begin{equation}
v_x = \frac{\partial p_l}{\partial x} \frac{z^2}{2} +f_1 z +f_2.
\label{vx1}
\end{equation}

Integrating the continuity equation for the liquid, i.e. Eq. (\ref{l-continuity d}), and using Eq. (\ref{vx1}) we get an expression for $v_z$:
\begin{equation}
v_z = -f_3- \int_{0}^{z} \frac{\partial v_x}{\partial x} \,dz.
\label{vz1}
\end{equation}

In Eq. (\ref{vx1}), $f_1(x,t)$ is determined by taking the derivative of $v_x$ with respect to $z$ and by setting $z=\zeta(x,t)$. Using the expression for $\frac{\partial v_x}{\partial z}\vert_{\zeta}$ from the tangential stress balance in the liquid phase, i.e. Eq. (\ref{l tang force bal d}), we get:
\begin{equation}
f_1 = \left( \epsilon^2 C_l \right)^{-1}\frac{\partial \sigma}{\partial x} - \zeta \frac{\partial p_l}{\partial x}.
\label{f1}
\end{equation}

$f_2(x,t)$ is determined by setting $z=\xi(x,t)$ on Eq. (\ref{vx1}) and using Eqs. (\ref{cont velocity x d}), (\ref{parab prof u}) and (\ref{b1t}):
\begin{equation}
f_2 = H \frac{\partial b_2}{\partial t} - H^2 \frac{\partial b_1}{\partial t} - \frac{\partial p_l}{\partial x}(\frac{\xi^2}{2} -\xi \zeta)- \left( \epsilon^2 C_l \right)^{-1} \xi \frac{\partial \sigma}{\partial x}. 
\label{f2}
\end{equation}
Consequently, the $x$-component of the liquid velocity, $v_x$, taking into account Eqs. (\ref{f1}) and (\ref{f2}) in Eq. (\ref{vx1}), equals to:
\begin{equation}
v_x = \frac{\partial p_l}{\partial x} \bigg(\frac{z^2}{2} - z \zeta - \frac{\xi^2}{2} +\zeta \xi\bigg) + \left( \epsilon^2 C_l \right)^{-1} \frac{\partial \sigma}{\partial x}(z-\xi) + H \frac{\partial b_2}{\partial t} - H^2 \frac{\partial b_1}{\partial t}.
\label{vx2}
\end{equation}

$f_3(x,t)$ is determined by setting $z=\xi(x,t)$ on Eq. (\ref{vz1}) and using Eqs. (\ref{cont velocity z d}), (\ref{uz0}) and (\ref{vx2}):
\begin{equation}
\begin{split}
f_3 = & -\frac{2H^3}{3} \frac{\partial^2 b_1}{\partial x \partial t} + \frac{H^2}{2} \frac{\partial^2 b_2}{\partial x \partial t} -\frac{\partial^2 p_l}{\partial x^2} \bigg (\frac{-\xi^3}{3} + \frac{\zeta \xi^2}{2} \bigg ) -\frac{\partial p_l}{\partial x} \bigg (\frac{\xi^2}{2} \frac{\partial \zeta}{\partial x} -\xi^2 \frac{\partial \xi}{\partial x} + \xi \zeta \frac{\partial \xi}{\partial x} \bigg ) \\ & + \left( \epsilon^2 C_l \right)^{-1} \left( \frac{\xi^2}{2} \frac{\partial^2 \sigma}{\partial x^2} + \xi \frac{\partial \xi}{\partial x} \frac{\partial \sigma}{\partial x} \right) - H \xi \frac{\partial^2 b_2}{\partial x \partial t} + H^2 \xi \frac{\partial^2 b_1}{\partial x \partial t}.
\end{split}
\label{f3}
\end{equation}

Substituting Eq. (\ref{f3}) in Eq. (\ref{vz1}) and using Eq. (\ref{vx2}) we finally get for $v_z$:
\begin{equation}
\begin{split}
v_z = & -\frac{\partial^2 p_l}{\partial x^2} \bigg (\frac{z^3}{6} -\frac{\zeta z^2}{2} - \frac{\xi^2 z}{2} + \zeta \xi z + \frac{\xi^3}{3} - \frac{\zeta \xi^2}{2} \bigg ) -\frac{\partial p_l}{\partial x} \bigg (-\frac{z^2}{2} \frac{\partial \zeta}{\partial x} -\xi z \frac{\partial \xi}{\partial x} \\ & + \xi z \frac{\partial \zeta}{\partial x} + \zeta z \frac{\partial \xi}{\partial x} - \frac{\xi^2}{2} \frac{\partial \zeta}{\partial x} +\xi^2 \frac{\partial \xi}{\partial x} - \xi \zeta \frac{\partial \xi}{\partial x} \bigg )- \left( \epsilon^2 C_l \right)^{-1} \frac{\partial^2 \sigma}{\partial x^2} \bigg(\frac{z^2}{2} - \xi z + \frac{\xi^2}{2} \bigg) \\ & + \left( \epsilon^2 C_l \right)^{-1} \frac{\partial \sigma}{\partial x} \bigg( z \frac{\partial \xi}{\partial x} - \xi \frac{\partial \xi}{\partial x} \bigg) + \frac{\partial^2 b_2}{\partial x \partial t} \bigg(-H z -\frac{H^2}{2} +H \xi \bigg) \\ & + \frac{\partial^2 b_1}{\partial x \partial t} \bigg( H^2 z+ \frac{2H^3}{3} - H^2 \xi \bigg).
\end{split}
\label{vz2}
\end{equation}

Taking the material derivative of both sides of $\xi=u_z(x,0,t)$ allows us to derive an evolution equation for $\xi(x,t)$. The material derivative of $\xi$ is derived using Eqs. (\ref{cont thermal flux d}) and (\ref{parab prof u}), while the material derivative of $u_z(x,0,t)$ is derived using Eq. (\ref{uz2}):
\begin{equation}
\frac{D\xi}{D t} = \frac{\partial \xi}{\partial t}+\frac{\partial \xi}{\partial x} \bigg( H \frac{\partial b_2}{\partial t}- H^2 \frac{\partial b_1}{\partial t}\bigg),
\label{mat der ksi}
\end{equation}
\begin{equation}
\frac{D u_z}{D t}\bigg \vert_{z=0}  = \frac{\partial u_z}{\partial t}\bigg \vert_{z=0}=\frac{\partial}{\partial t} \bigg(\frac{2H^3}{3} \frac{\partial b_1}{\partial x}- \frac{H^2}{2} \frac{\partial b_2}{\partial x}\bigg).
\label{mat der uz}
\end{equation}

\section{Effect of domain size and boundary conditions}
\label{appendix:raw2}

\renewcommand\thefigure{\thesection\arabic{figure}}    
\setcounter{figure}{0}    

\begin{figure}
	\centering
	\vspace{1cm}
    (a) \hspace{0.47\textwidth} (b) \\
	\includegraphics[width=0.47\textwidth]{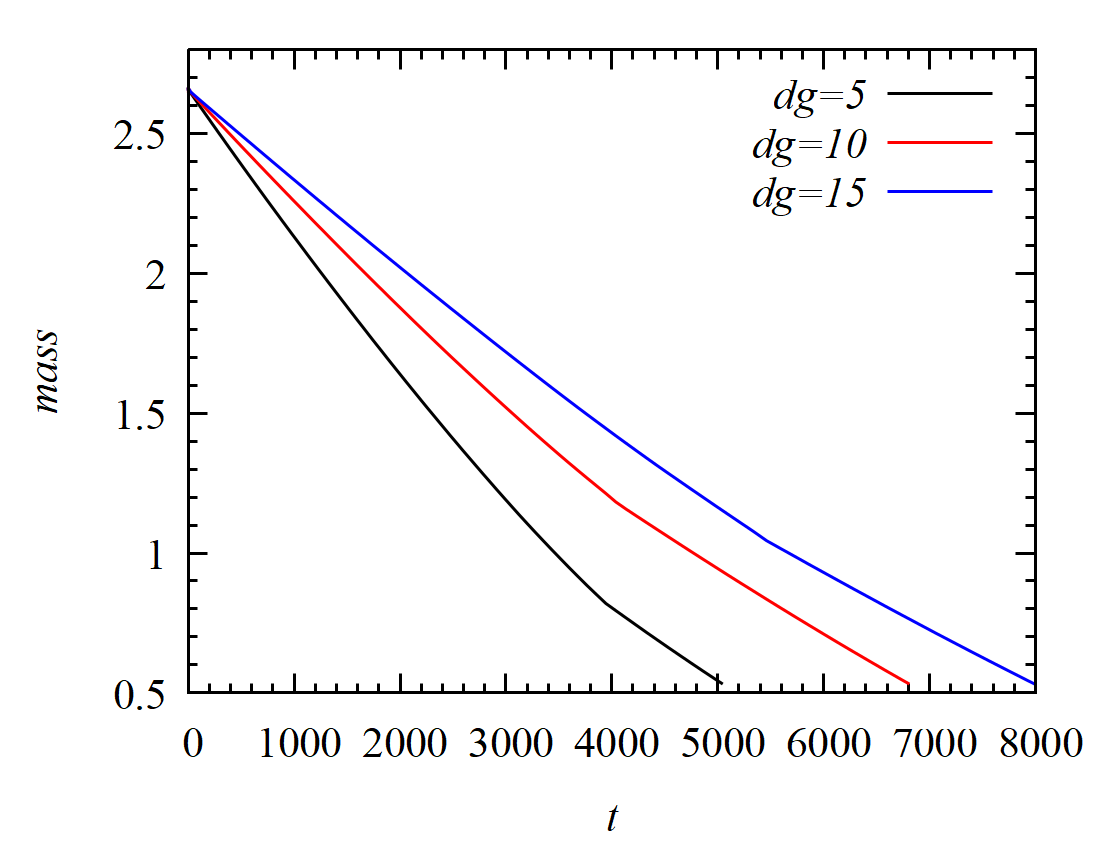} \hspace{0.5 cm}
	\includegraphics[width=0.47\textwidth]{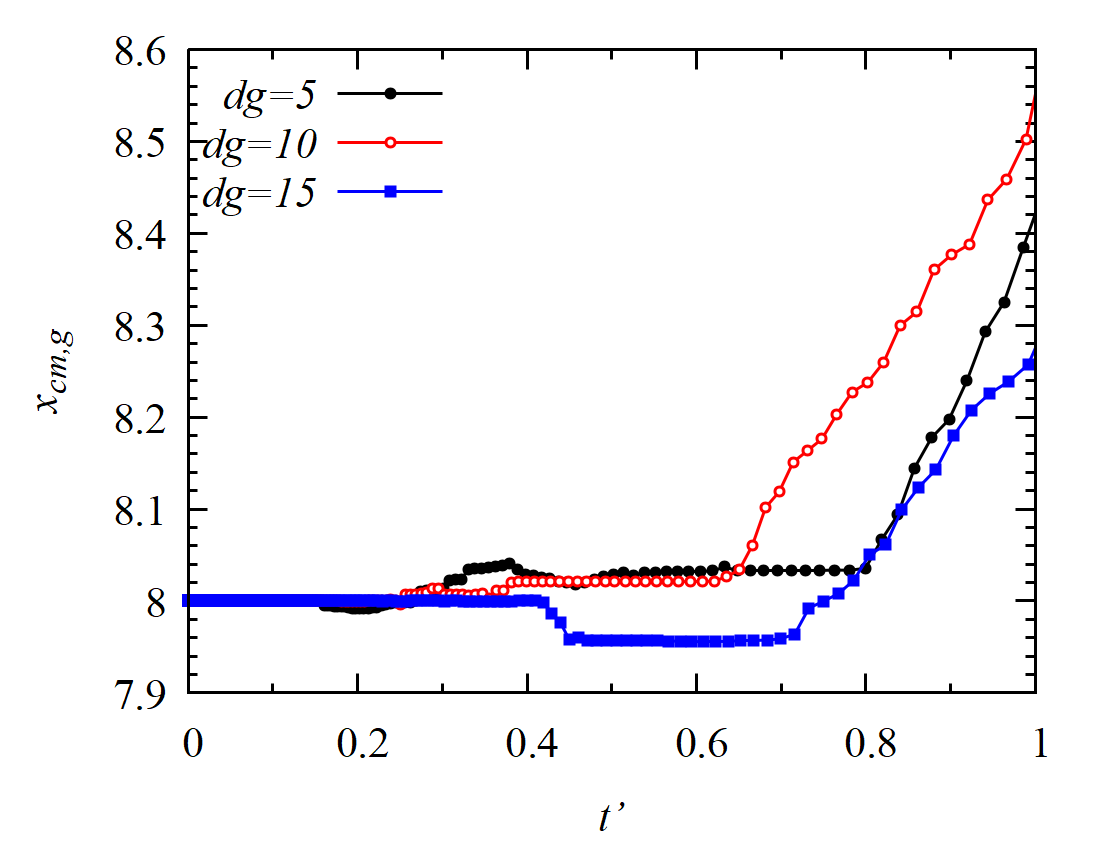}\\ 
    (c) \hspace{0.47\textwidth} (d) \\
	\includegraphics[width=0.47\textwidth]{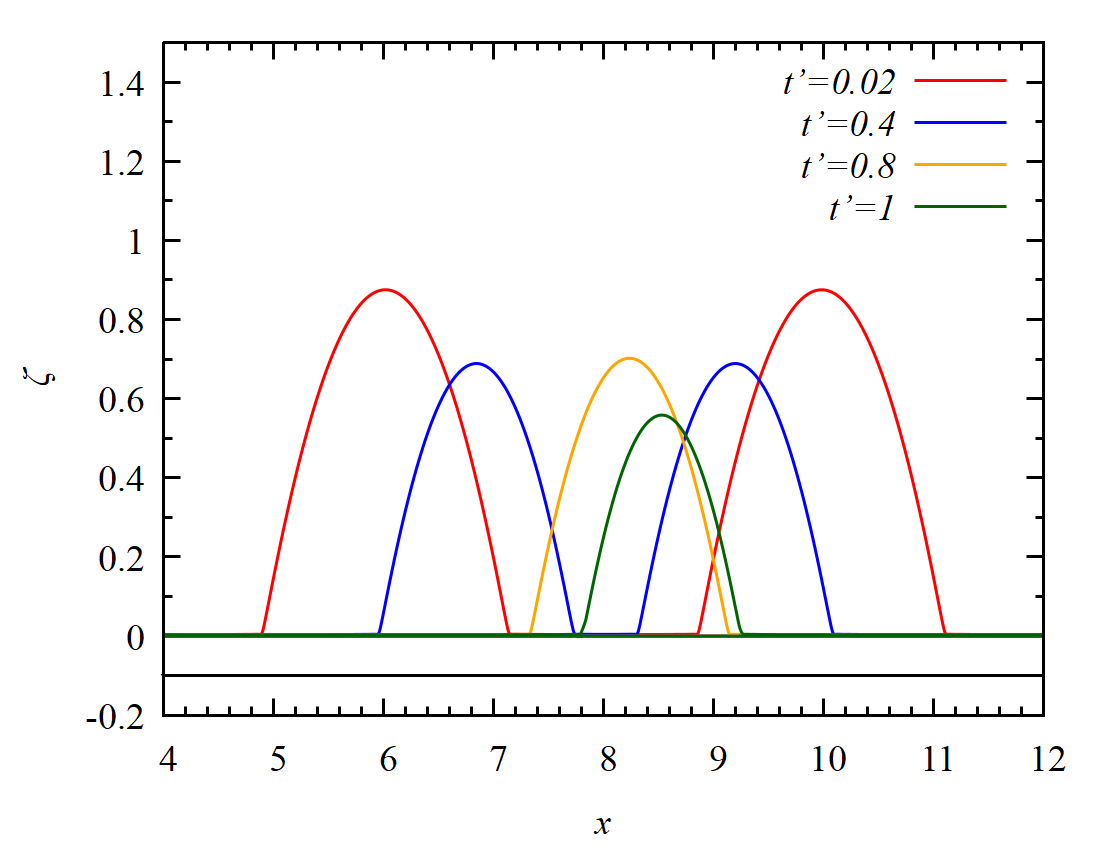}
    \hspace{0.5 cm}
	\includegraphics[width=0.47\textwidth]{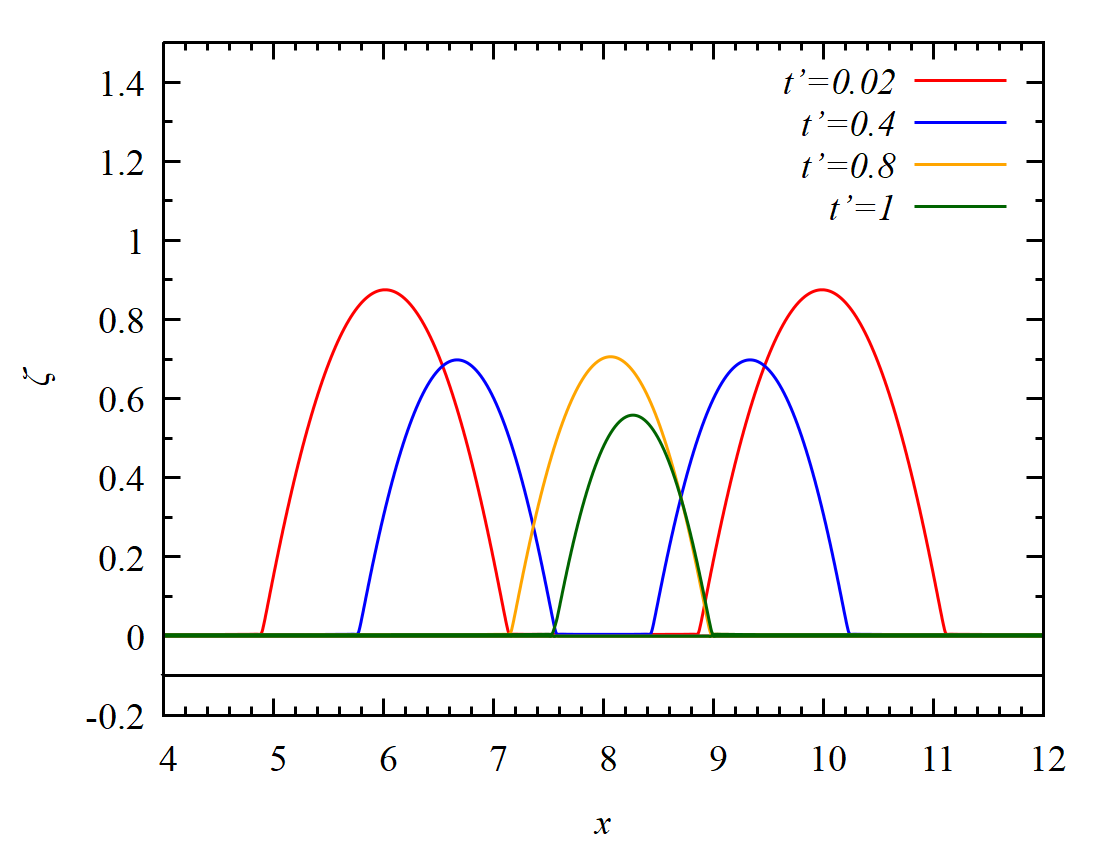} 
     \caption{Time evolution of (a) the system mass and (b) the center of mass of the system $x_{cm,g}$ varying the height $d_g$ of the gas phase. Time evolution of the liquid-air ($\zeta$) and the liquid-solid ($\xi$) interfaces for 2 droplets with (c) $dg=10$ ($t_{ev}= 6810$) and (d) $d_g=15$ ($t_{ev}=7993$) respectively. In all panels $M_a=0$ and $G=10^5$. The rest of the system parameters are the same with the 'base' case.}
     \label{fig:16}
\end{figure}
\begin{figure}
	\centering
	\vspace{1cm}
    (a) \hspace{0.47\textwidth} (b) \\
	\includegraphics[width=0.47\textwidth]{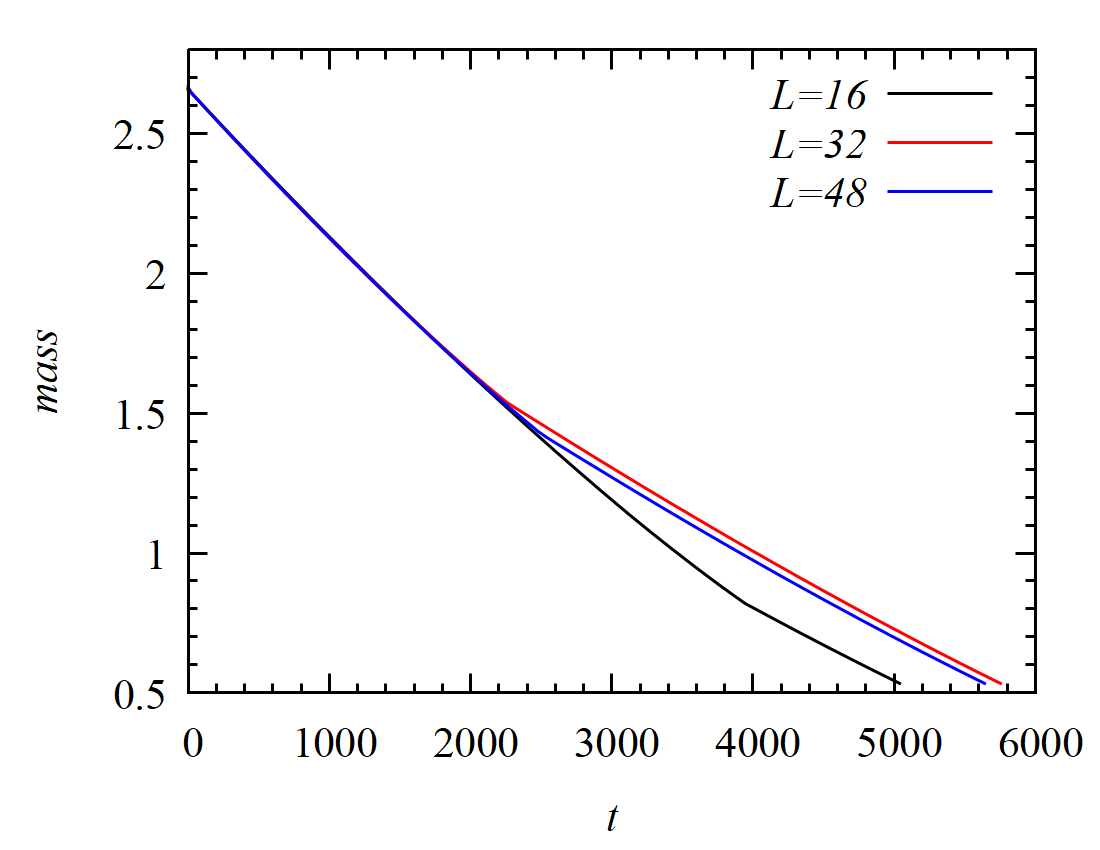} \hspace{0.5 cm}
	\includegraphics[width=0.47\textwidth]{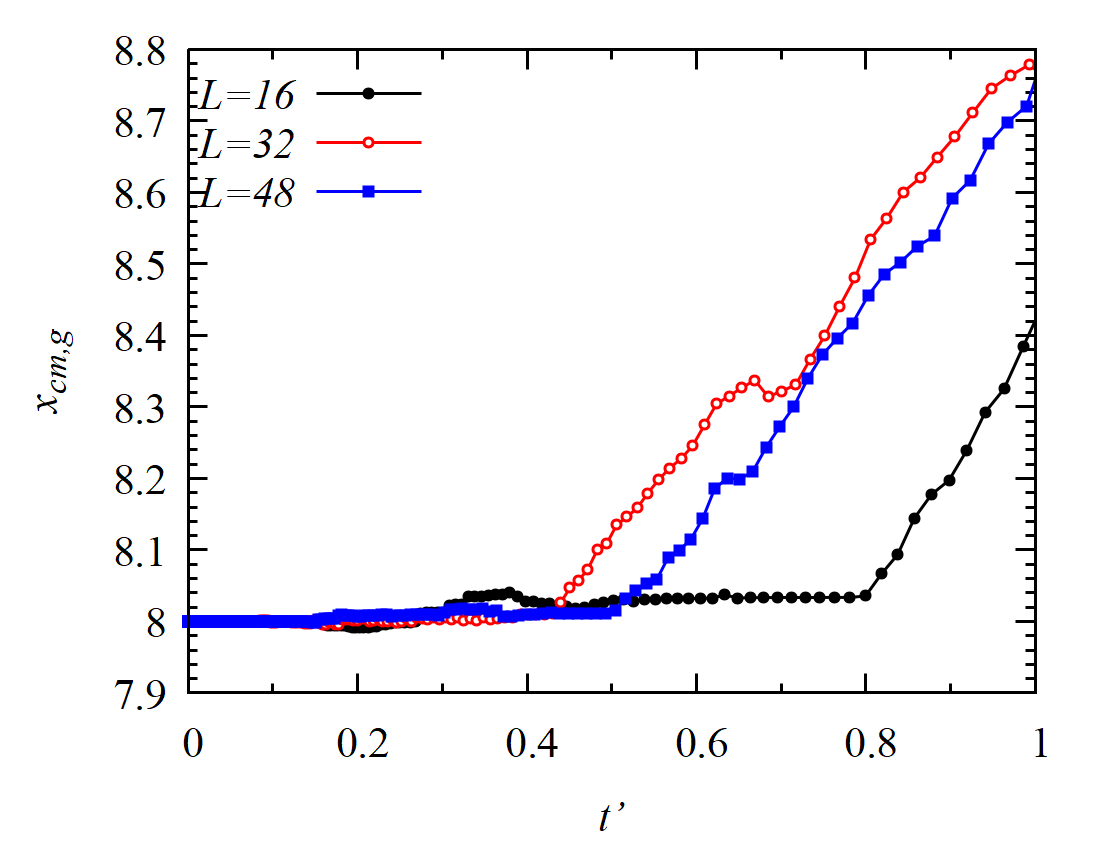}\\ 
     \caption{Time evolution of (a) the system mass and (b) the center of mass of the system $x_{cm,g}$ varying the length $L$ of the solid substrate, for $M_a=0$ and $G=10^5$. The initial value of $x_{cm,g}$ has been moved to $L=8$ for all cases for presentational purposes. The rest of the system parameters are the same with the 'base' case.}
    \label{fig:17}
\end{figure}
\begin{figure}
	\centering
	\vspace{1cm}
    (a) \hspace{0.47\textwidth} (b) \\
	\includegraphics[width=0.47\textwidth]{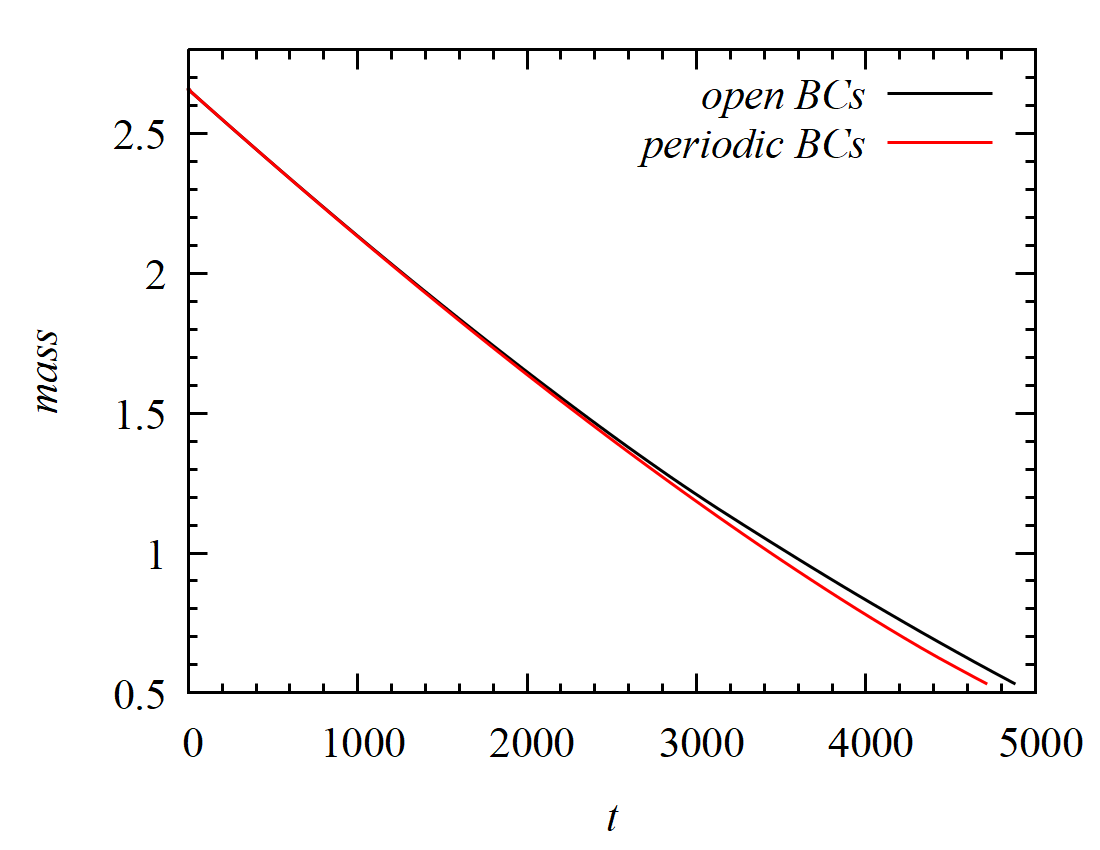} \hspace{0.5 cm}
	\includegraphics[width=0.47\textwidth]{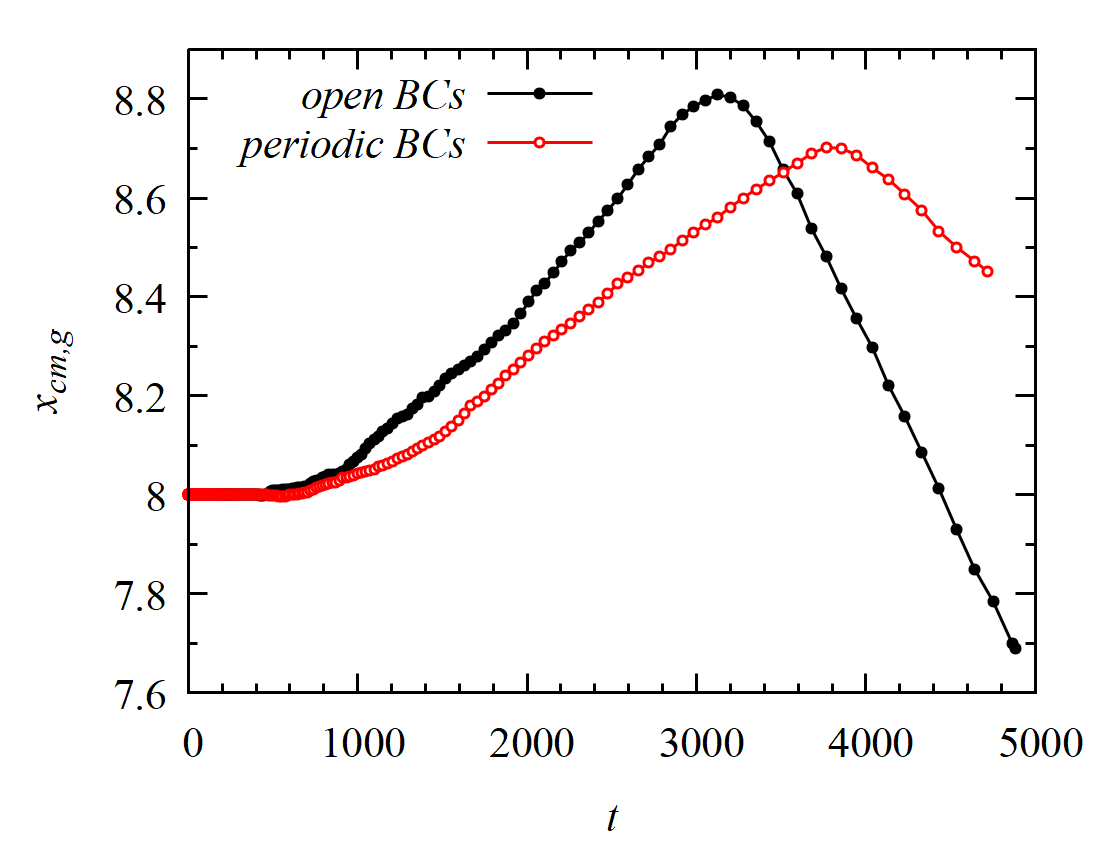}\\ 
    (c) \hspace{0.47\textwidth} (d) \\
	\includegraphics[width=0.47\textwidth]{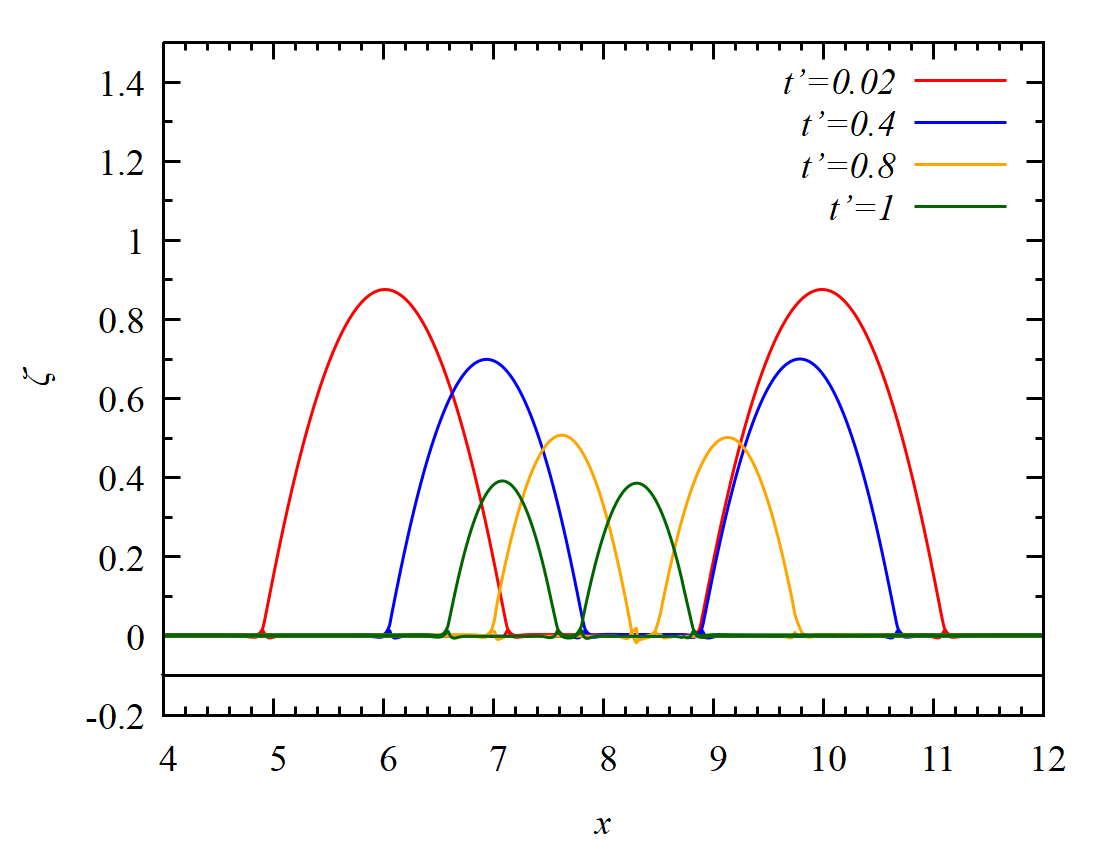}
    \hspace{0.5 cm}
	\includegraphics[width=0.47\textwidth]{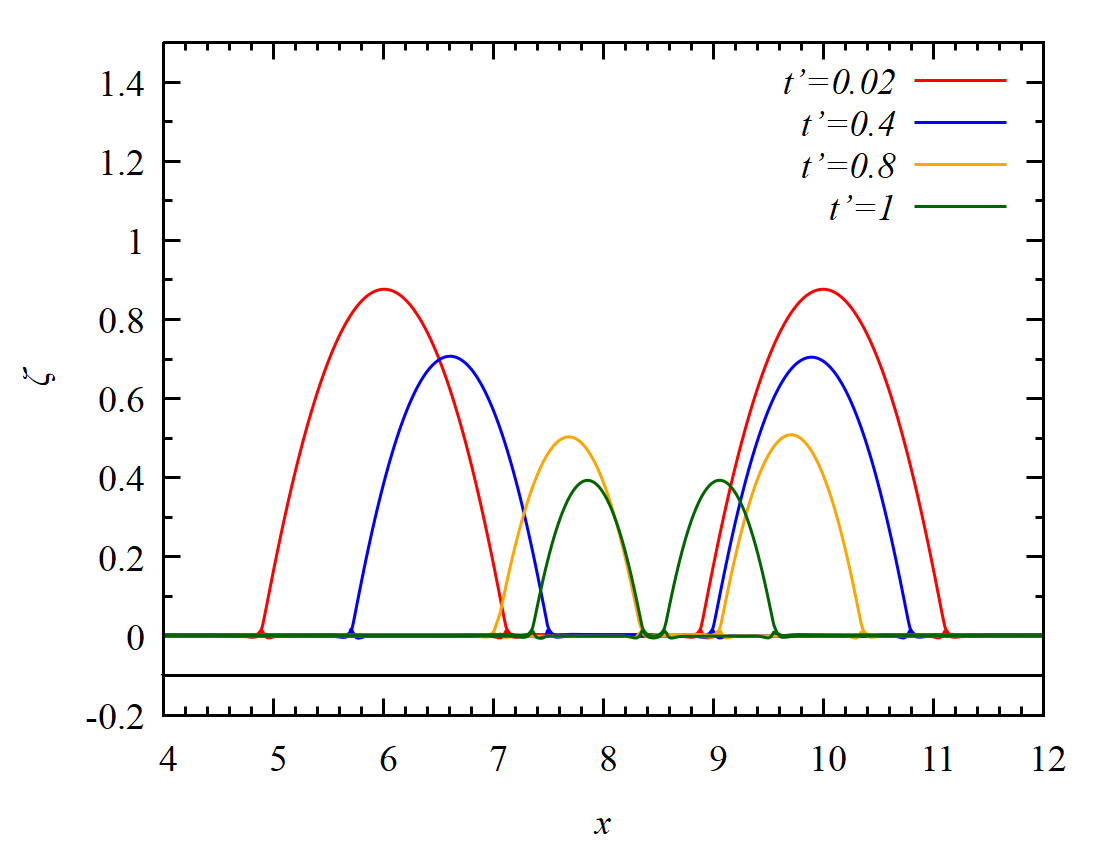}  
     \caption{Time evolution of (a) the system mass and (b) the center of mass of the system $x_{cm,g}$ applying open and closed boundary conditions. Time evolution of the liquid-air ($\zeta$) and the liquid-solid ($\xi$) interfaces for 2 droplets with (c) open boundary conditions ($t_{ev}= 4882$) and (d) closed boundary conditions ($t_{ev}=4716$) respectively. In all panels $M_a=0$ and $G=300$.The rest of the system parameters are the same with the 'base' case.}
    \label{fig:18}
\end{figure}

Here, we examine the effect of the boundary conditions on the observed system dynamics. To this end, we vary the size of the domain both in $x$ and $z$ directions, as well as considering the application of periodic boundary conditions in the $x$-direction.

In Fig. \ref{fig:16} we examine the effect of the position of the far-field boundary condition in $z$-direction. As explained in section \ref{Gas phase}, the most natural choice would be to impose the boundary condition of constant vapour concentration at $z \rightarrow \infty$. Here, however, we follow a similar approach to \cite{Schofield_2020} considering a finite domain of the gas phase and the far-field condition is replaced by a similar Dirichlet condition at a distant, but finite, boundary. To examine the effect of this simplification, we vary the distance, $d_g$, where the Dirichlet condition is imposed. In Fig. \ref{fig:16} we depict the effect on the evolution of both the total droplet mass and the position of the system center of mass. As shown in Fig. \ref{fig:16}a, decreasing $d_g$ results in faster evaporation due to the higher concentration gradient near the liquid-gas interface that is  implicitly imposed and the fact that evaporation rate depends on the rate of diffusion; we observe a rather slow convergence of the total evaporation time with increasing values of $d_g$.

In order to examine the effect of $d_g$ on the dynamics of a pair of droplets, we focus on the case presented in Fig. \ref{fig:10}c for $M_a=0$ and $G=10^5$ and $d_g=5$. In Fig. \ref{fig:16}c and \ref{fig:16}d we depict the same case for $d_g=10$ and $15$, respectively, at times that correspond to the same scaled time, $t'$;  we note the great similarity between the three cases, which is also reflected on the evolution of the position of the center of mass, presented in Fig. \ref{fig:16}b. Clearly, the position of the far-field condition does not affect the qualitative characteristics of either droplet coalescence or the observed symmetry breaking.  %

To investigate the effect of the boundary condition in the $x$-direction we follow two different routes. The first is to simply examine the effect of the domain length, $L_x$ for the same case examined in Fig. \ref{fig:10}c. In Fig. \ref{fig:17} we see that the length mildly affects the predicted evaporation time but does not have a significant impact on the the rest qualitative characteristics of the flow (e.g. see Fig. \ref{fig:17}b). Secondly, we solve a case for $M_a=0$ and $G=300$ imposing periodic boundary conditions at the edges of the domain to check whether in cases where a symmetry breaking appears this might be due to the symmetry being already broken by the boundary conditions; the remaining parameters are kept the same with the 'base' case. As shown in Fig. \ref{fig:18}, imposing either open or closed boundary conditions does not affect significantly the qualitative characteristics of the system dynamics and the symmetry breaking persists irrespective of the applied boundary conditions. 

\section{Mean velocity}
\label{appendix:raw3}

In order to compute the mean velocity of each droplet we first computed the average $x$-velocity, $v_{x,ave}$ (using Eq.(\ref{vx2})), as follows
\begin{equation}
\begin{split}
v_{x,ave} =\frac{1}{h} \int_{0}^{h} v_x \,dz = & \frac{\partial p_l}{\partial x} \bigg(\frac{h^2}{6} - \frac{h \zeta}{2} - \frac{\xi^2}{2} +\zeta \xi\bigg) + \left( \epsilon^2 C_l \right)^{-1} \frac{\partial \sigma}{\partial x}(\frac{h}{2}-\xi) \\ & + H \frac{\partial b_2}{\partial t} - H^2 \frac{\partial b_1}{\partial t}.
\end{split}
\end{equation}

The total mean velocity in $x$-direction of each droplet can be evaluated by the following expression:

\begin{equation}
\begin{split}
\overline{v} =\frac{\int h v_{x,ave}  \,dx}{\int h \,dx}  = & \frac{1}{\int h \,dx} \int \bigg[ h \frac{\partial p_l}{\partial x} \bigg(\frac{h^2}{6} - \frac{h \zeta}{2} - \frac{\xi^2}{2} +\zeta \xi\bigg) + \left( \epsilon^2 C_l \right)^{-1} \frac{\partial \sigma}{\partial x} h \bigg( \frac{h}{2}-\xi \bigg) \\ & + h \bigg(H \frac{\partial b_2}{\partial t} - H^2 \frac{\partial b_1}{\partial t}\bigg) \bigg] \,dx =\overline{v}_{cap,l}+\overline{v}_{cap,s}+\overline{v}_{Ma}+\overline{v}_{el}.
\end{split}
\end{equation}

The above integrals are solved between the two contact lines of each droplet. The total mean velocity of each one of the two droplets is the sum of the contribution of the capillary forces, the Marangoni stresses and the forces due to the elastic substrate. The contribution of the solid substrate, both in terms of capillary forces and in terms of elasticity is $O(10^{-4})$. More specifically, using Eq.(\ref{b1t}) and Eq.(\ref{b2t}), we get:
\begin{equation}
\overline{v}_{cap,l} =\frac{1}{\int h \,dx} \int h \frac{\partial p_l}{\partial x} \bigg(\frac{h^2}{6} - \frac{h \zeta}{2} - \frac{\xi^2}{2} +\zeta \xi - \frac{H}{m} h \bigg) \,dx,
\end{equation}
\begin{equation}
\overline{v}_{cap,s} =\frac{1}{\int h \,dx} \int \frac{-h H^2}{2m} \frac{\partial p_s}{\partial x} \,dx,
\end{equation}
\begin{equation}
\overline{v}_{Ma} =\frac{1}{\int h \,dx} \int \left( \epsilon^2 C_l \right)^{-1} \frac{\partial \sigma}{\partial x} h \bigg( \frac{h}{2}-\xi+\frac{H}{m} \bigg)  \,dx,
\end{equation}
\begin{equation}
\overline{v}_{el} =\frac{1}{\int h \,dx} \int h \frac{H G}{m} (H b_1 -b_2)  \,dx.
\end{equation}

\bibliography{Exported}
\bibliographystyle{jfm}
\end{document}